\documentclass[aps,prb,twocolumn,english,superscriptaddress,nolongbibliography]{revtex4-2}
\usepackage{amsmath}
\usepackage{amssymb,scalerel}
\usepackage{graphicx}
\usepackage[colorlinks,bookmarks=false,citecolor=red,linkcolor=blue,urlcolor=blue]{hyperref}
\usepackage{lipsum, color}
\usepackage{wasysym}
\usepackage{orcidlink}
\usepackage{ulem}
\def\Hhat{\hat{H}}

\def\r{{\ve r}}

\def\S{{\boldsymbol S}}
\def\Q{{\boldsymbol Q}}

\newcommand{\be}{\begin{equation}}
\newcommand{\ee}{\end{equation}}
\newcommand{\bea}{\begin{eqnarray}}
\newcommand{\eea}{\end{eqnarray}}
\newcommand{\beas}{\begin{eqnarray*}}
\newcommand{\eeas}{\end{eqnarray*}}
\newcommand{\ve}[1]{\boldsymbol{#1}}

\begin{document}
\title{Phases and phase transitions of an $S=3/2$ chain on metallic and semi-metallic surfaces}%
\author{Bimla Danu \orcidlink{0000-0002-5210-1511}}
\affiliation{Institut f\"ur Theoretische Physik und Astrophysik, Universit\"at W\"urzburg, 97074 W\"urzburg, Germany}
\author{Fakher F. Assaad \orcidlink{0000-0002-3302-9243}}
\affiliation{Institut f\"ur Theoretische Physik und Astrophysik and W\"urzburg-Dresden Cluster of Excellence ct.qmat, Universit\"at W\"urzburg, 97074 W\"urzburg, Germany}
\date{\today}
\begin{abstract}
Motivated by recent scanning tunneling microscopy experiments on chains of Co adatoms on Cu surfaces, we investigate the physics of a spin-$3/2$
 Heisenberg chain with single-ion anisotropy ($D$) on metallic and semi-metallic surfaces. In the strong Kondo coupling ($ J_k$) limit, a perturbative 
 analysis maps the system onto a Haldane spin-1 chain with single-ion anisotropy, ferromagnetically coupled to the metallic surface. This Haldane state, arising from the underscreening of the $S = 3 / 2$ chain, is stable against small values of $D$ and is characterized by topological edge modes.
The nature of the $D$-driven transitions out of this state depends on the metallic environment. 
Coupling to a metal (semi-metal) constitutes a relevant (irrelevant) perturbation at the decoupled fixed point between the spin-1 chain and the two-dimensional electron gas. 
 In the large positive $D$
 limit, the system maps onto an anisotropic spin-$ 1 / 2 $ Kondo system that has been studied. For large negative $D$, in the Ising phase, the spins are frozen.
For small values of $J_k$, the nature of the metallic phase plays a dominant role. On a two-dimensional semi-metal, the Kondo coupling is irrelevant at the decoupled fixed point, $J_k = 0$, leading to a Kondo breakdown phase at weak coupling, irrespective of $D$. In contrast, on a two-dimensional metal, the resulting dissipative Ohmic bath acts as a marginally relevant perturbation at the decoupled fixed point, inducing antiferromagnetic ordering along the spin chain. In this case, $D$ drives a spin-flop transition between Ising and XY ordered phases. At $D = 0$, we observe continuous transitions between the Kondo breakdown or dissipation-induced long-range ordered phases and the underscreened Haldane phase. This understanding of the phase diagrams is supported by scaling arguments as well as  by unbiased, sign-free auxiliary-field quantum Monte Carlo 
simulations.
\end{abstract}
\maketitle

\tableofcontents

\section{Introduction}

Spin chains provide a fundamental platform for exploring exotic phases of quantum matter, where quantum fluctuations and topology give rise to phenomena that transcend classical expectations~\cite{giamarchi2003}. Topology provides a key distinction  between half-integer and integer spin chains due to presence of 
a $\vartheta$-term at $ \vartheta = 2 \pi S $ in the low-energy field theory~\cite{PRLHaldane1983, Haldane1982PLA, AffleckHaldanePRB1987}. 
For half-integer spins, the system remains gapless and is described by a Luttinger liquid with algebraic correlations. In contrast, integer-spin chains possess a gapped excitation spectrum. The former state is realized exactly in the Affleck-Kennedy-Lieb-Tasaki (AKLT) model~\cite{AKLT1987}. This dichotomy underscores the profound role of topology in low-dimensional magnets and the emergence of symmetry-protected topological (SPT) phases~\cite{AKLT1987, Pollmann2012, Chen2011}.

Canonical heavy fermion (HF) systems consist of lattices of spin-$1/2$
 magnetic impurities embedded in a metallic host~\cite{Coleman_2001,Qimiao2001,Qimiao_Si2014}. The key interaction in these systems is the antiferromagnetic Kondo coupling between the spins and conduction electrons. In the HF phase, the spins are screened by the conduction electrons via the Kondo effect~\cite{hewson1993,WilsonRMP1975,Yosida1991,Kondo1964,Doniach1987}. This entanglement process generates a new itinerant electronic degree of freedom--the so-called composite fermion~\cite{RaczkowskiPRL2019,Danu2021,RaczkowskiPRB2022}--which participates in the Luttinger volume count~\cite{Yamanaka1997,OshikawaPRL2000}. This phase corresponds to the heavy fermion Fermi liquid (HFL), with a coherence temperature set by the energy scale at which the composite fermion emerges~\cite{Georges00,Assaad04a,Beach08}. Below the coherence temperature, the heavy fermion quasiparticles can form various states of matter, such as topological Kondo insulators~\cite{Dzeroprl2010,MezioPRB2015,Nakajima20016,JWerner13,JWerner14} and Weyl-Kondo semimetals~\cite{Dzsaber21}.

Heavy fermion criticality deals with the instabilities of the HFL phase. For instance, the Ruderman-Kittel-Kasuya-Yosida (RKKY) interaction between impurity spins can lead to magnetic ordering \textit{without} destroying the composite fermion~\cite{Hertz1976, Millis1993,Frank22,LiuZH22a}. This magnetic transition in metallic environments leads to non-Fermi-liquid behaviour at the critical point. Another route to destabilise the HFL is by breaking up the composite fermion. The corresponding Kondo breakdown (KBD) transition is a non-symmetry-breaking quantum phase transition (QPT) that falls outside the traditional Landau-Ginzburg scenario and is often interpreted as an orbital-selective Mott transition~\cite{Vojta10}, where the hybridization between localized $f$-electrons and conduction electrons vanishes, rendering the composite fermion incoherent. This leads to a sudden reconstruction of the Fermi surface and a collapse of heavy quasiparticles. One of the central theoretical tools in understanding such phases is the Oshikawa-Luttinger sum rule, which links the Fermi volume to the total number of electrons, including both localized spins and itinerant electrons~\cite{Yamanaka1997, OshikawaPRL2000}. This framework helps distinguish between the conventional HFL, where local moments are Kondo screened and included in the Fermi surface, and the fractionalized Fermi liquid (FL$^*$)~\cite{TSenthil2003}. For an odd number of spin impurities per unit cell, the FL$^*$ phase violates Luttinger’s theorem. The way out of this dilemma is to assume that the impurity spins form a topological phase of matter, as realized by a deconfined gauge theory~\cite{TSenthil2003}. Although the KBD transition is relatively well understood in single-impurity models~\cite{PRLWithoff1990, PRBFritz2004, Vojta10}, its extension to Kondo lattices presents a richer and more intricate landscape. Theoretical approaches, including large-N methods~\cite{Coleman_2001}, extended dynamical mean-field theory (EDMFT)~\cite{Qimiao_Si2014}, and advanced numerical simulations~\cite{JHofmann2019, Danu2020}, have made progress, but a full understanding of KBD quantum criticality in Kondo lattice models remains a fundamental and ongoing challenge.  KBD  leads to strange-metal behaviour~\cite{Qimiao2001,Mazza24,Chen23}, which is a generic feature of many correlated electron systems, including high-temperature superconductors~\cite{Bonetti25}.

Spin chains on metallic surfaces provide a model system to study various phenomena pertaining to heavy fermion physics~\cite{LobosPRB2012,Danu2022,Danu2020,Danu2019}. While in these low-dimensional systems the interaction does not lead to long-range order, the metallic surface provides a source of Ohmic dissipation that can stabilize long-range magnetic order~\cite{Weber2021}. Recent experimental advances have made the theoretical investigation of artificial spin lattices in metallic environments particularly timely. In particular, scanning tunneling microscopy (STM) has enabled  atom-by-atom construction of quantum spin chains by manipulating magnetic adatoms--such as cobalt (Co)--on two-dimensional (2D) metallic surfaces like Cu(100) or Au(111)~\cite{Otte2008, Spinelli2015, MTernes2015, Toskovic2016, Moro2019, Danu2019}. Co adatoms on a Cu$_2$N/Cu(100) surface carry a spin-$3/2$ moment, characterized by strong uniaxial hard-axis anisotropy ($ D \gg 0$), along with tuneable magnetic exchange interactions. Since each Co adatom is also locally Kondo-coupled to the conduction electrons of the Cu substrate, this system provides a prototypical realization of a Kondo-Heisenberg lattice with dimensional mismatch: a one-dimensional (1D) spin chain embedded in a 2D electron bath. In the large positive $D$ limit, the Schrieffer-Wolff transformation (SWT) maps the spin-$3 /2$  Heisenberg chain onto an effective spin-$1 / 2$
 XXZ chain. Recently, there have been intriguing claims of a novel \textit{spinaron} quasiparticle emerging from a single Co adatom on Cu(111) and Au(111) surfaces~\cite{Bouaziz2020, Friedrich2024}. Other promising avenues for building artificial Kondo systems include organic radicals on metallic surfaces~\cite{LiE24}.

In this paper, our main focus is to investigate the physics of  an $S=3/2$ chain of Co adatoms on 2D metallic and semi metallic surfaces.  We consider the following Hamiltonian to investigate this setup,
\begin{eqnarray}
\hat{H}=&&-t \sum_{\langle  {\boldsymbol i}, {\boldsymbol j}\rangle } \Big( {\boldsymbol{\hat c}}^{\dagger}_ {\boldsymbol i} \,  e^{\frac{2 \pi i}{\Phi_0} \int_{\boldsymbol i}^{\boldsymbol j} {\boldsymbol A} \cdot {\boldsymbol d} {\boldsymbol l}} {\boldsymbol{\hat c}}^{\phantom\dagger}_{\boldsymbol j}   +\text{H.c.} \Big) 
\nonumber \\ 
& & + \frac{J_k}{2} \sum^L_{\boldsymbol r=1} {\boldsymbol {\hat c}}^{\dagger}_{\boldsymbol r} {\boldsymbol \sigma}  {\boldsymbol {\hat c}}_{\boldsymbol r} \cdot  {\boldsymbol{\hat S}}_{\boldsymbol r} \nonumber \\&&+J_h\sum^L_{{\boldsymbol r}=1} {\boldsymbol {\hat S}}_{\boldsymbol r} \cdot {\boldsymbol {\hat S}}_{\boldsymbol r + \Delta {\boldsymbol r}}  + D \sum^L_{\boldsymbol r=1} \big({\hat S}^z_{\boldsymbol r}\big) ^2.
\label{ham_1}
\end{eqnarray}

In the above $t$ is the nearest neighbour hopping parameter of the electrons,  $\hat{c}^\dagger_{\boldsymbol i}= \{\hat{c}^\dagger_{\boldsymbol i, \uparrow}, \hat{c}^{\dagger}_{\boldsymbol i, \downarrow}\}$ is a  spinor creating electrons in a  Wannier state centered at $\ve{i}$, $J_k$ is the antiferromagnetic Kondo interaction between $S=3/2$ local moments and spins of conduction electrons, $J_h$ is a Heisenberg  interaction,  $D$ is the single ion anisotropy on localised spins, and we set $\Delta \r =(a,0)$ with unit interatomic distance $a=1$ between localised spins. Periodic boundary conditions (PBC) are imposed both along the spin-3/2 chain as well as on  2D electronic substrate. Translation by one lattice site preserve  crystal momentum along the spin chain.  Furthermore, ${{\Phi}_0}$ denotes the flux quantum, $\ve{A}=B(-y,0,0)$ is  the vector potential where $B$ is an external magnetic field along the $z$-direction. By tuning the orbital magnetic  field in the hopping term one can tune the density of states of the conduction electrons.  Specifically, a  $\pi$-flux per square plaquette generates Dirac fermions with vanishing density of states at the Fermi energy. This efficiently allows to analyse the Kondo physics for both finite and zero  density of states at the Fermi energy corresponding to  zero flux and  $\pi$-flux, respectively.  We set $t=1$ and $J_h=1$ throughout  in our numerical simulations.

The decoupled spin-3/2 Heisenberg  chain has been extensively  studied both theoretically and numerically~\cite{AffleckHaldanePRB1987, LiangPRL1990, NgPRB1994, QinPRB1995, HallbergPRL1996,  FathPRB2006}.  Unlike the spin-1/2 chain, higher half-integer spin chains, such as spin-3/2, exhibit edge modes with effective spin $(S - 1/2)/2$. However, the nature of these edge modes differs significantly from gapped systems. For example, the emergent spin-1/2 edge modes in a gapped spin-1 chain are exponentially localized at the two ends. In contrast, in the critical spin-3/2 Heisenberg chain, the spin-1/2 edge modes are  delocalized \cite{NgPRB1994,QinPRB1995,FathPRB2006}.

For Kondo systems with higher spins  $S>\frac{1}{2}$, the low-energy physics is determined by the number of conduction electron channels $\mu$ to which the local moment is coupled~\cite{Nozieres1980, ParcolletPRL1997}. When $\mu=2S$, the spin-$S$ local moments are exactly screened, resulting in a local Fermi liquid ground state. For $\mu>2S$, the moments are over-screened, leading to a non-Fermi liquid state. Conversely, for  $\mu<2S$, the moments are underscreened, leaving the residual spin-$(S-1/2)$ local moments. Such a state is referred to  a singular Fermi liquid~\cite{ColemanPRB2003, Mehta2005}. In our case, we consider a 1D lattice of spin-3/2 local moments coupled to a single-channel ($\mu=1$) 2D fermionic bath. In the strong coupling regime, this setup leads to a 1D lattice of residual spin-1 local moments.

The spin-$S$ Kondo impurity problem has been extensively studied using various analytical and numerical methods, including Wilson’s numerical renormalization group (NRG)~\cite{WilsonRMP1975, PRBZitko2008, Zitko_2009, Zitko_2010, Blesio2019}, Bethe ansatz techniques\cite{Bethe1931,NAndrei1980, NAndrei1983,NAndrei1984}, perturbative $T$-matrix approaches~\cite{Kondo1964}, conformal field theories~\cite{AffleckPRL1991, AffleckPRB1993, AffleckNuclearPhysB, AffleckNuclearPhysB1}, and large-$S$ expansions~\cite{Krishnan24}. However, most numerical  investigations of higher-spin Kondo lattices have been limited to 1D systems. A recent study~\cite{RikuMasui2022} examined the phase diagram of a 1D spin-1 Kondo-Heisenberg model using a combination of perturbative analysis and density matrix renormalization group (DMRG) simulations.

Despite these advances, simulating higher-spin electronic models--even in 1D--remains computationally challenging due to the rapid growth of the local Hilbert space and the increasing complexity of quantum entanglement. Developing reliable and efficient numerical methods for such systems, particularly in higher-dimensions and within context of Kondo lattice models, remains an open and important problem in condensed matter physics and related fields. In this regard, our work represents the first numerical effort employing unbiased auxiliary-field quantum Monte Carlo (AFQMC) simulations to study the spin-3/2 Kondo-Heisenberg lattice model.

By tuning the $J_k$ in the Hamiltonian of Eq.~(\ref{ham_1}), we expect a QPT from the critical spin-3/2 Heisenberg chain at small $J_k$ to a residual Haldane spin-1 chain at large $J_k$. This suggests that Co adatom chains on metallic surfaces could provide  a promising platform for realizing SPT states. Such states are particularly significant due to their potential applications in measurement-based quantum computation~\cite{Else2012, Raussendorf2003, Briegel2009}. In this context, the spin-3/2 AKLT model on a honeycomb graph has also been proposed as a robust framework for universal quantum computation~\cite{WeiPRL2011, WeiPRA2012}.
 
The paper is organized as follows. Section~\ref{perturb_cal} presents large $J_k$  and large $D$ perturbative analysis. Section~\ref{RG_analysis} provides a broad overview of the possible phases based on power counting within a renormalization group (RG) framework. Section~\ref{AFQMC_method} outlines the details of the AFQMC algorithms. Section~\ref{AFQMC_results} discusses the AFQMC results and their physical interpretation. Finally, Section~\ref{summary_outlook} summaries the main findings and offers an outlook for future directions.

\section{Strong coupling perturbative analysis }  \label{perturb_cal}
In the following, we first present the perturbative calculations in the $J_k \rightarrow \infty$ limit, followed by the large-$D$  perturbative analyses in the $D \rightarrow \infty$ and  $D\rightarrow -\infty$ limits. 
\subsection{$J_k \gg 1 $}
\label{sec:strong_coupling}
Consider the large  $J_k$ limit.  Here, the Kondo exchange  favors the formation of a triplet state. In particular and with $\ve{\hat{s}}_{\ve{r}} =
\frac{1}{2} \boldsymbol{\hat{c}}^{\dagger}_{\ve{r}} {\boldsymbol \sigma} \boldsymbol{\hat{c}}_{\ve{r}}$, the  Kondo coupling term in Eq.~(\ref{ham_1}) can be written as, 
\begin{equation}
  J_k \ve{\hat{s}}_{\ve{r}} \cdot \ve{\hat{S}}_{\ve{r}} =  \frac{J_k}{2}
  \left( \left(\ve{\hat{s}}_{\ve{r}} + \ve{\hat{S}}_{\ve{r}}\right)^2 - \frac{3}{4} - 
  \frac{15}{4}  \right),
\end{equation}
such that the $S=1$ states have energy  $-5J_k/4 $  and the $S=2$ states energy $3 J_k/4$.  The conduction electron states with even parity (doubly or singly occupied) have vanishing energy.   Here  we will consider  the case where $- 5J_k/4 $ lies well  below the Fermi energy so  that projection onto this  multiplet is justified.  Let $\hat{P}_{\ve{r}} = \sum_{m=-1}^1 \big|\ve{r},1,m\big\rangle \big\langle \ve{r}, 1, m \big| $ be the projection operator onto the $S=1$ subspace,  $\hat{P} =  \otimes_{\ve{r}}\hat{P}_{\ve{r}}$,  and $\hat{Q} =  1 - \hat{P} $  the  projection onto the  orthogonal space. To determine the energy spectrum we will solve: 
\begin{eqnarray}
& & \det \begin{bmatrix} 
  \hat{Q}\hat{H}\hat{Q} - E &   \hat{Q}  \hat{H} \hat{P}  \\
  \hat{P} \hat{H} \hat{Q} & \hat{P} \hat{H} \hat{P} - E
\end{bmatrix} = \det \left( \hat{Q}\hat{H}\hat{Q} - E \right)   \times  \\ 
& & \det \left( \hat{P} \hat{H} \hat{P} - E - \hat{P} \hat{H} \hat{Q} \left( \hat{Q}\hat{H}\hat{Q} - E \right)^{-1} \hat{Q} \hat{H} \hat{P} \right) = 0. \nonumber
\end{eqnarray}
We are targeting energies of the order  $ E_T = E_\text{FS} + L (-5J_k/4) $ where $E_\text{FS}$ is the energy of the conduction  electrons  with open boundary 
conditions in the direction perpendicular to the chain.  $L (-5J_k/4) $  corresponds to the energy of the $L$ triplet states. The first excitations occur when the parity of  a  conduction electron site coupled to a spin  $3/2$  degree of freedom changes from odd to even.  The excitation energy with  respect to the target energy reads  $5J_k/4$. We will hence replace   
$\hat{Q}\hat{H}\hat{Q} - E $  by $ 5J_k/4$. Within this approximation, the effective Hamiltonian reads: 
\begin{eqnarray}
  \hat{H}_\text{eff} = \hat{P} \hat{H} \hat{P}  - \frac{4}{5J_k}\hat{P} \hat{H} \hat{Q} \hat{H} \hat{P}.  
\end{eqnarray}
A calculations gives: 
\begin{eqnarray}
  \hat{H}_\text{eff} = & & \hat{P}\hat{H}_t \hat{P} + \Big(\frac{1}{10} \frac{t^2}{J_k} + \frac{25}{16} J_h  \Big) \sum_{\ve{r}=1}^L \ve{\hat{K}}_{\ve{r}} \cdot \ve{\hat{K}}_{\ve{r} + \ve{\Delta r}} \nonumber \\
  & & + \frac{25}{16} D \sum_{\ve{r}=1}^L \big(\hat{K}^z_{\ve{r}}\big)^2  \nonumber \\ 
    & &  -\frac{4t^{2}}{5 J_k} \sum^L_{\ve{r}=1} \ve{\hat{K}}_{\ve{r}} \cdot \left( \ve{\hat{s}}_{\ve{r} + \ve{a}_y} +
  \ve{\hat{s}}_{\ve{r} - \ve{a}_y} \right) \nonumber \\
  & & - \frac{2t^{2}}{5 J_k}  \sum_{\ve{r},\sigma} \big( \hat{c}^{\dagger}_{\ve{r}+ \ve{a}_y} \frac{\ve{\sigma}}{2} \hat{c}^{\phantom\dagger}_{\ve{r}- \ve{a}_y} \cdot \ve{\hat{K}}_{\ve{r}}  + \text{H.c.} \big).
  \label{largeJkef}
\end{eqnarray}
Here, $\ve{\hat{K}}_{\ve{r}} $ denotes the spin-1 operator. In the first term, the projection forbids  conduction electrons to hop onto Wannier states that couple to spins. This corresponds to open boundary conditions for the 2D electrons.  The second term in the above equation describes a spin-1 chain with renormalized antiferromagnetic  Heisenberg coupling  $ J^\text{eff}_h =\big(\frac{1}{10} \frac{t^2}{J_k}  +\frac{25}{16} J_h \big)$. The  fourth term accounts  for a \textit{ferromagnetic} coupling between the spin-1 chain and the electron bath. The sign of this coupling is important: consider open boundary conditions with spin-1/2  edge modes of the spin-1 chain. Owing to the ferromagnetic coupling, these modes will not be screened by the conduction electrons such that we can foresee that the spin-1 chain remains stable in this limit. This statement is independent on the nature, semi-metal or  metal, of the conduction electrons. The last term accounts for three body assisted hopping term that allows for electronic transport across the  chain. 

\subsection{$D \gg 1 $}
We now consider the large  positive $D$ limit.  In this case we  project onto the  $S_z = \pm 1/2 $ sates of the impurity spin. Hence 
$\hat{P}_{\ve{r}} = \sum_{m=\pm 1/2} \big|\ve{r},\frac{3}{2},m \big\rangle \big\langle \ve{r}, \frac{3}{2},m \big| $. %
A second  order perturbative calculation in $1/D$  give rise the following effective Hamiltonian,
\begin{eqnarray}
  \Hhat^{D>0} _\text{eff}=  \hat{H}_t  +  \sum_{\ve{r}=1}^{L} \Big\{ J^{\perp}_{h}  \big( {\hat \tau}^x_\r {\hat \tau}^x_{\r+1}  + {\hat \tau }^y_\r {\hat \tau}^y_{\r+1}  \big)\nonumber \\ + J^z_{h} {\hat  \tau}^z_\r {\hat \tau}^z_{\r+1} \Big\} +  J^{\perp}_{2}  \sum_{\ve{r}=1}^{L}  \big( {\hat \tau}^x_\r {\hat \tau}^x_{\r+2}  + {\hat \tau}^y_\r {\hat \tau}^y_{\r+2}  \big) \nonumber \\+ J_k \sum_{\ve{r}=1}^{L} \Big\{\hat{\tau}^z_{\ve{r}} \hat{s}^z_{\ve{r}} + 2\big( \hat{\tau}^x_{\ve{r}} \hat{s}^x_{\ve{r}} +   \hat{\tau}^y_{\ve{r}} \hat{s}^y_{\ve{r}} \big)\Big\},
\label{model_efham_PD}
\end{eqnarray}
with the effective  couplings $J^{\perp}_{h}=4 J_h, ~J^{z}_{h} =J_h - \frac{39 J^2_h}{8D},~  J^{\perp}_{2}=-\frac{3J^2_h}{4D}$.  In the above the spin-1/2  operator  $\boldsymbol{\hat \tau}_{\ve{r}}$  acts on the the  doublet $\left\{\big|\frac{3}{2}, \pm \frac{1}{2}  \big\rangle\right\} $. 
As apparent the SU(2) spin symmetry is reduced to U(1), and the spin chain to an XXZ chain with Kondo coupling to the bath.  Since  the effective  next-nearest neighbour coupling $J^{\perp}_{2}$ along the chain is ferromagnetic it does not trigger frustration, and hence we  do not expect it to alter the underlying physics. 
The XXZ  chain is in the Luttinger  liquid regime.  In the decoupled phase, $J_k = 0$, and  in leading order in $D$  where $J_h^{\perp}/J_h^{z} = 4$ the scaling dimension of $\tau^x$  reads \cite{Peschel75}:
\begin{equation}
  \label{scaling_Dinfty.eq}
  \Delta_{\perp} = \frac{1}{4} + \frac{1}{2}\frac{\sin^{-1}(1/4)}{\pi} \approx 0.29.
\end{equation}
As a consequence, and as discussed in more details below,  the coupling to a semi-metallic bath is irrelevant, such that we  expect as in Ref.  \cite{Danu2020}  a KBD  phase in the weak $J_k$ limit. In contrast for a metallic bath, the coupling is relevant and we expect dissipation induced long ranged order \cite{Weber2021}.  In the strong coupling limit $ J_k  \ll  D $  but $J_k \gg 1$  the system is in  a Kondo  screened phase. 
 
\subsection{$D \ll -1 $}
In large negative $D$, we project onto the $S_z = \pm 3/2 $ states of the impurity spin. 
Hence
$\hat{P}_{\ve{r}} = \sum_{m=\pm 3/2} \big|\ve{r}, \frac{3}{2},m \big\rangle \big\langle \ve{r}, \frac{3}{2},m| $. %
  In this case up to the first order perturbation in $1/D$ there are no single spin flip terms that connect the states  $\big|{i}\big\rangle= \Big|\frac{3}{2}, -\frac{3}{2}\Big\rangle_\r \otimes  \Big|\frac{3}{2}, \frac{3}{2}\Big\rangle_{\r+1}$ and $\big|{i^\prime}\big\rangle= \big|\frac{3}{2}, \frac{3}{2}\big\rangle_{\r} \otimes  \big|\frac{3}{2}, -\frac{3}{2}\big\rangle_{\r+1}$.  The  second order perturbative calculation in $1/D^2$ give rise the non vanishing matrix elements $\frac{J^3_h}{32D^2}$ $\big\langle {i^\prime}\big|  { \hat S}^+_\r {\hat S}^-_{\r+1}  {\hat S}^+_\r {\hat S}^-_{\r+1}  { \hat S}^+_\r {\hat S}^-_{\r+1} \big|{i}\big\rangle=  \frac{J^3_h}{32D^2} \times 4  \big({\sqrt 3}\big)^4$.  The effective Hamiltonian derived  using the SWT up to third order in $J_h$  takes  the following form,
\begin{eqnarray}
  \Hhat^{D<0} _\text{eff}&= &  \hat{H}_t +  J^z_h  \sum_{\ve{r}=1}^{L} {\hat \tau}^z_\r {\hat \tau}^z_{\r+1} + J^\perp_h  \sum_{\ve{r}=1}^{L} \big( {\hat \tau}^x_{\r} {\hat \tau}^x_{\r+1}  + {\hat \tau}^y_{\r} {\hat \tau}^y_{\r+1}  \big)   \nonumber \\
   &&+ \frac{3J_k}{2} \sum_{\ve{r}=1}^{L}  \hat{\tau}^z_{\ve{r}} \hat{s}^z_{\ve{r}},
\label{model_ham_nxz_ND}
\end{eqnarray}
with effective couplings $J^z_h =\frac{9 J_h}{4} +  \frac{6 J^2_h}{4D}$ and $J^\perp_h = \frac{9J^3_h}{4D^2}$ and the $\boldsymbol{\hat \tau}_{\ve{r}}$  operators  acts on the  doublet $\left\{\big|\frac{3}{2}, \pm \frac{3}{2}\big\rangle \right\}$.

The key observation is that there are no single spin-flip terms of local Kondo interactions, $(\hat{\tau}^+ {\hat s}^- + \hat{\tau}^- {\hat s}^+)$,  at any order of perturbation that connect the states $\big\{\big|\frac{3}{2}, \pm\frac{3}{2}\big\rangle \otimes \big|{\hat c}^\dagger_{\downarrow (\uparrow)}\big|0 \big\rangle,$ $\big|\frac{3}{2}, \mp\frac{3}{2}\big\rangle \otimes  {\hat c}^\dagger_{\uparrow(\downarrow)}\big|0\big\rangle \big\}$. 

\section{Scaling analysis and phases} \label{RG_analysis}
\begin{figure}[htbp]
 \includegraphics[clip, width=9.1cm]{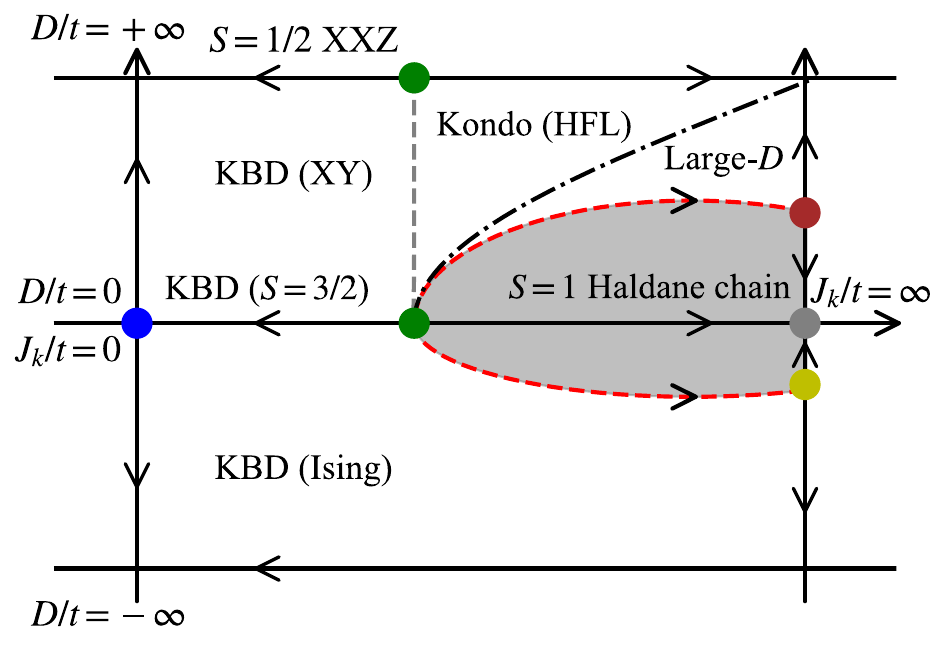}
 \caption{Phase diagram of a $S=3/2$ chain on a 2D semi-metal in the $D/t$ versus $J_k/t$ plane at fixed $J_h/t$. At the $J_k/t = 0$ limit, the vertical line along the $D$-axis characterizes the physics of the spin-3/2 chain as a function of $D$. At $D/t = 0$, the isotropic spin-3/2 chain undergoes a QPT  from a KBD phase to a topological Kondo phase as a function of $J_k$. The green dot marks the KBD quantum critical point. The grey-shaded region indicates a coexistence regime where partial Kondo screening occurs alongside residual spin-1 chain physics. The  brown (yellow) dots mark the Gaussian (Ising) critical points of an $S=1$ chain.}
 \label{fig:RG_cartoon_sm}
\end{figure}

\subsection{$J_k/t \ll 1$}
\label{sec:Jk_small}
To at best understand this limit, we  formulate the partition function using  spin and fermion coherent  state representation.   Expanding  the fermions determinant up to second order in the Kondo coupling gives $ {\cal S} = {\cal S}_{\text{Heisenberg}}^{(3/2)} + {\cal S}_{\text{Diss}} $ with ${\cal S}_{\text{Heisenberg}}^{(3/2)}$   the action of the  $S=3/2$ Heisenberg model with single ion anisotropy,
 \begin{eqnarray}
 {\mathcal S}_\text {Diss}  =  & &  - \frac{S^2J^2_k}{2} \int d\tau d\tau^\prime \sum_{\r, \r^\prime} \sum_{\alpha,\alpha'=1}^{3} \nonumber \\  & & \ve{n}_{\r, \alpha}(\tau)  \chi_{\alpha,\alpha'}(\r-\r^\prime,\tau-\tau^\prime)   \ve{n}_{\r',\alpha'}(\tau^\prime).
\end{eqnarray}
In the above, $\chi_{\alpha,\alpha'} (\ve{r},\tau) = \left< \hat{s}_{\ve{r},\alpha}(\tau)
\hat{s}_{\ve{0},\alpha'}(0)  \right>_0$  corresponds to the spin susceptibility of the  conduction electrons.   Owing to the SU(2) spin symmetry of the metallic surface, this quantitiy is diagonal and independent on $\alpha,\alpha'$.  
The $S=3/2$ Heisenberg model has a very rich phase diagram \cite{NChepiga2020}.  For  nearest neighbour coupling  it  maps onto a Wess-Zumino-Witten (WZW) level-1 model and  hence shares the properties  of the spin-1/2 chain. Upon switching on the easy axis anisotropy one will  obtain  Luttinger liquid   (Ising) phases for  $D>0$  ($D<0$).  In the Luttinger liquid phase, the scaling dimension of the spin 
operator satisfies $\Delta_{\perp} \leq 0.5 $.  In particular at  $D = 0$, $\Delta_{\perp}   =0.5$  and is given by Eq.~(\ref{scaling_Dinfty.eq}) in the large positive $D$ limit.  For  $D < 0 $ we expect long-range Ising order.   For the Lorentz invariant semi-metalic state,  the spin susceptibility scales as  
$\frac{1}{\sqrt{ |\ve{r}|^2 + \left(v_f \tau \right)^2}^4} $ with $v_f$ the Fermi velocity. Thereby, at the decoupled fixed point  the  local  time displaced retarded interaction  will  scale  as  $b^{3 - 2 \Delta_{\perp} - 4}$ with   
$b>1$  the scaling factor. Since the Kondo coupling is irrelevant  we foresee, as in Ref.~\cite{Danu2020}, a KBD phase in the  small $J_k$ limit. For $D<0$ where long ranged Ising order is present, we can repeat the above argument  with $\Delta_{\perp} = 0$ to again obtain a decoupled phase.  To summarize, the  small $J_k$  region of  Figure~\ref{fig:RG_cartoon_sm} corresponds to a KBD  phase where spin and conduction electrons are effectively decoupled.  

In the Kondo breakdown phase (KBD phase of Figure~\ref{fig:RG_cartoon_sm}), the  spin-3/2 Heisenberg chain in the presence of a finite single-ion anisotropy $D \neq 0$ behaves as an isolated one.  Using bosonization techniques, Schulz predicted that the 
system exhibits a QPT at the isotropic Heisenberg point and the critical point $D_c = 0$ separates the two distinct XY critical phases--referred to as XY1 and XY2--for $D>0$ and $D<0$, respectively~\cite{SchulzPRB1986}. In the XY1 phase ($D>0$), both the transverse $\langle \hat{S}^+_0 \hat{S}^-_r + \text{H.c.} \rangle$ and longitudinal $\langle \hat{S}^z_0 \hat{S}^z_r \rangle$ spin-spin correlations decay algebraically. For sufficiently large positive $D$, the higher-energy $S_z =\pm 3/2$ states are suppressed, and the system effectively behaves like a spin-1/2 XXZ chain in the subspace of the low-energy $S_z=\pm 1/2$ doublet. Thus, the $D>0$ regime remains gapless with quasi-long-range order. In the XY2 phase ($D<0$), the anisotropy favors the $S_z=\pm 3/2$ doublet. In the large negative $D$ limit, the higher $S_z=\pm 1/2$ states are separated by a large spin gap $\Delta_D= 2|D|$, and the system again maps onto an effective spin-1/2 XXZ model within the low energy $S_z=\pm 3/2$ subspace.  However, two-point transverse spin-flip processes are suppressed, as flipping $2S=3$ spins requires higher order processes.  As a result, the transverse correlation $\big\langle \hat{S}^+_0 \hat{S}^-_r +\text{H.c.}\big\rangle$ decays exponentially, while the longitudinal correlation $\big\langle \hat{S}^z_0 \hat{S}^z_r\big\rangle$ acquires a long range order.  In this large negative $D$ regime, higher order transverse spin-spin  correlations such as $\big\langle (\hat{S}^+_0 \hat{S}^-_r + \text{H.c.})^{2S}\big\rangle$ are expected to show power-law decay. 

\begin{figure}[htbp]
 \includegraphics[clip, width=9.1cm]{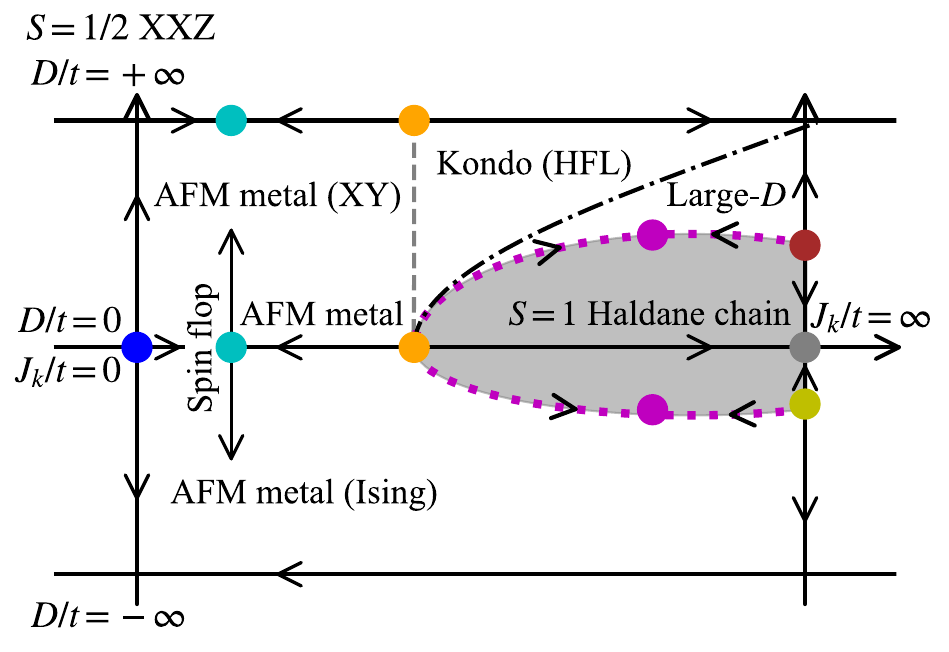}
  \caption{Phase diagram of a $S=3/2$ chain on a 2D metal in the $D/t$ versus $J_k/t$ plane at fixed $J_h/t$. At the $J_k/t=0$ point, the vertical line along the $D$-axis corresponds to the physics of a spin-3/2 chain as a function of $D$. At $D/t=0$, the chain undergoes an order-disorder QPT between an AFM metallic phase and a topological Kondo phase as a function of $J_k$. The orange dot marks the order-disorder quantum critical point. The cyan dot represents the dissipation-induced AFM ordering phase at finite but weak $J_k/t$. The dissipation induced AFM phase undergoes a spin-flop transition to XY, for $D>0$, and to Ising, for $D<0$, phases at finite anisotropies. The relevance of the metallic surface at the transitions out of the Haldane phase will generate new  dissipation induced criticality corresponding to the magenta dots. The  brown (yellow) dots mark the Gaussian (Ising) critical points of an $S=1$ chain.}
 \label{fig:RG_cartoon_m}
\end{figure}
 
For the metallic surface (See Figure~\ref{fig:RG_cartoon_m}) the  local  imaginary time displaced spin correlation function  scales as 
$\chi(\ve{r}=0,\tau) \propto 1/{\tau^2}$.   A  similar scaling argument as discussed above shows that the Kondo coupling is marginal at  the SU(2) symmetry point.   Since at this point the $S=3/2$ spin-chain  with nearest neighbour interactions is described by the same level-1 WZW model as the spin-1/2 chain, we can take over previous results for the spin-1/2 chain.  The Kondo coupling is marginally relevant  and leads to dissipation induced long ranged order \cite{Weber2021, Danu2022}. 

The long ranged, dissipation induced, magnetically ordered state is characterized by spontaneous symmetry breaking. At the SU(2) symmetric point, the  direction of the staggered magnetization is arbitrary and defines the 
vacuum around which one will expand  within a spin-wave theory. The single-ion anisotropy $D$ breaks down the symmetry from SU(2) to U(1), and places energetic constraints on the direction of the quantization axis. For $D<0$, the $z$-quantization axis  is favoured  whereas for $D>0$  spontaneous symmetry breaking occurs in the $xy$-plane. This sudden change in the quantization axis  and  thereby 
change  of the vacuum state bears  similarities to a spin-flop transition~\cite{Fisher75}, where the sudden change of the quantization axis is driven by a time reversal symmetry breaking magnetic field.  

\subsection{$J_k/t \gg 1$}
\label{sec:Jk_big}
At $J_k = \infty $,  partial Kondo screening leads to a spin-1 Haldane chain subject to single ion-anisotropy (see Eq.~(\ref{largeJkef})).  This applies for both metallic and semimetallic surfaces depicted in   Figures~\ref{fig:RG_cartoon_sm} and~\ref{fig:RG_cartoon_m}.

The spin-1 chain exhibits distinct quantum phases depending on the sign and magnitude of $D$~\cite{SchulzPRB1986,WChen2003}. The underlying phases and associated transitions are related to pre-roughening surface phase transitions~\cite{NijsPRB1989}. When $D=0$, the ground state is in the well known Haldane phase--a gapped topological phase characterized by non-local string order and spin-1/2 edge modes. The Haldane phase remains stable for small $|D|$. According to Schulz bosonization results \cite{SchulzPRB1986}, unlike the spin-3/2 chain, the spin-1 chain exhibits transitions into the XY1 and XY2 phases at finite values of critical anisotropy $D^\pm_c$~\cite{SchulzPRB1986}. For $D>0$, the system undergoes a continuous QPT from the Haldane phase to a large-$D$ trivial phase, where the ground state is dominated by local $S_z=0$ configurations on each site. This transition occurs at $D^+_c \approx 0.97$ and belongs to the Gaussian universality class with central charge $c =1$ \cite{Hu11}. On the other hand, for $D<0$, the Haldane phase becomes unstable and the system undergoes another continuous transition at $D^-_c \approx -0.3$~\cite{AlbuquerquePRB2009, LeporiPRB2013} into an Ising phase characterized by N\'eel order of alternating $S_z=\pm1$ doublet.
This transition belongs to the Ising universality class, with central charge $c=1/2$. Thus, the Haldane phase is sandwiched between these two critical points and exists only in the intermediate regime $-0.3\lesssim D\lesssim0.97$. The large negative $D$ regime corresponds to the XY2 phase, where higher transverse spin-spin correlations, 
$\big\langle (\hat{S}^+_0 \hat{S}^-_r+\text{H.c.})^2\big\rangle$ 
are expected to exhibit power-law decay. 

We now consider small deviations from the large $J_k$ limit. Importantly, 
the coupling to the conduction electron is ferromagnetic (see Eq.~(\ref{largeJkef})).  
This implies that even 
for open boundary conditions where the spin-1 chain hosts spin-1/2 edge states, the Kondo coupling of magnitude set by $t^2/J_k$ can be treated perturbatively. 
Neglecting correlated hopping terms in Eq.~(\ref{largeJkef}), we can integrate out the fermions and expand the fermion determinant up to second order in $4t^2/5J_k$ to obtain: 
$ {\cal S} = {\cal S}_{\text{Haldane}} + {\cal S}_{\text{Diss}}$ with ${\cal S}_{\text{Haldane}}$ the action of the Haldane spin-1 chain with single-ion anisotropy $D$ and
\begin{eqnarray}
  \label{eq:haldane_diss}
 {\mathcal S}_\text {Diss}  =  & &  - \frac{8 S^2t^4}{25J_k^2} \int d\tau d\tau^\prime \sum_{\r, \r^\prime} \sum_{\delta,\delta'}  \\  & & \ve{n}_{\r +\ve{\delta}}(\tau)  \chi(\r + \ve{\delta}-\r^\prime- \ve{\delta}',\tau-\tau^\prime)   \ve{n}_{\r' + \delta'}(\tau^\prime) \nonumber.
\end{eqnarray}
Here, $\ve{\delta}$ runs over $\pm \ve{a}_y$. 
As a function of the single-ion anisotropy, the Haldane model undergoes quantum phase transitions between Haldane and large-$D$ ($D>0$) as well as  Haldane and Ising ($D<0$) phases. At these points, coupling to the semi-metallic (metallic) surface will be relevant provided that $3 - 2 \Delta_n - 4 > 0$ ($3 - 2 \Delta_n - 2 > 0$), where $\Delta_n$ is the scaling dimension of the spin-1 operator. Hence, for the semi-metallic surface and since $\Delta_n > 0$, the 
coupling to the fermions is irrelevant, and the aforementioned universality class of the quantum 
phase transitions in the Haldane model will not be altered. In contrast, for the metallic surface, the condition for the relevance of the metallic surface reads, $\Delta_n < 1/2$. This condition is 
satisfied \cite{Hu11,SchulzPRB1986} for the Haldane model. Thereby, the metallic surface introduces a relevant coupling at the phase transition between the Haldane and trivial large-$D$ phase as well as between the Haldane and Ising phase.  The magneta points in Figure~\ref{fig:RG_cartoon_m} indicate new fixed points for these transtions. 

\subsection{Kondo and Kondo breakdown phases}
\label{sec:kondo_phases}
We now consider Kondo and Kondo breakdown phases. A sharp definition of the Kondo phase is based on the
electron count, via the Luttinger volume \cite{OshikawaPRL2000}. Our model has translation symmetry along the chain direction. Hence, for an $L\times L$ lattice of conduction electrons, our model maps onto 
a 1D chain with $L$ conduction and a spin degree of freedom per unit cell. On any finite lattice, we can hence define a Luttinger volume as 
\begin{equation}
  V = \pi \mod( L + n_s,2),
\end{equation}
where $n_s$ corresponds to the number of spin degrees of freedom that participate in the Luttinger volume. 
In the above, the volume of the Brillouin zone corresponds to $2 \pi /a $, and we have set $a=1$.  The Kondo phase corresponds to $n_s=1$, and Kondo breakdown to $n_s=0$. 
Given this definition, the $J_k = \infty$ phase corresponds to a Kondo phase, the  rationale being the following. At this point, the spin chain and conduction electrons that couple to the spin chain via the Kondo interaction are effectively decoupled from the other conduction electrons. The Luttinger count hence reads $V = \pi \mod( L - 1, 2) \equiv 2 \pi \mod( L + n_s, 2)$ with $n_s = 1$. Given this, we conjecture that there is no sharp distinction between the Kondo phase and the trivial large-$D$ phase in the large $J_k$ limit.  The same argument leads to the conclusion that the $J_k = \infty$  line for  both the semi-metallic and metallic surfaces 
are Kondo phases.    

Generically, the Kondo phase is characterized by  the emergence of a composite fermion excitation $ \hat{\psi}^{\dagger}_{\r,\sigma} = \sum_{\sigma^\prime}{\hat c}^\dagger_{\r,\sigma^\prime} {\boldsymbol \sigma}_{\sigma,\sigma^\prime} \cdot \hat {\S}_{\r}$ \cite{RaczkowskiPRL2019,Danu2021} at the Fermi energy.
We note that in our setup, since the spin-chain is subintensive, the Fermi energy is set by that of the conduction electrons. 
Following Ref.~\cite{Danu2021}, 
one expects that when the Kondo effect  partially screens the spin  $S=3/2$ down to $S=1$, the remaining spin-1/2 degree of freedom is absorbed by the conduction electrons via the emergence of the composite fermion. The composite fermion should show up as a low lying mode in the corresponding single particle spectral function. 
This need not be the case.  In particular, in the limit $J_k \rightarrow \infty$ the composite fermion excitation has an energy threshold set by $J_k$.   We will
hence introduce the notion of  Type I and Type II Kondo phases. In the Type I Kondo phase, the composite fermion excitation is a well defined low lying excitation  whereas in the Type II Kondo phase, the composite fermion excitation is not well defined. 

With the above definition of the Kondo phase, we cannot sharply distinguish dissipation induced AFM phases with and without well defined composite fermion excitations. The reason is that due to the antiferromagnetic ordering, the size of the unit cell doubles such that Kondo breakdown and Kondo phases have the same Luttinger volume.

\section{Auxiliary Field Quantum Monte-Carlo}\label{AFQMC_method}
We numerically simulate the model of Eq.~(\ref{ham_1}) using the  Algorithms for Lattice Fermions (ALF) \cite{ALF_v2_1} implementation of the AFQMC algorithm~\cite{BlankenbeclerPRD1981,AssaadPRL1999, PRBCapponi2001,Assaad2008}.  
We use both finite-temperature \cite{Hirsch85,White89}  and  zero-temperature  AFQMC algorithms 
\cite{Sugiyama86,Sorella89} for our numerical investigations.

To formulate the AFQMC algorithm for spin-$S$ Kondo problems, we use the recent approach proposed in Ref.~\cite{SchwabPRB2023}. To simulate a generic spin $S$ degree, we enlarge the Hilbert space to that of $2S$ spin-1/2 degrees of freedom and energetically project onto the symmetric states by adding a ferromagnetic Hund's rule term to the Hamiltonian.  The Hamiltonian of Eq.~(\ref{ham_1}) can be rewritten in terms of the spin-1/2 degrees of freedom as follows:
\begin{eqnarray}
& & \hat{H}= \hat{H}_t
 + \frac{J_k}{2} \sum^L_{\boldsymbol r=1} \sum_{n=1}^{2S}{\boldsymbol {\hat c}}^{\dagger}_{\boldsymbol r} {\boldsymbol \sigma}  {\boldsymbol {\hat c}}_{\boldsymbol r} \cdot  {\boldsymbol{\hat S}}_{\boldsymbol r, n}\nonumber \\&&+
 J_h\sum^L_{{\boldsymbol r}=1} \sum_{n,m=1}^{2S} {\boldsymbol {\hat S}}_{\boldsymbol r,n} \cdot {\boldsymbol {\hat S}}_{\boldsymbol r + \Delta {\boldsymbol r},m}  +  D \sum^L_{\boldsymbol r =1} 
 \left(\sum_{n=1}^{2S}{\hat S}^{z}_{\boldsymbol r, n}\right) ^2 \nonumber \\
 & & - J_F \sum_{\ve{r}=1}^{L} \left( \sum_{n=1}^{2S} \hat{\ve{S}}_{\boldsymbol r,n}\right)^2.
\label{ham_2}
\end{eqnarray}
In the above, 
${\boldsymbol{\hat S}}_{\boldsymbol r,n}$ is a spin-1/2 degree of freedom. $\left( \sum_{n=1}^{2S} \hat{\ve{S}}_{\boldsymbol r,n}\right)^2$ corresponds to the Casimir operator of the total spin and commutes with each component of the total spin.  Hence,  the ferromagnetic constraint, $J_F > 0$, commutes with the Hamiltonian and the projection onto the physical Hilbert space is implemented very efficiently \cite{SchwabPRB2023}.

To proceed, we use an Abrikosov representation of the spin-1/2 degree of freedom:
$\ve{\hat{S}}_{{\boldsymbol r}, n} = \frac{1}{2}\sum_{\sigma, \sigma^\prime} {\hat f}^\dagger_{{\boldsymbol r}, \sigma, n}   {\boldsymbol \sigma}_{\sigma \sigma^\prime} {\hat f}_{{\boldsymbol r}, \sigma^\prime, n}$ with the constraint $\sum_{\sigma} {\hat f}^\dagger_{{\boldsymbol r},\sigma,n} {\hat f}_{{\boldsymbol r},\sigma,n} = 1$. Here, ${\boldsymbol \sigma}$ are the Pauli matrices. In the Monte Carlo simulation, we relax the constraint on the Hilbert space and impose it energetically by adding a Hubbard $U$-term on the $f$-orbitals.
Using the operator  identity:  $ -\frac{1}{4}  \left( \ve{\hat{f}}^{\dagger}_{\ve{r},n} \ve{\hat{f}}^{\phantom\dagger}_{\ve{r}',n}  + \text{H.c}  \right)^2 + \frac{1}{4} = \ve{\hat{S}}_{\ve{r},n} \cdot \ve{\hat{S}}_{\ve{r}',n} $   on the physical Hilbert space, we can rewrite, up to a constant, 
the Hamiltonian of Eq.~(\ref{ham_2}) in terms of the fermionic operators as follows:

\begin{eqnarray} 
& & \hat{H}_\text{AFQMC} 
  ={\hat H}_t- \frac{J_k}{4} \sum^{2S}_{{\boldsymbol r},{n=1}} \Big({\boldsymbol{\hat c}}^{\dagger}_{\boldsymbol r} {\boldsymbol {\hat f}}^{\phantom\dagger}_{\boldsymbol r,n} + \text{H.c.}  \big)^{2}   \nonumber \\  
& & - \frac{J_h}{4}\sum_{\langle {\boldsymbol r}, {\boldsymbol r^\prime} \rangle} \sum_{n,m}^{2S}  \Big({\boldsymbol {\hat f}}^{\dagger}_{{\boldsymbol r},n}{\boldsymbol {\hat f}}^{\phantom\dagger}_{{\boldsymbol r}^\prime,m} + \text{H.c.}  \Big)^2   \nonumber \\
& &  
-|D|  \sum_{\boldsymbol r}  \sum_{n>m}^{2S} 
\left( S^{z}_{\ve{r},n} - \frac{D}{|D|} S^{z}_{\ve{r},m} \right)^2
\nonumber\\
& & 
+ \frac{J_F}{2} \sum_{\boldsymbol r}  \sum_{n>m}^{2S} \Big(\boldsymbol{{\hat f}}^{\dagger}_{\boldsymbol r,n}{\boldsymbol {\hat f}}^{\phantom\dagger}_{\boldsymbol r,m} + \text{H.c.}  \Big)^2 \nonumber\\
& & 
+  U_f \sum^{n=2S}_{{\boldsymbol r},n=1} \Big({\boldsymbol{\hat f}}^{\dagger}_{{\boldsymbol r},n} {\boldsymbol  {\hat f}}^{\phantom\dagger}_{{\boldsymbol r},n} -1\Big)^2. 
\label{H_QMC}
 \end{eqnarray}
 In the above $J_F$ is a ferromagnetic coupling that projects onto the triplet sector and $U_f$ is the on-site Hubbard interaction. 
We note that for the SU(2) case various representations of the Heisenberg term are  possible. In particular,  one could consider the current instead of the 
kinetic energy  to obtain:  $ -\frac{1}{4}  \left( i \ve{\hat{f}}^{\dagger}_{\ve{r},n} \ve{\hat{f}}^{\phantom\dagger}_{\ve{r}',n}  + \text{H.c}  \right)^2 + \frac{1}{4} = \ve{\hat{S}}_{\ve{r},n} \cdot \ve{\hat{S}}_{\ve{r}',n} $. Again, this equation is valid only in the physical Hilbert space. 

 A key point is that the  Hubbard  term $\hat{H}_\text{Hubbard} = U_f \sum_{{\boldsymbol r},n} \big({\boldsymbol{\hat f}}^{\dagger}_{{\boldsymbol r},n} {\boldsymbol  {\hat f}}^{\phantom\dagger}_{{\boldsymbol r},n} -1\big)^2 $ on $\hat f$ orbitals  commutes with the Hamiltonian,  $\big[\hat{H}_{\text{AFQMC}} ,\hat{H}_\text{Hubbard}\big]=0$ and thus the doubly occupied  and empty unphysical   states with even parity on the localized $f$ orbitals  are suppressed by a factor $e^{-\beta U_f}$ at large $\beta U_f$.  Here, $\beta = 1/(k_B T)$ is the inverse temperature where $k_B$ denotes the Boltzmann constant.
 Moreover, $\left[\hat{H}_{\text{AFQMC}},\left( \sum_{n=1}^{2S} {\hat \S}_{\boldsymbol r}^{(n)} \right)^2 \right] = 0$, 
 such that the projection onto the spin-$S$  Hilbert space is very efficient.  Finally, in this terminology  the $S=3/2$ model  corresponds to $2S=3$.

 All interactions in Eq.~(\ref{H_QMC}) are written in terms of perfect squares such that the Hamiltonian complies to the standard Hamiltonian of the ALF-library \cite{ALF_v2_2}. Consequently, in the path integral formulation of the partition function $Z=\big[ e^{-\beta \hat{H}}\big]$, we can utilise the Hubbard-Stratonovich (HS) transformation to decompose the two-body operators into one-body operators. To separate the imaginary-time propagators of non-interacting and interacting terms we employ the Trotter-Suzuki decomposition~\cite{Trotter1959,SUZUKI1976,SUZUKI1977}. The absence of the negative sign problem follows from particle-hole symmetry of the  Hamiltonian. We refer  the reader to Ref.~\cite{SatoT17_1}  for a detailed account of this point.  Both equal-time and time-displaced observables in finite-temperature grand canonical AFQMC algorithms can be measured using Metropolis Monte Carlo (MC) sampling, $ \big\langle {\hat O} \big\rangle = \text{Tr} \big[ e^{-\beta \hat{H}} \hat{O}\big]/\text{Tr} \big[ e^{-\beta \hat{H}}\big]$.  In MC sampling, we sum over all space- and time-dependent HS fields,  the resulting outcomes are bias free.

 To maintain the hermicity  of the imaginary time propagators a second-order symmetric Trotter-Suzuki decomposition is used in our Monte-Carlo sampling approach,
 \begin{eqnarray}
  e^{-\Delta_\tau \hat{H}_\text{AFQMC}} = e^{-\frac{\Delta_\tau}{2} \hat{H}_t}    e^{-\Delta_\tau \hat{H}_\text{int}}   e^{-\frac{\Delta_\tau}{2} \hat{H}_t}  + \mathcal{O}(\Delta_\tau^3).
  \end{eqnarray} 
  Here, the  Trotter step is given by:  $\Delta_\tau = \beta/L_\tau$ with  $L_\tau$ the number of Trotter slices.

To generate the trial wave function $| \psi_0 \rangle = \lim\limits_{\Theta\rightarrow \infty}  e^{-\Theta {\hat H} } \big| \psi_T \big\rangle $ in projector algorithm we use the following Hamiltonian,
 \begin{eqnarray}
\hat{H}_\text{trial}=\hat{H}_t  - \frac{J_k}{4} \sum^{2S}_{{\boldsymbol r},n=1} \big({\boldsymbol{\hat c}}^{\dagger}_{\boldsymbol r} {\boldsymbol {\hat f}}^{\phantom\dagger}_{\boldsymbol r,n} + \text{H.c.}  \big)\nonumber \\  - \frac{J_h}{4}\sum_{\langle {\boldsymbol r}, {\boldsymbol r^\prime} \rangle} \sum_{n=1}^{2S}  \big({\boldsymbol {\hat f}}^{\dagger}_{{\boldsymbol r},n}{\boldsymbol {\hat f}}^{\phantom\dagger}_{{\boldsymbol r}^\prime,n} + \text{H.c.}  \big).
 \end{eqnarray}
   Here, $2\Theta = L_\tau \Delta_\tau$ is the projection parameter with Trotter step $\Delta_\tau$. Observables in the projective algorithm  are evaluated through,
\begin{eqnarray}
 \big\langle  {\hat O} \big\rangle  =\lim\limits_{\Theta \rightarrow \infty} \frac{\big\langle \psi_T \big| e^{-\Theta {\hat H}}  {\hat O}  e^{-\Theta {\hat H}}\big |\psi_T \big\rangle}{\big\langle \psi_T \big|e^{-2\Theta {\hat H}}\big|\psi_T \big\rangle}.
 \end{eqnarray}

 \begin{figure}[htbp]
 \includegraphics[clip, width=8.9cm]{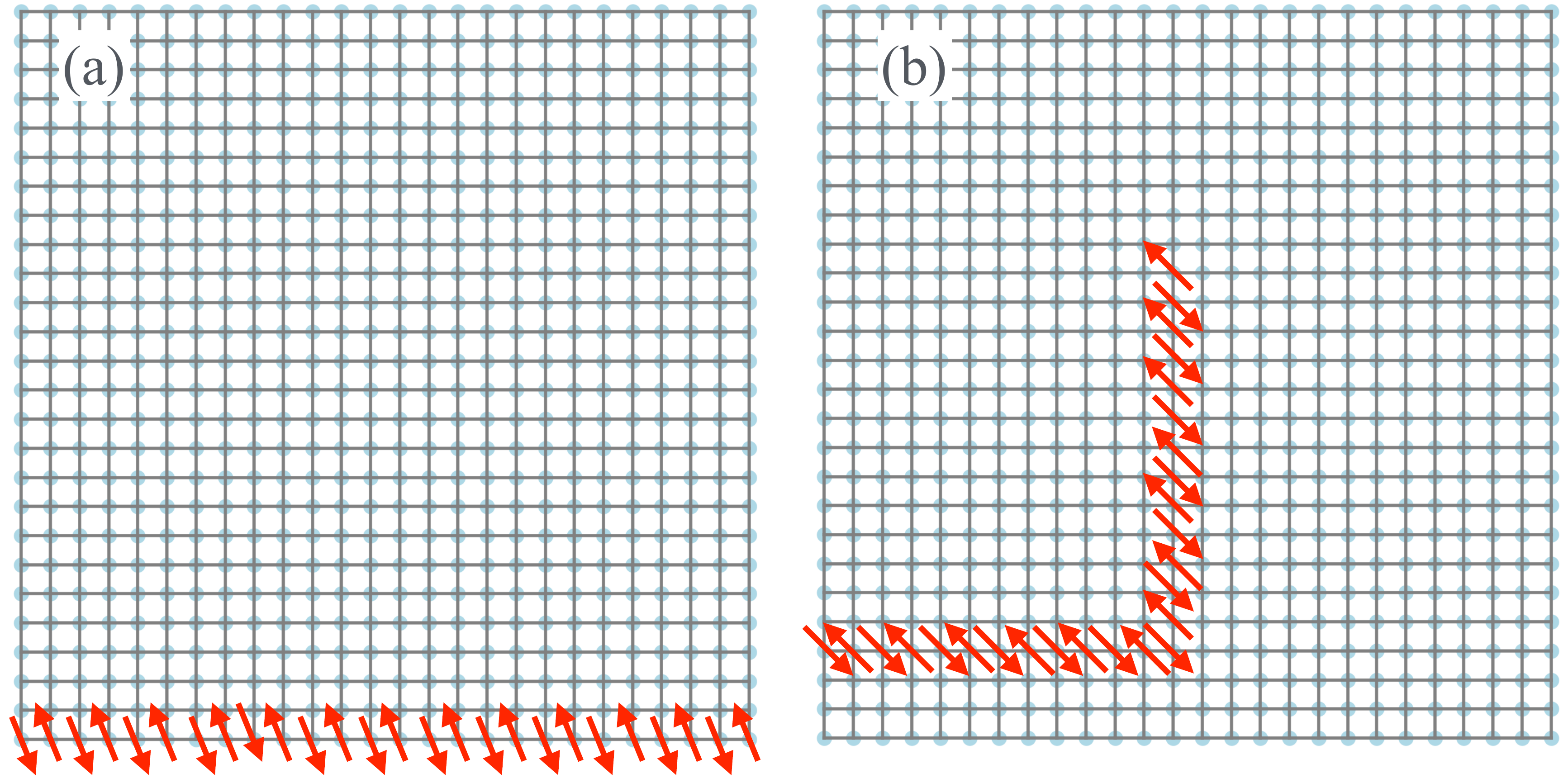}
 \caption{Sketch of a  26-site spin-3/2 chain embedded in  a  $26\times26$ periodic graph of  2D electrons, as used  in AFQMC simulations.  Lattice arrangement (a) is  used   to simulate a periodic spin-3/2 chain.  Arrangement (b) is designed to prob massless spin-1/2 edge modes in an open spin-3/2 chain at  $J_k/t\rightarrow\infty$ limits. This  open geometry in (b) is specifically chosen to avoid the RKKY interactions between local moments at edges. }%
 \label{fig:Sketch_opbc_latt}	
\end{figure}
To analyse the emergent topological spin-1/2 edge modes in the strong Kondo coupling limit where we expect 
the emergence of a spin-1 Haldane phase, 
 we  simulate the Hamiltonian in Eq.(\ref{ham_1}) using  open boundary conditions (OBC).
 The lattice geometry used in our AFQMC simulations are shown in Figure \ref{fig:Sketch_opbc_latt}. Importantly, in arrangement (b) the 
 Manhattan distance between the two edge sites is odd  such that  edge spins are aligned antiferromagnetically.

\begin{figure}[htbp]
 \includegraphics[clip, width=8.8cm]{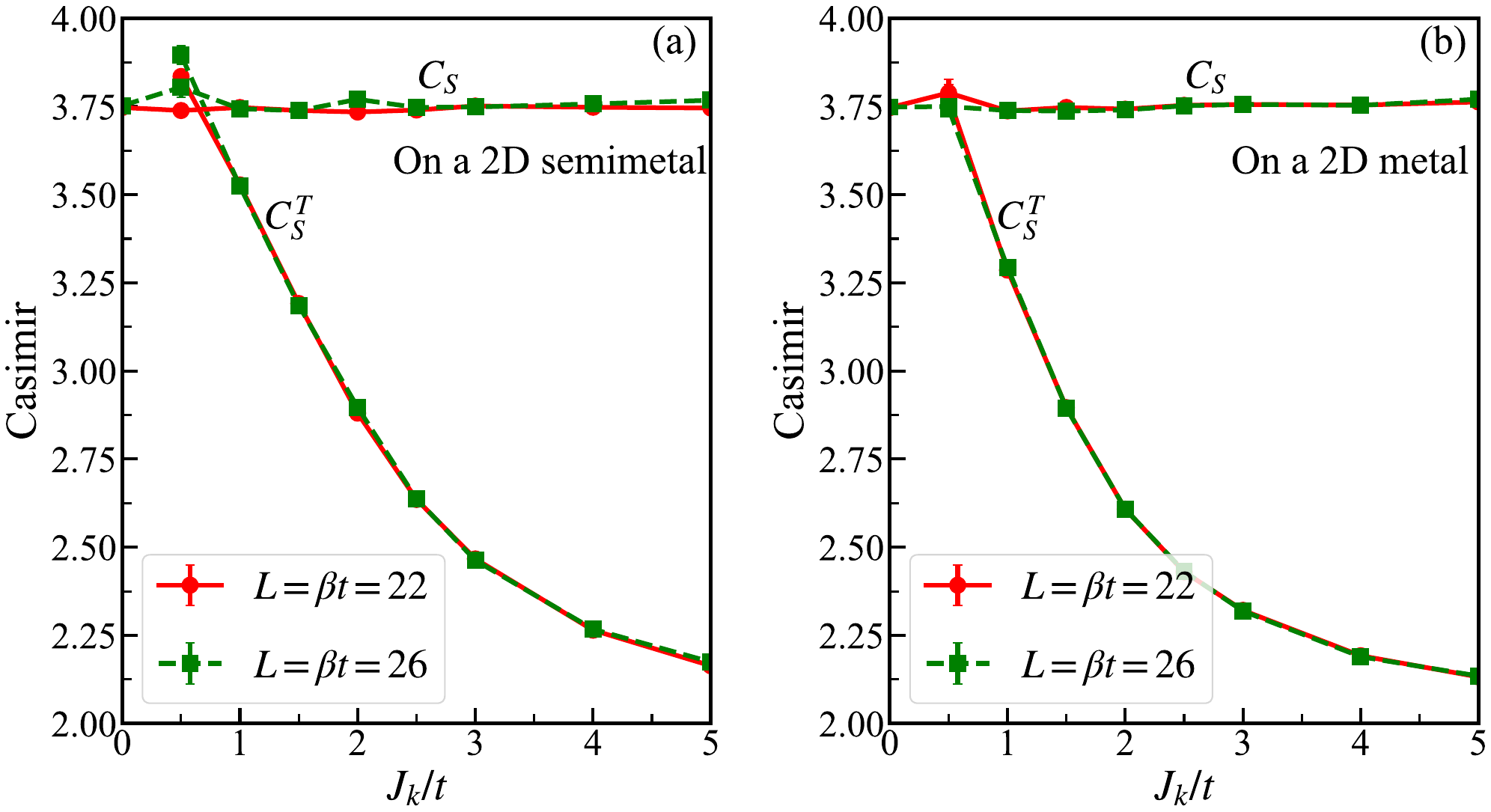}
 \caption{AFQMC results for Casimir operators  as a function of $J_k/t$ along the spin chain  with OBC at $J_h/t=1, D/t=0, \beta t =L$.  (a) $S=3/2$  Heisenberg chain on a 2D semi-metal. (b) The same on a 2D metal.}
 \label{fig:casimir_vs_Jk_BetaL_sm_m}
\end{figure}

We benchmark our simulations by analysing  the Casimir operator,
\begin{equation}
  C_S = \left< \left( \sum_{n=1}^{2S}  \ve{\hat{S}}_{\ve{r},n} \right)^2 \right>,
\end{equation}
as a functions of $J_k/t$.  As mentioned above, this operator commutes with the  Hamiltonian   and the value of $J_F$ has to be chosen large enough so as to guarantee projection on the symmetric state. 
Figure~\ref{fig:casimir_vs_Jk_BetaL_sm_m}  provides numerical justification for this statement since this quantity is pinned to $C_S = S(S+1)$ with $S=3/2$  for all considered values of $J_k/t$ and for  both  semi-metallic and metallic phases. We also consider the local total spin,
\begin{equation}
  C_S^T= \left< \left( \frac{1}{2}\ve{\hat{c}}^{\dagger}_{\ve{r}} \ve{\sigma} 
  \ve{\hat{c}}^{\phantom\dagger}_{\ve{r}}  + \sum_{n=1}^{2S}  \ve{\hat{S}}_{\ve{r},n} \right)^2 \right>. 
\end{equation}
This quantity can take any value  between $C_S^T = 2(2+1) $ and $C_S^T = 1(1+1) $.  However, in the strong coupling limit, $J_k/t \rightarrow \infty$, $C_S^T = 1(1+1) $   corresponding to the emergence of the  spin-1 degree of freedom.  This corresponds to partial Kondo screening.  This behaviour is confirmed in Figure~\ref{fig:casimir_vs_Jk_BetaL_sm_m}.

\section{Numerical results} \label{AFQMC_results}
Below, we present our AFQMC results in four parts. First in Sec.~\ref{sec:semi-metal}, we investigate the physics of a spin-3/2 Heisenberg chain coupled to a 2D semi-metallic surface. Next, in Sec.~\ref{sec:metal} we extend our analysis to the case of a 2D  metallic surface. 
Both Sections \ref{sec:semi-metal} and \ref{sec:metal} concentrate on the $D=0$ case. In Sec.~\ref{sec:edge} we pin down signatures of the Haldane phase by investigating chains with open boundary conditions. In the last section, we  discuss the overall effect of the single-ion anisotropy.

To investigate the spin dynamics of the spin-3/2 Heisenberg chain, we measure the space and time displaced  spin-spin correlation 
function,
\begin{equation}
    S(\ve{r},\tau) = \left< \ve{\hat{S}}_{\ve{r}}(\tau) \cdot \ve{\hat{S}}_{\ve{0}}(0) \right>.
\end{equation}
Using the ALF \cite{ALF_v2_2} implementation of the stochastic  analytical continuation 
\cite{KSDBeach2004,ASandvik1998}  we can extract 
the   dynamical spin susceptibility: 
\begin{equation}
     S(\ve{k},\tau)  =  \frac{1}{\pi} \int d \omega \frac{e^{-\omega \tau}}{1-e^{-\beta \omega}} \chi''(\ve{k},\omega),
\end{equation}
with 
\begin{equation}
 S(\ve{k},\tau) =   \sum_{\ve{r}}e^{i \ve{k}\cdot \ve{r}} S(\ve{r},\tau).
\end{equation}
From the imaginary part of the  dynamical spin susceptibility, we can obtain the dynamical spin structure factor, 
\begin{equation}
  S(\ve{k},\omega)  = \frac{\chi''(\ve{k},\omega)}{ 1-e^{-\beta \omega}},
\label{eq:Skomega}
\end{equation}
as  well as 
\begin{equation}
  \chi(\ve{k},\omega) =  \frac{1}{\pi} \int  d \omega' \frac{\chi''(\ve{k},\omega')}{\omega-\omega' - i0^{+}}.
\end{equation}

For the spin-1/2  Kondo model, we defined in previous works \cite{RaczkowskiPRL2019,Danu2021}  the composite fermion operator: 
\begin{equation}
 \hat{\psi}^{\dagger}_{\r,\sigma} = \sum_{\sigma^\prime}{\hat c}^\dagger_{\r,\sigma^\prime} {\boldsymbol \sigma}_{\sigma^{\prime},\sigma} \cdot  \sum^{2S}_{n=1}\ve{\hat{S}}_{\r,n}.
\end{equation} 
Here, we will generalise  this quantity  to the spin-3/2  case, 
merely by replacing the spin-1/2  operator  by  the  spin-3/2 one.  $\hat{\psi}^{\dagger}_{\r,\sigma} $  carries  the  same  quantum 
numbers  as the electron \cite{Danu2021} and  we will use  this quantity  to probe for  signatures of Type I and Type II Kondo phases.

\subsection{$S=3/2$  Heisenberg chain on a 2D semi-metal}
\label{sec:semi-metal}
\begin{figure}[htbp]
 \includegraphics[clip, width=8.9cm]{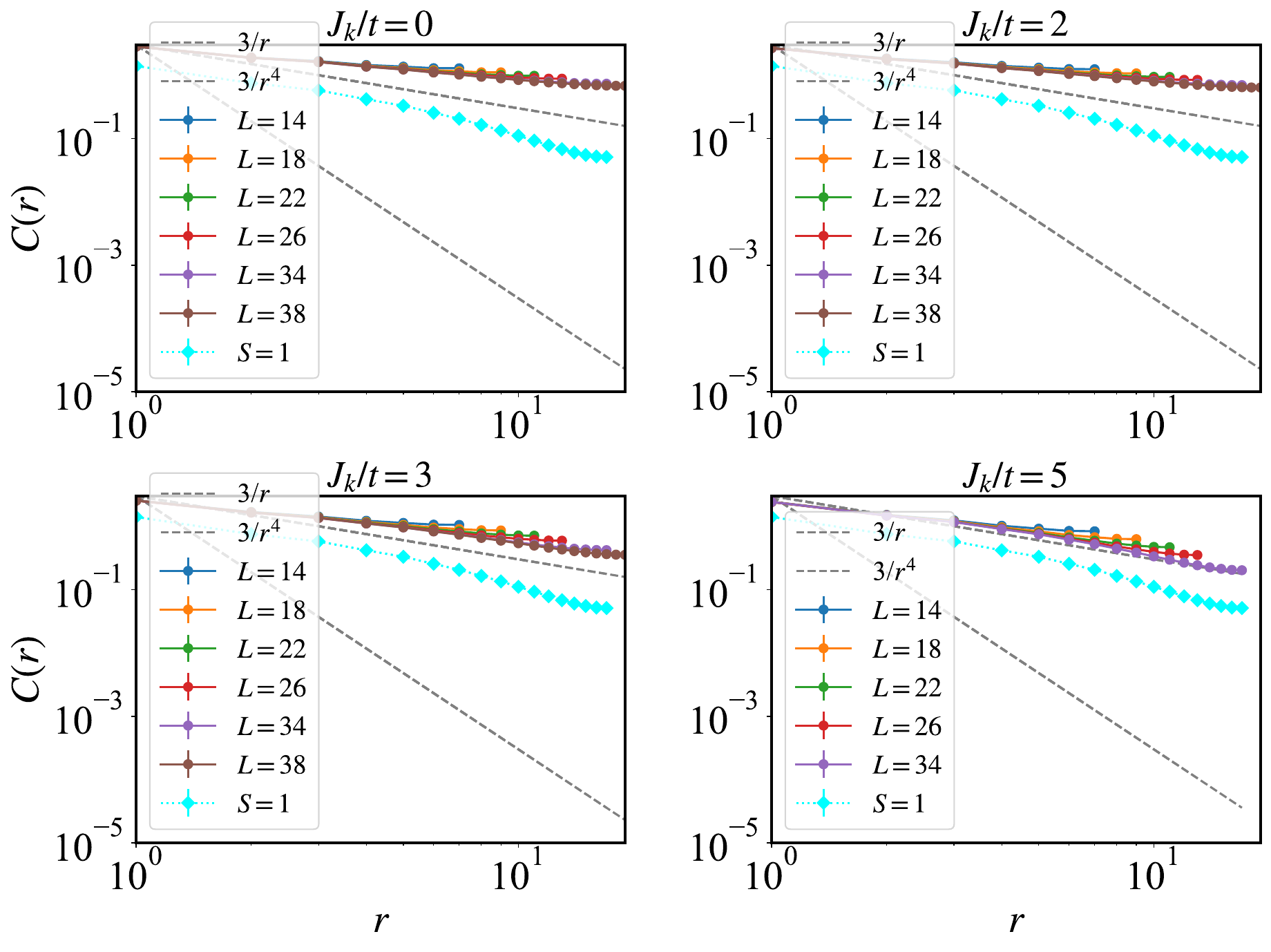}
 \caption{$S=3/2$ Heisenberg chain on a 2D semi-metal with PBC at $\beta t =L, J_h/t=1, D/t=0$.  Real space spin-spin correlations $C(r)$ as a function of distance $r$  along the spin chain  at various $J_k/t$. The cyan colour represents the decoupled spin-1 Heisenberg chain  data at $L=34$. The grey dashed  lines correspond to $3/r$ and $3/r^4$ decays, which represent the  decoupled spin-3/2 chain and 2D  Dirac  electrons spin-spin correlations decays, respectively.}
 \label{fig:Cr_vs_r_BetaL_sm}	
\end{figure}
\begin{figure}[htbp]
 \includegraphics[clip, width=8.9cm]{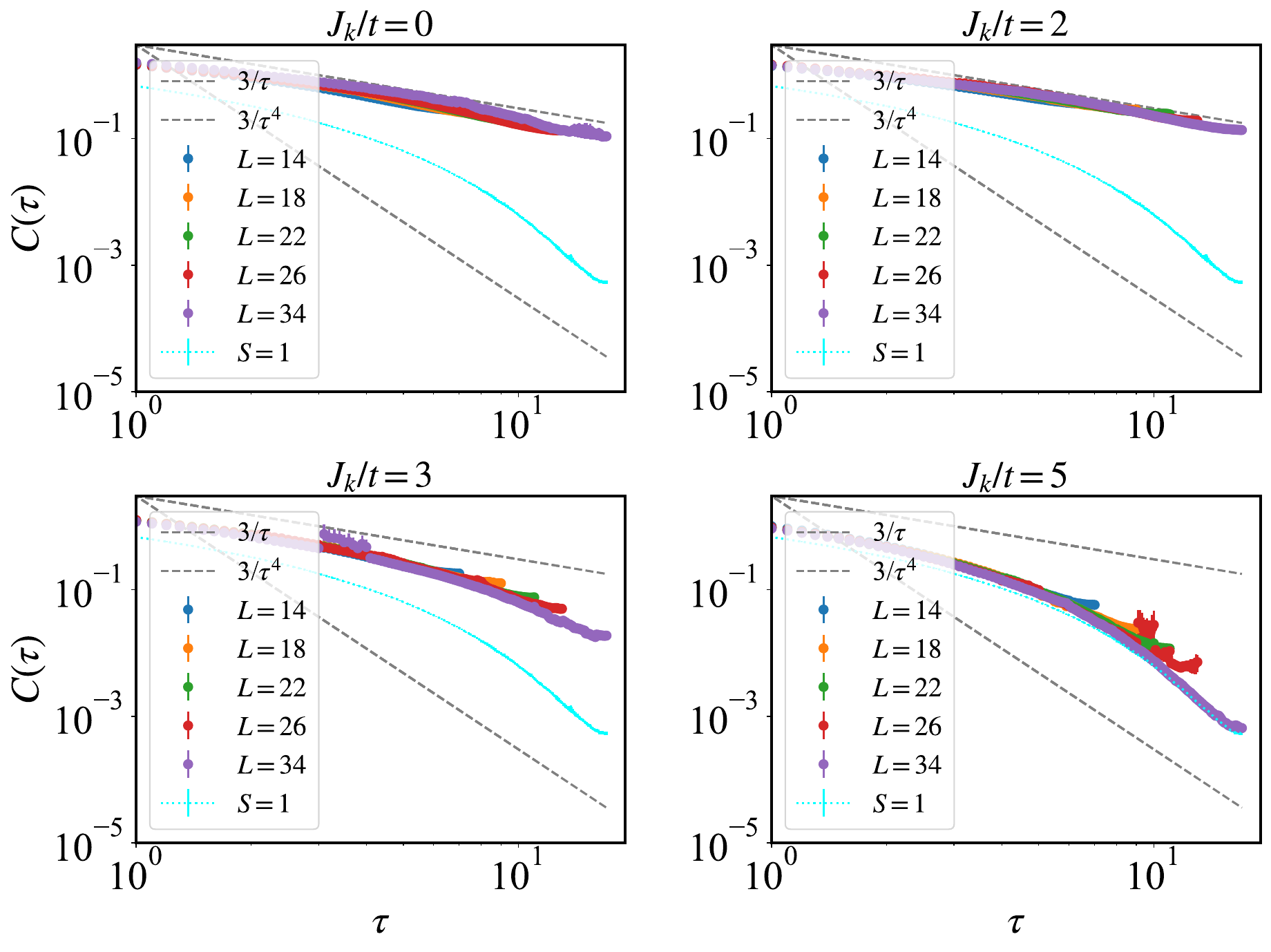}%
  \caption{$S= 3/2$ Heisenberg  chain on a 2D semi-metal  with PBC  at $\beta t =L, J_h/t=1, D/t=0$. Imaginary time  spin-spin correlations $C(\tau)$ as a function of  imaginary time  $\tau$  along  the spin  chain  at various $J_k/t$. The cyan color represents the decoupled spin-1 Heisenberg chain data  at $L=34$. The grey dashed  lines correspond to $3/\tau$ and $3/\tau^4$ decays.} 
 \label{fig:Ct_vs_t_BetaL_sm}
\end{figure}
\begin{figure}[htbp]
 \includegraphics[clip, width=8.9cm]{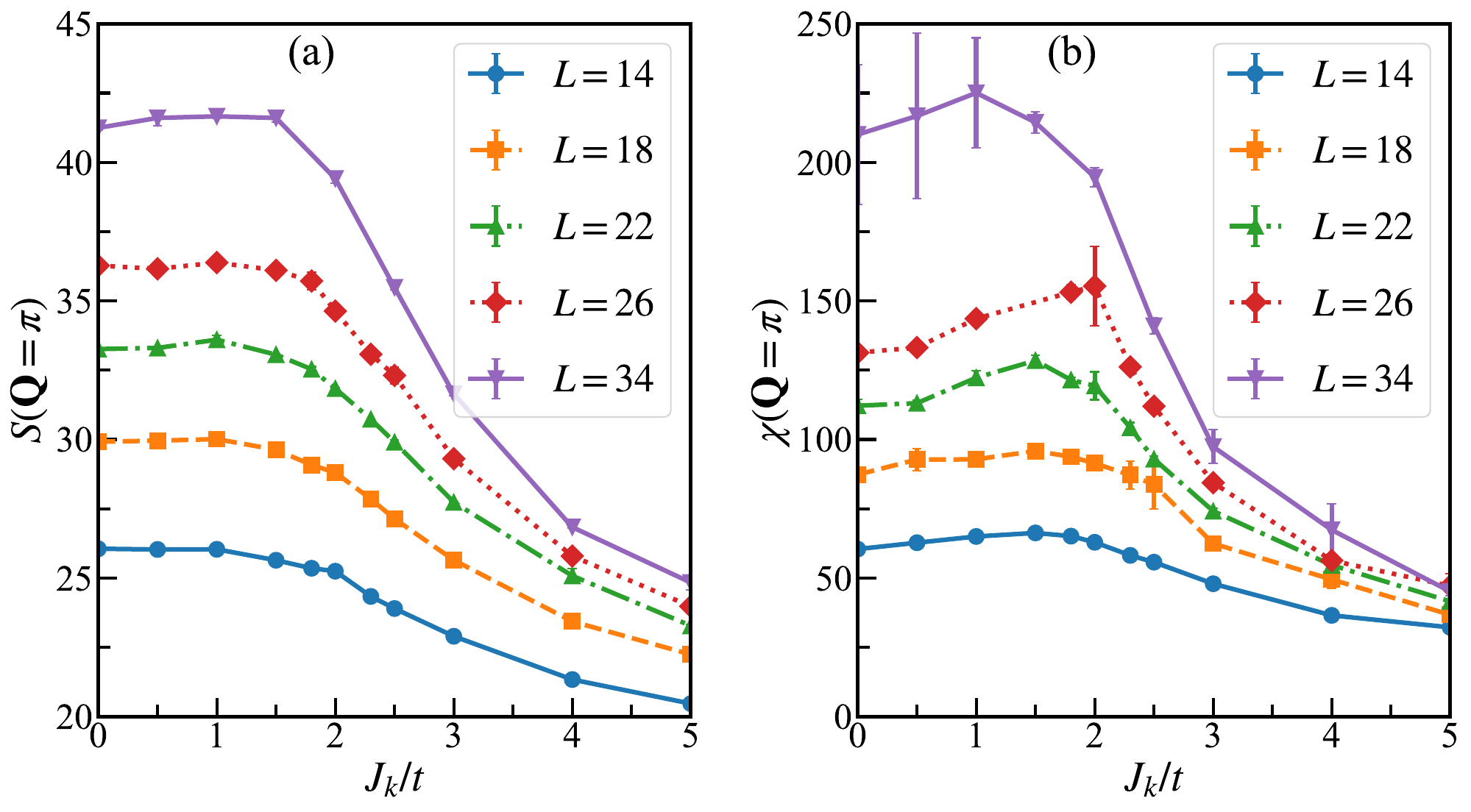}
 \caption{ $S=3/2$ Heisenberg chain on a  2D semi-metal  with PBC  at $\beta t =L, J_h/t=1, D/t=0$.  (a) Static structure factor $S(\Q =\pi)$ along  the spin  chain  at ordering vector $\Q=\pi$  as a function of $J_k/t$. (b)  Spin-susceptibility $\chi(\Q =\pi)$ along  the spin chain  as a function of $J_k/t$.  }
  \label{fig:chik_Sk_vs_k_BetaL_sm}
\end{figure}
Figure \ref{fig:Cr_vs_r_BetaL_sm} show the equal-time ground state spin-spin  correlation  function, $C(\ve{r})= (-1)^{\ve{r}} S(\ve{r},0)$, as a function of distance $\ve{r}$ along  the spin-3/2 chain on a 2D semi-metal, for various values of $J_k/t$ at $\beta t =L$.  For $J_k/t\le 2$,  the correlations follow  the form  $C(r)\propto 1/r$, which is characteristic of the SU(2)-invariant spin-3/2 Heisenberg chain with nearest neighbour couplings. Here we omit logarithmic corrections. 
 For $J_k/t> 2$,  the Kondo coupling becomes effective, leading to a marked deviation from this power law behaviour, signalling the partial Kondo screening of spin-3/2 local moments.   Similarly,  Figure \ref{fig:Ct_vs_t_BetaL_sm} shows the time-displaced correlation function, $C(\tau) =  S(\ve{r}=0,\tau)$, as a function of  imaginary time $\tau$. In the decoupled KBD phase at $J_k/t\le 2$ the correlations decay as $C(\tau) \propto 1/\tau$,  consistent with  Lorentz-invariance of 1D Luttinger liquids. For  $J_k/t>2$, the temporal  decay becomes significantly faster than in the $J_k\le2$ limit, indicating the onset of Kondo screening.

\begin{figure}[htbp]
 \includegraphics[clip, width=8.9cm]{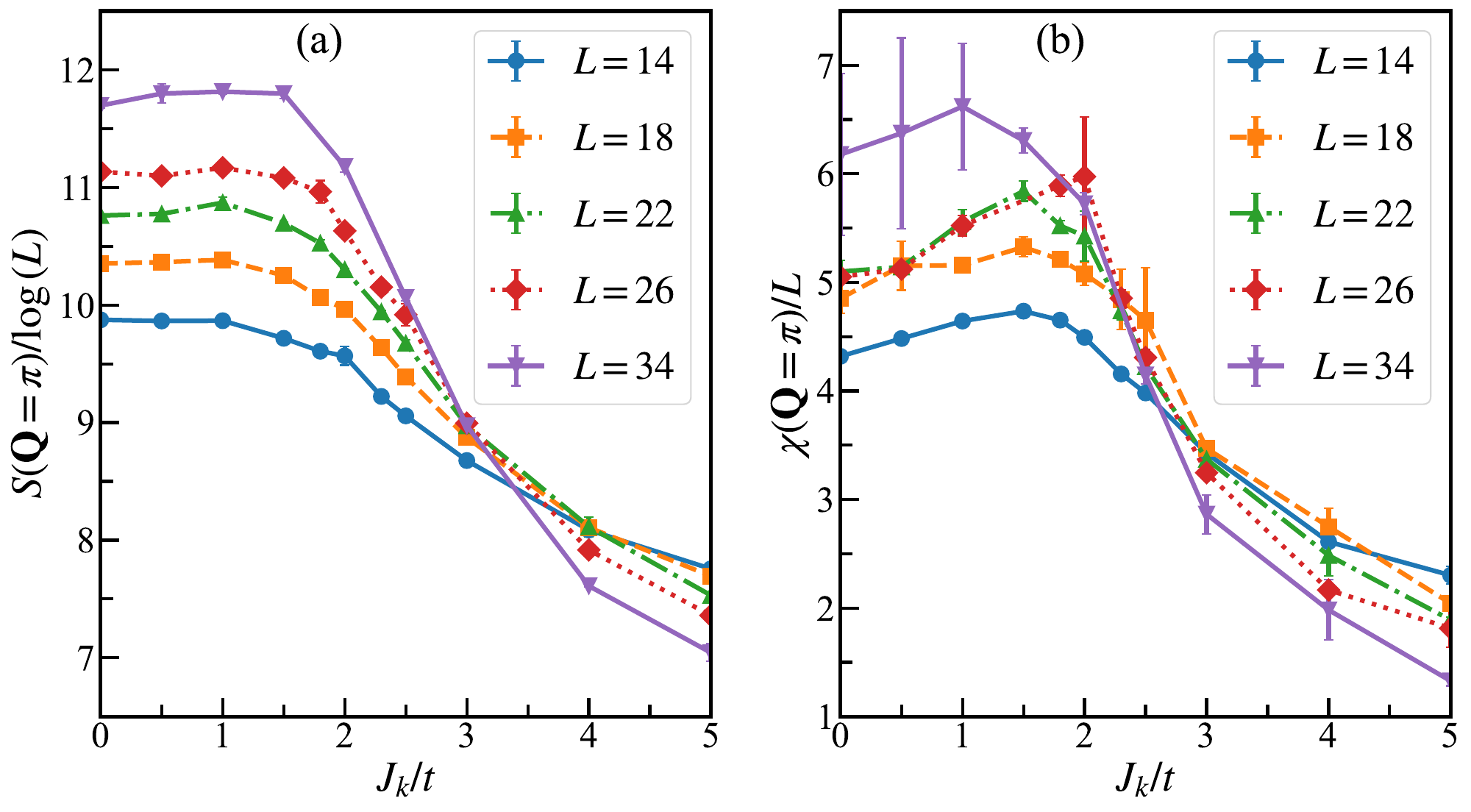}
 \caption{ Normalised spin  structure factor (a)   and  spin susceptibility (b) along the spin-3/2 chain on a semi-metallic surface. }
  \label{fig:chik_Sk_vs_k_BetaL_sm_normalized}
\end{figure}

Figures \ref{fig:chik_Sk_vs_k_BetaL_sm}(a) and \ref{fig:chik_Sk_vs_k_BetaL_sm_normalized}(a) show the static structure factor, $S(\ve{k})=S(\ve{k},\tau=0)$, evaluated at the antiferromagnetic wave vector $\Q=(\pi,0)$ along the spin-3/2 chain, as a function of $J_k/t$. For $J_k/t \leq 2$, $S(\Q)$ exhibits a logarithmic divergence with system size $L$, consistent with the algebraic power-law decay observed at the decoupled point $J_k/t=0$. In contrast, for $J_k/t>2$, the growth of $S(\ve{Q})$ with $L$ slows considerably, reflecting the onset of partial Kondo screening.

At low energies,  Lorentz invariance of the isolated SU(2) spin symmetric  Heisenberg chain implies  that  space-time correlations decay $1/\sqrt{r^2 + (v_s\tau)^2}$ with  $v_s$  being the spin velocity along the chain.  To confirm Kondo breakdown at small $J_k$  we compute the static spin susceptibility $\chi(\ve{k})= \chi(\ve{k},\omega = 0)$ along the spin-3/2 chain. 
By setting $\beta t =L$ we expect the $\chi(\Q)$  to  scale as  $L$.  Figures \ref{fig:chik_Sk_vs_k_BetaL_sm}(b)  and 
\ref{fig:chik_Sk_vs_k_BetaL_sm_normalized}(b) plot $\chi(\Q)$  versus  $J_k/t$ for various system sizes. The data is consistent with a $L$ divergence. For $J_k>2$, the magnitude of  $\chi(\Q)$  is significantly reduced as compared to the $J_k\le2$ limit, indicating the onset of  partial Kondo screening  of the spin-3/2 local moments.

The aforementioned spin data, suggest a  quantum phase transition between  a  KBD phase at \textit{small} values of $J_k$, in  which the low 
energy  features  of the spin-3/2 chain are identical to those of the isolated chain,   and  a large $J_k$  
phase  where the spin correlations are  suppressed. At  $J_k =  \infty $,  the spin correlations  inherit those  of  the 
spin-1 Haldane chain (see Eq.~(\ref{largeJkef})) and are  characterized by exponential decay.   Away   from 
this point, Eq.~(\ref{eq:haldane_diss}) implies that the spin-spin correlations will inherit those of the host metal in space and time, characterized by a $ \frac{1}{\left[ \ve{r}^2 + (v_s \tau)^2 \right]^2 } $ law. Upon inspection of the data, one will see that our lattice sizes are not large enough to capture this asymptotic behaviour unambiguously.
 \begin{figure}[htbp]
\includegraphics[clip, width=9.1cm]{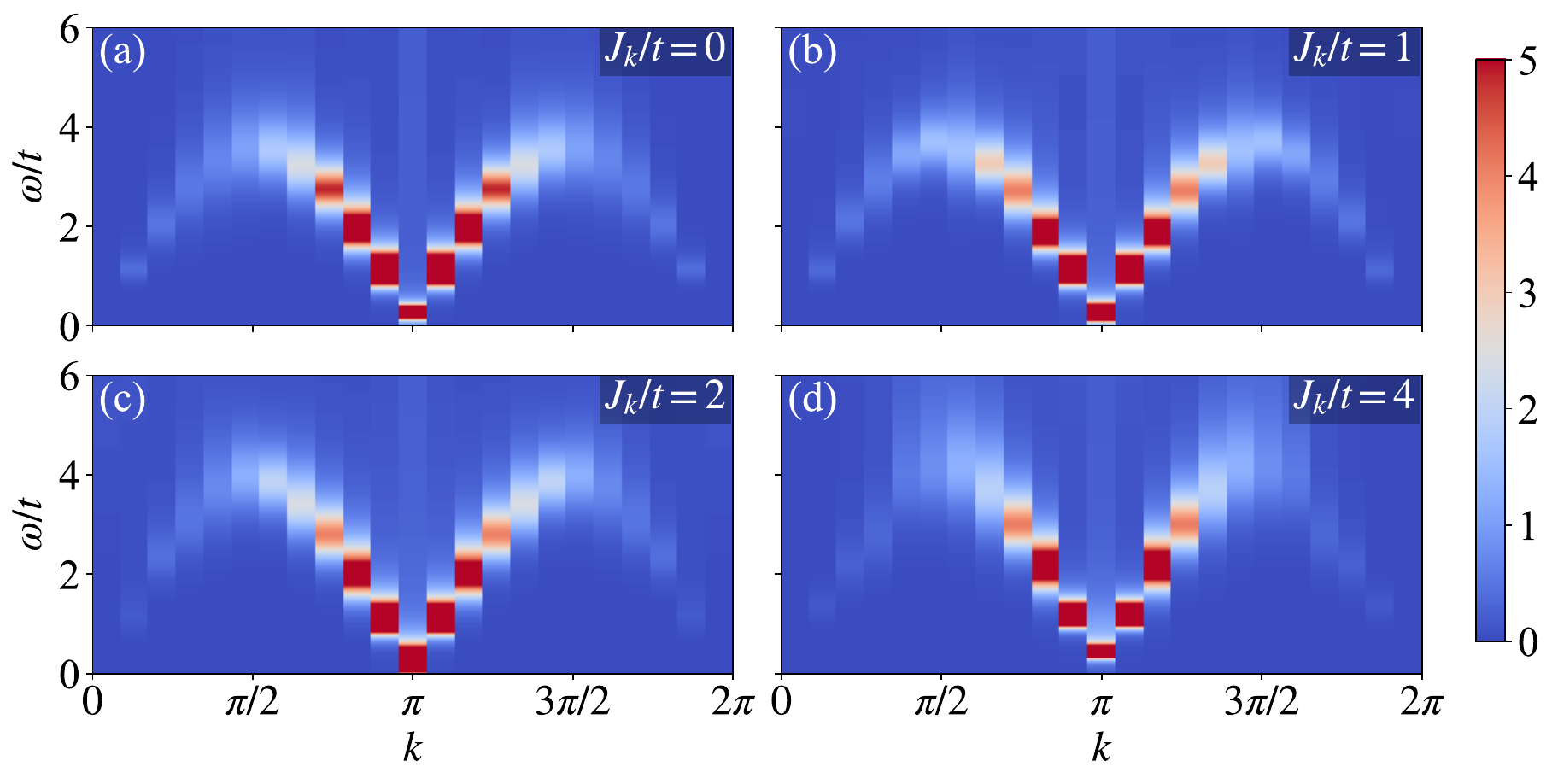}
 \caption{$S=3/2$ Heisenberg chain on a 2D semi-metal with PBC  at $\beta t =L =22$, $J_h/t=1, D/t=0$ for various $J_k/t$ values.  Dynamical structure factor  $S(k,\omega)$  along the spin chain as a function of energy ($\omega/t$) and momentum ($k$).  (a) $J_k/t=0$, (b) $J_k/t=1$, (c) $J_k/t=2$, (d) $J_k/t=4$.} 
 \label{fig:Skomega_vs_k_BetaL_sm_L22_D0}	
\end{figure}

To investigate the spin excitations along the spin-3/2 chain, we  compute the dynamical structure factor as defined by Eq.~(\ref{eq:Skomega}).  Figure~\ref{fig:Skomega_vs_k_BetaL_sm_L22_D0} shows the  $S(\ve{k}, \omega)$ as a function of energy ($\omega$) and momentum ($\ve{k}$) along the spin-3/2 chain for various values of $J_k/t$.   To set the stage, we consider the $J_k=0$ case. As mentioned above, the low energy physics  of the  nearest neighbour spin-3/2 chain maps onto the level-1  WZW  model, such that we expect  the very same spectrum  as  for the  spin-1/2 case~\cite{Danu2020}.   In comparison to the spin-1/2  case of  Ref.~\cite{Danu2020},  we  see  an  enhancement of  the spin-wave velocity  and  a considerable  suppression of the spectral weight  as a function of the two-spinon continuum.  Both of these features  are consistent  with the enhancement  of $S$  from $S=1/2$   to $S=3/2$ and associated suppression of quantum fluctuations.  In the KBD phase at  $J_k = 1$ and $J_k = 2$ we observe  very similar spin  dynamics as for the $J_k = 0$  case.  This is consistent  with the notion that spin and electronic  degrees of  freedom decouple at low energies in the KBD  phase.  

At $J_k=4$  we observe a depletion of low-lying spectral weight near the antiferromagnetic point. Precisely at $J_k = \infty$, and owing to Eq.~(\ref{largeJkef}), we expect a gap of magnitude $\Delta_S\sim0.41 \frac{25}{16}J_h$~\cite{Todo01}, akin to spin-1 Haldane chains. Away from this limiting point, however, the coupling between the gapless fermion modes and spin degrees of freedom will fill the gap with spectral weight. 
 \begin{figure}[htbp]
 \includegraphics[clip, width=8.9cm]{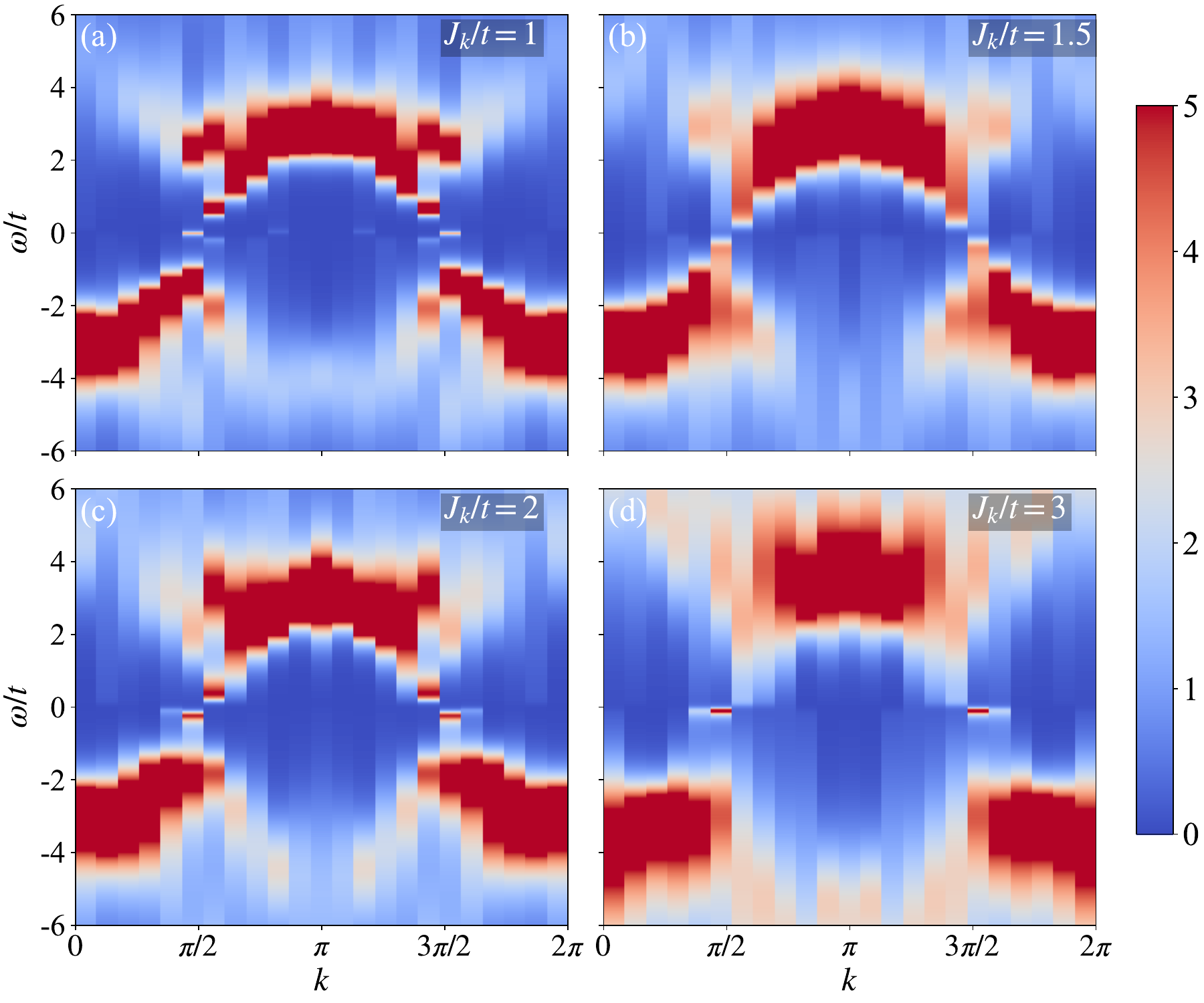}
 \caption{Composite fermion spectral function $A^\psi(k,\omega)$  as a function of  $k$ and $\omega/t$  along the $S=3/2$ Heisenberg chain on a 2D semi-metal with PBC  at $\beta t =L =22$, $J_h/t=1, D/t=0$ for various $J_k/t$ values:  (a) $J_k/t=1$, (b) $J_k/t=1.5$, (c) $J_k/t=2$, (d) $J_k/t=3$.}  
 \label{fig:Apsi_vs_kL_sm_L22_D0}	

 ~
  
  \includegraphics[clip, width=8.9cm]{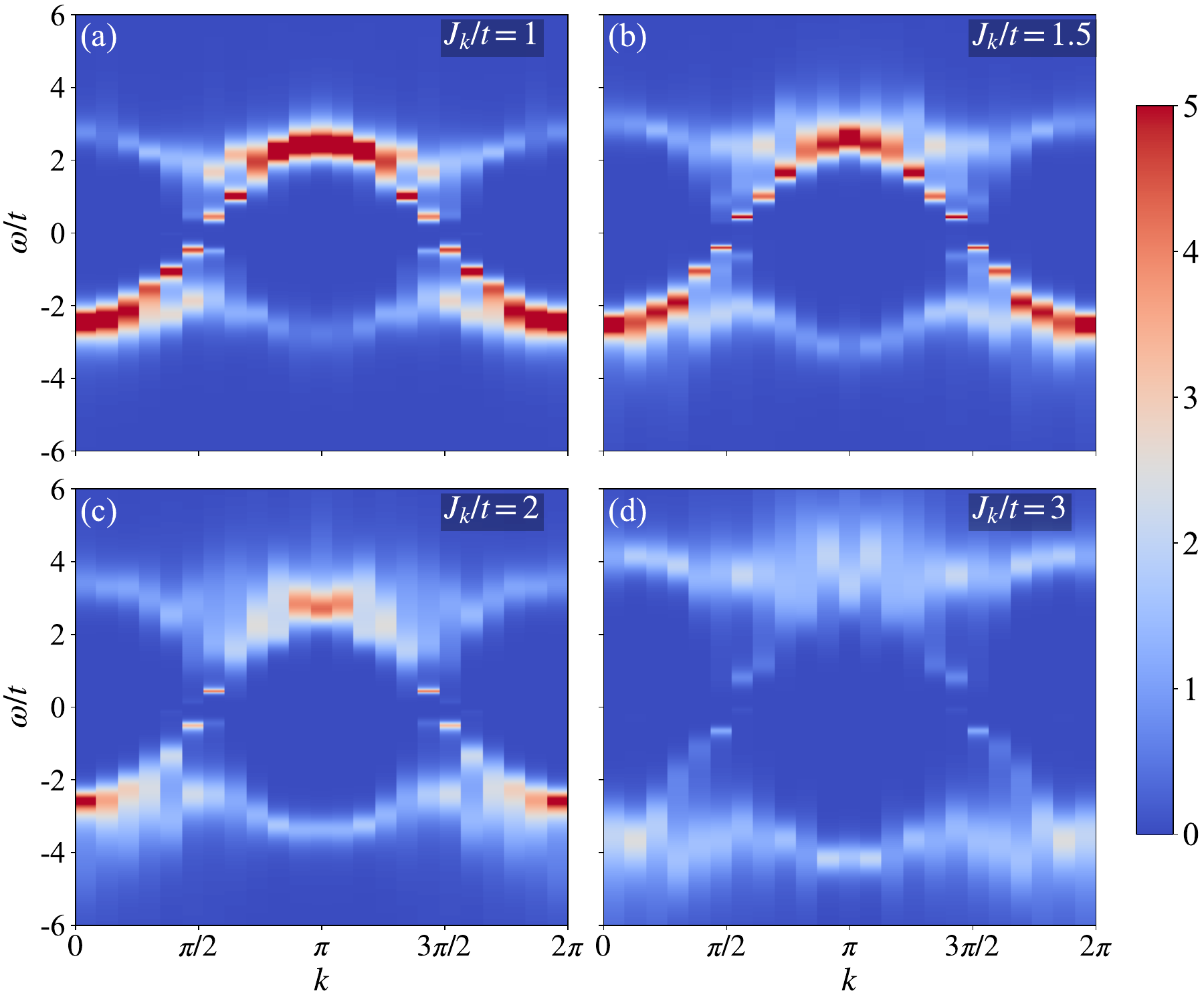}
 \caption{Conduction electrons spectral function $A^c(k,\omega)$  as a function of  $k$ and $\omega/t$  along the Kondo coupled row of 2D Dirac electrons with PBC  at $\beta t =L =22$, $J_h/t=1, D/t=0$ for various $J_k/t$ values:  (a) $J_k/t=1$, (b) $J_k/t=1.5$, (c) $J_k/t=2$, (d) $J_k/t=3$.}%
 \label{fig:Ac_vs_kL_sm_L22_D0}	
\end{figure}

As mentioned in Sec.~\ref{sec:kondo_phases}, the $J_k=\infty$ point is an instance of a Type II Kondo phase, where the composite fermion mode is not a well-defined low-lying electronic excitation. In this limit, the composite fermion excitation involves an energy cost set by $J_k$, since it breaks the low-lying $S=1$ degree of freedom, $|\ve{r},1,m\rangle$, defined in Sec.~\ref{sec:strong_coupling}, to an $S=3/2$ one. To address this point, we compute the composite fermion spectral function $A^\psi(\ve{k},\omega) = -\frac{1}{\pi} \text{Im} G^{\psi} (\ve{k},\omega)$ from the retarded Green's function, 
\begin{eqnarray}
G^{\psi}(\ve{k},\omega) = -i\int^\infty_0  d t  e^{i \omega t} \sum_\sigma \Big\langle\big\{ \hat{\psi}_{\ve{k},\sigma} (t),  \hat{\psi}^\dagger_{\ve{k},\sigma}\big\} \Big\rangle,
 \end{eqnarray}
where $\hat{\psi}^\dagger_{\ve{k},\sigma}   = \frac{1}{{\sqrt L}} \sum_{\ve{r}} e^{i\ve{k}\cdot \ve{r}} \hat{\psi}^\dagger_{\ve{r},\sigma}$.
The zero frequency signal of composite fermion spectral function, $A^\psi(\omega=0)=\frac{1}{L} \sum_{\ve{k}} A^\psi(\ve{k},\omega=0)$,  corresponds to the zero-bias  differential conduction $\left. dI/dV\right|_{V=0} $  in STM measurements.     We can 
estimate this quantity directly from the Monte Carlo data  using the following: 
 \begin{eqnarray}
  \label{eq:dIdV}
\left. dI/dV \right|_{V=0} =A^\psi (\omega=0)  \simeq \frac{1}{\pi} \frac{\beta}{L}  \sum_{\ve{k}} G^\psi(\ve{k}, \tau=\beta/2). \nonumber\\
\end{eqnarray}
Figure~\ref{fig:Aomega0_vs_k_BetaL_m_L22}(a) plots this quantity  for  various values of  the Kondo coupling.  Interestingly,  we see that 
for the semi-metallic surface,  this quantity  remains vanishingly small across the  KBD  transition. Within our resolution, our results  
support a type II Kondo phase  as discussed in Sec.~\ref{sec:kondo_phases}.

In  the KBD  phase, it is instructive to consider the momentum  dependence of both the composite fermion  spectral  function 
as  well as that of the conduction electrons  that couple  to the impurity spins: $A^c(\ve{k},\omega) = - \frac{1}{\pi}\text{Im} G^{c}
(\ve{k},\omega)$   with,
\begin{eqnarray}
G^{c}(\ve{k},\omega) = -i\int^\infty_0  d t  e^{i \omega t} \sum_\sigma \Big\langle\big\{ \hat{c}_{\ve{k},\sigma} (t),  \hat{c}^\dagger_{\ve{k},\sigma}\big\} \Big\rangle.
\end{eqnarray}
Since both the composite fermions and conduction electrons carry the same quantum numbers, we expect the support of the two spectral functions to be identical. Figures~\ref{fig:Apsi_vs_kL_sm_L22_D0} and \ref{fig:Ac_vs_kL_sm_L22_D0} confirm this. In the KBD phase, $J_k=1, 1.5$ the conduction electron spectral function 
exhibits  a cosine band,  with shadow features stemming from the coupling to the spin 3/2 chain (see Figures~\ref{fig:Apsi_vs_kL_sm_L22_D0} (a) and (b)). In contrast 
the composite fermion spectral function, Figures~\ref{fig:Ac_vs_kL_sm_L22_D0} (a) and (b),  shows broader features and  very little low energy spectral weight. We understand this incoherent behaviour of the composite spectral function to be a characteristic of the KBD phase~\cite{RaczkowskiPRB2022}.   In particular, in the KBD phase the spin 3/2  and conduction electrons decouple at low energies such that one can omit vertex corrections: 
\begin{eqnarray}
 \big\langle {\hat \psi}^{}_{\ve{r},\sigma}  {\hat \psi}^\dagger_{\ve{r}',\sigma}(\tau)  \big\rangle \simeq 
 \big\langle {\hat \S}_{\ve{r}}  {\hat \S}_{\ve{r}'}(\tau)  \big\rangle  \big\langle  {\hat c}^{}_{\ve{r},\sigma} {\hat c}^\dagger_{\ve{r}',\sigma} (\tau)  \big\rangle. 
\end{eqnarray}
As a consequence, the composite fermion spectral function is given by the convolution of the spin and conduction electron spectral functions \cite{Danu2021}. 
In the strong coupling limit,  both the composite  fermion (Figures~\ref{fig:Apsi_vs_kL_sm_L22_D0}(c)-(d)) and conduction electron (Figures~\ref{fig:Ac_vs_kL_sm_L22_D0}(c)-(d)) show a shift of spectral weight to higher energies. As mentioned above, removing or adding an electron breaks  the $S=1$  degree of freedom and reveals the $S=3/2$ impurity spin.   In other words,  the Kondo hole corresponds to a spin 3/2  degree of freedom and involves an energy scale set by $J_k$. The dispersion
relation tracks  the propagation of the  Kondo hole \cite{Tsunetsugu97_rev}. 
\subsection{$S=3/2$  Heisenberg chain on a  2D metal}
\label{sec:metal}
\begin{figure}[htbp]
 \includegraphics[clip, width=8.9cm]{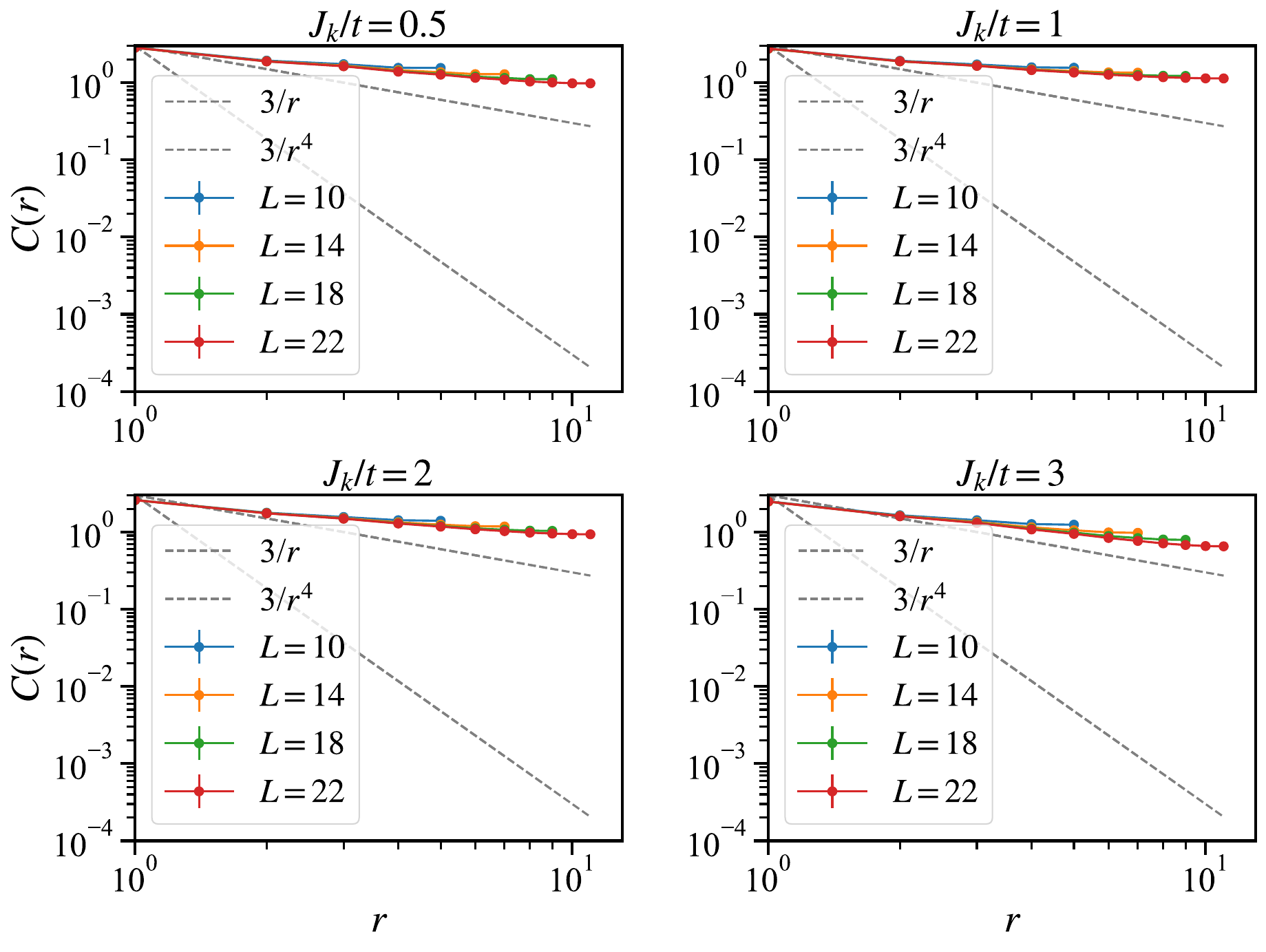}
  \caption{$S=3/2$ Heisenberg chain on a 2D metal  with PBC.  Real space spin-spin correlations $C(r)$ as a function of $r$ along  the spin chain  at various $J_k/t$ at $\beta t =L^2/4, J_h/t=1, D/t=0$. The grey dashed  lines correspond to $3/r$ and $3/r^4$ decays.}
  \label{fig:Cr_vs_r_BetaL_m}
\end{figure}
\begin{figure}[htbp]
 \includegraphics[clip, width=8.9cm]{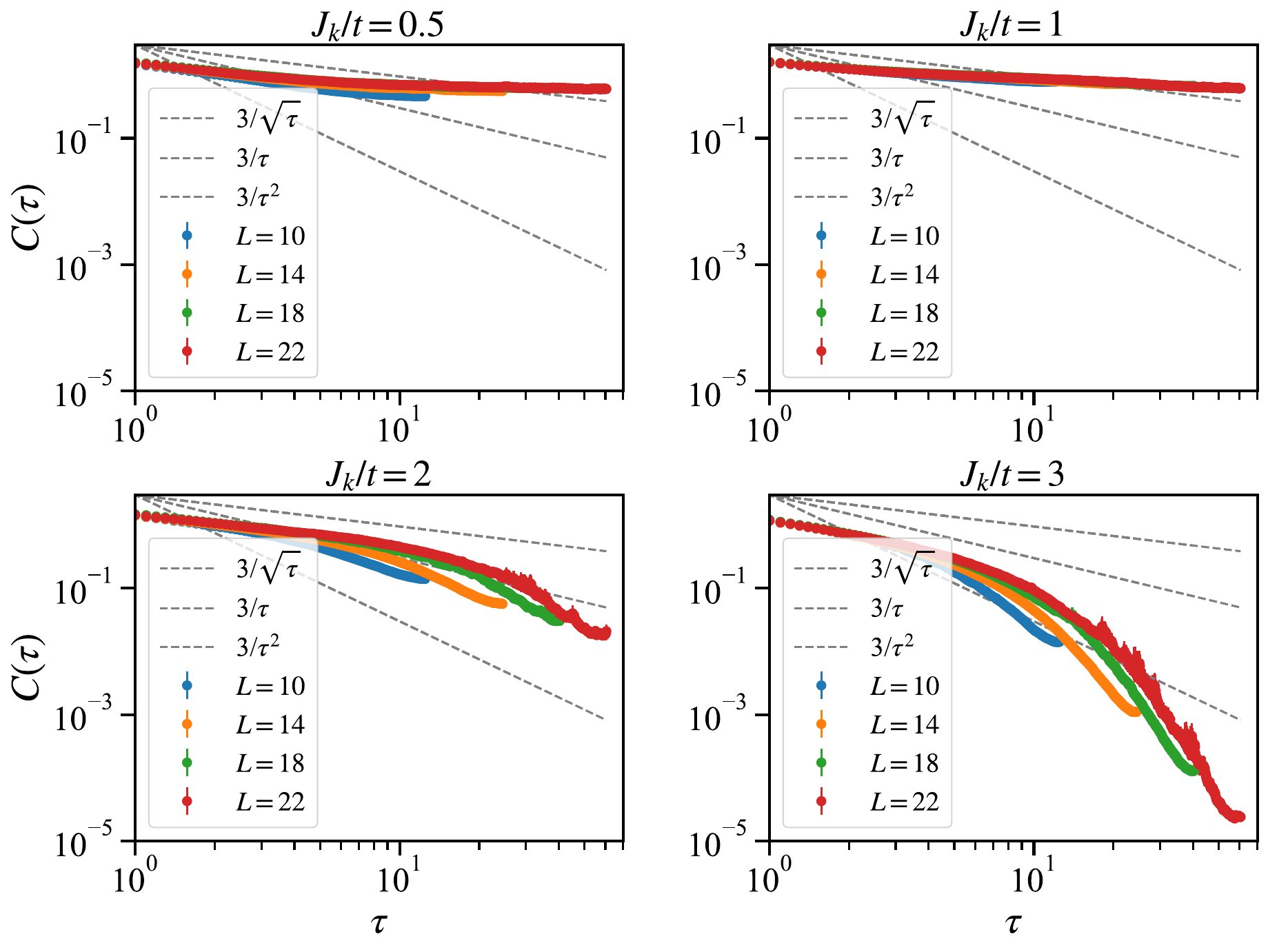}
  \caption{$S= 3/2$ Heisenberg  chain on a 2D metal with PBC.  Time displaced spin-spin correlations $C(\tau)$ as a function of  $\tau$   along the spin  chain   at various $J_k/t$ at $\beta t =L^2/4$ scaling and $J_h/t=1, D/t=0$. The grey dashed  lines correspond to the  $3/\sqrt{\tau}, 3/\tau$, and $3/\tau^2$ decays, which  respectively correspond to  the dissipative chain, the decoupled spin-3/2 Heisenberg chain  and  the 2D conduction electrons correlation decays.}
  \label{fig:Ct_vs_t_BetaL_m}
\end{figure}
As  mentioned  in Sec.~\ref{sec:Jk_small}  the metallic surface is described by an  Ohmic  bath that leads 
to dissipation induced long ranged order Ref.~\cite{Weber2021}. Before  discussing  the numerical data,  we provide 
an overview of the effective field theory description of this state.  Let $\ve{n}(x,\tau)$ be a space-time
dependent vector on  $S^2$ that captures the spin fluctuations of this ordered state. The effective field theory then reads: 
\begin{eqnarray}
  {\cal S}& &= \int dx d \tau \frac{\rho_s}{2} \left[  (\partial_{\tau} \ve{n})^2 + (\partial_{x} \ve{n})^2 \right] -  
  \nonumber \\ 
  & & \alpha \int dx d\tau d \tau' \frac{\ve{n}(x,\tau) \cdot \ve{n}(x,\tau')}{(\tau - \tau')^{2}},
\end{eqnarray}
with $\alpha$ a coupling constant set by $J_k^2$.
We will assume long range-order in the $z$-direction  and  consider  small  transverse fluctuations:  $\ve{n}(x,\tau) = (\ve{n}_{\perp},\sqrt{1-|\ve{n}_{\perp}|^2})$ with $\ve{n}_{\perp}$ an $O(2)$ vector  satisfying $|\ve{n}_{\perp}| \ll 1$.  The action  for the transverse fluctuations is  given by ${\cal S} = {\cal S}_0 + {\cal S}  _1 $   with,
\begin{eqnarray}
   {\cal S}_0 & &  =\int dx d \tau \frac{\rho_s}{2} \left[  (\partial_{\tau} \ve{n}_{\perp})^2 + (\partial_{x} \ve{n}_{\perp})^2 \right] - 
  \nonumber \\ 
  & & \alpha \int dx d\tau d \tau' \frac{\ve{n}_{\perp}(x,\tau) \cdot \ve{n}_{\perp}(x,\tau')}{(\tau - \tau')^{2}}.
\end{eqnarray} 
${\cal S}_0$ is scale invariant under the  transformation $x \rightarrow bx$, $\tau \rightarrow b^z  \tau$ with $z=2$   and scaling dimension 
of $\ve{n}_{\perp}$ set to  $\Delta n_{\perp} = 1/2$.  At this fixed point all terms in ${\cal S}_1$ are irrelevant.  This field theory captures 
the dissipation induced long ranged ordered phase.  Since $z=2$  we  have to scale $\beta$ as  $L^2$ so as to  capture  ground state properties.   Specifically, we have chosen  $\beta t=L^2/4$. 

In the Monte Carlo  simulations, we   can detect long ranged order  by computing spin correlations. 
Figures~\ref{fig:Cr_vs_r_BetaL_m} and~\ref{fig:Ct_vs_t_BetaL_m}  respectively show the spatial and temporal spin-spin correlations along the spin-3/2 Heisenberg chain for various values of $J_k/t$.
For $J_k/t \leq 2$, the spatial correlations  certainly  show  a  decay  that is much slower than $1/r$.  We note that the Ohmic bath is 
a marginally relevant perturbation  such that  at \textit{small}  values of $J_k$ very large lattices are required to observe long ranged 
order \cite{Weber2021}.   In the temporal  direction, the signal is much more clear.  Since  $z=2$  we  can probe much longer distances in imaginary time  as opposed to real  space before confronting finite size effects.  In particular, the data for  $J_k < 2$ shows next to no  decay in imaginary time.  At  $J_k \geq 2$ the temporal correlation  clearly  show the breakdown of  dissipation  induced long  ranged order.  
\begin{figure}[htbp]
 \includegraphics[clip, width=8.9cm]{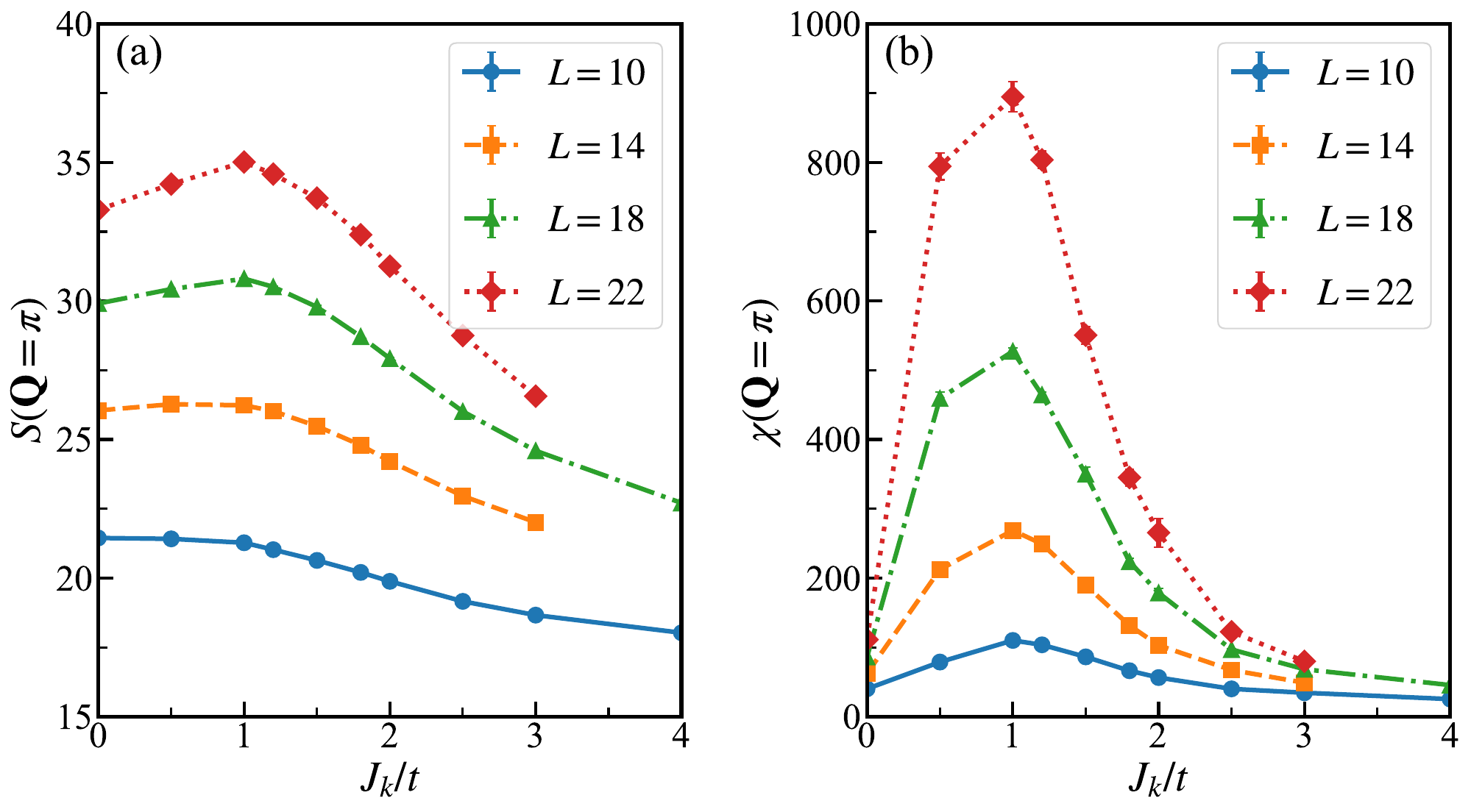}
 \caption{$S= 3/2$ Heisenberg chain on a  2D metal with PBC  at $\beta t =L^2/4, J_h/t=1, D/t=0$.  (a)  Static structure factor $S(\Q =\pi)$   along  the spin chain   as a function of $J_k/t$. (b) Spin-susceptibility $\chi(\Q =\pi)$  along the spin chain  as a function of $J_k/t$.}
 \label{fig:chik_Sk_vs_k_BetaL_m}
\end{figure}

This interpretation is further corroborated by the data of the static structure factor $S(\mathbf{Q})$ and the spin susceptibility $\chi(\mathbf{Q})$ at $\beta t=L^2/4$ (see Figure~\ref{fig:chik_Sk_vs_k_BetaL_m}(a) and~\ref{fig:chik_Sk_vs_k_BetaL_m}(b), respectively).
As  mentioned previously, at $J_k = 0$,  $S(\ve{Q})$ is  expected to grow logarithmically.   For \textit{small} lattices, $L=10$, and 
\textit{small}  values of  $ J_k  = 0.5, 1$   we see no major  deviations for the  $J_k = 0 $ value of $S(\ve{Q})$.  On the other hand, for large lattices, $L=22$,   $S(\ve{Q})$ at  $J_k = 1$  is greater than the $J_k=0$ values.   On the other hand, at   large values of $J_k$  the
antiferromagnetic spin-spin correlations are suppressed.  This overall behaviour is much more apparent in the staggered spin susceptibility. 

\begin{figure}[htbp]
 \includegraphics[clip, width=8.9cm]{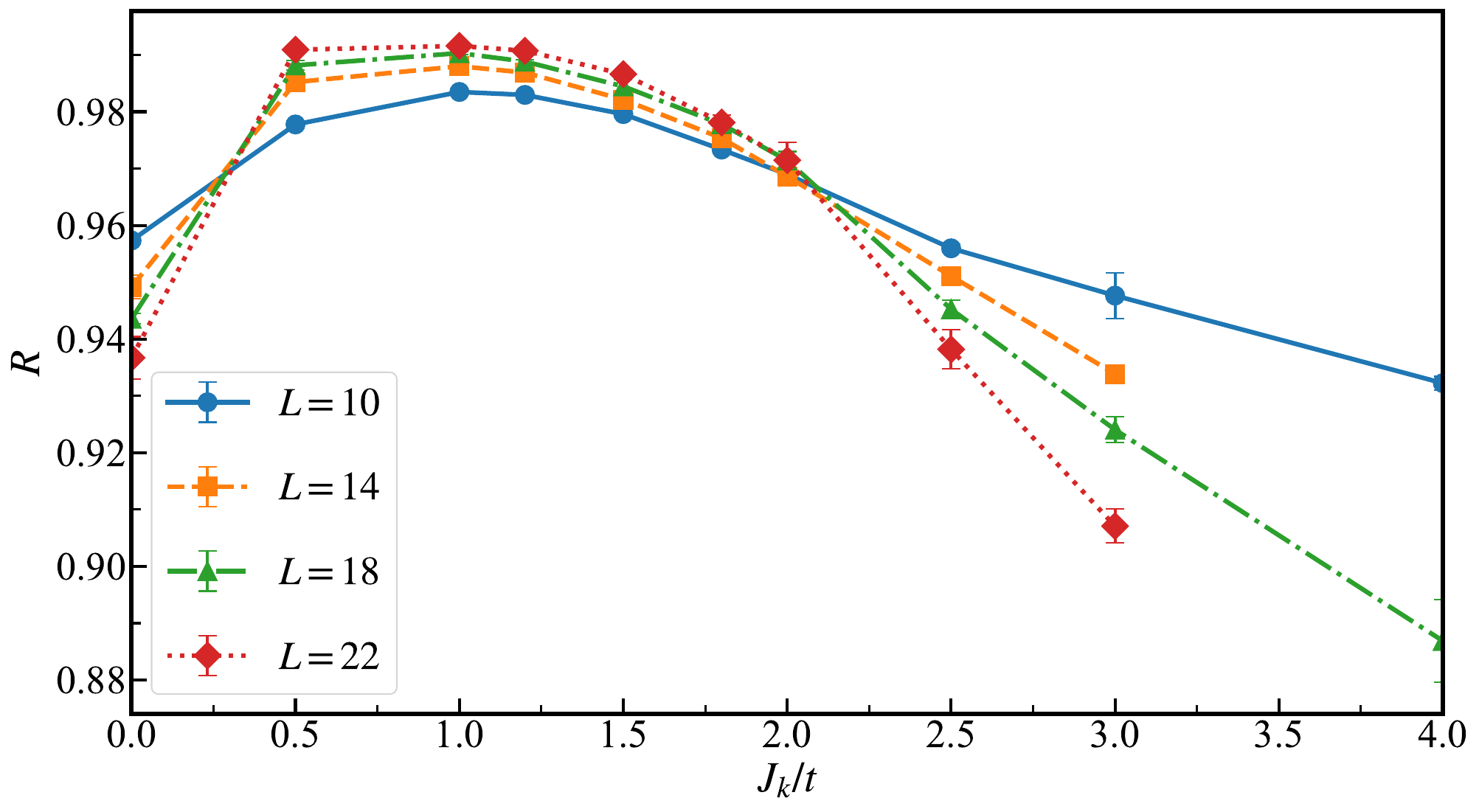}
 \caption{$S= 3/2$ Heisenberg chain on a  2D metal with PBC  at $\beta t =L^2/4, J_h/t=1, D/t=0$. Correlation ratio $R$  along the spin chain  as a function of $J_k/t$.}
  \label{fig:R_vs_Jk_BetaL_sm}
\end{figure}

To further characterize the quantum phases, we compute  the correlation ratio,
\begin{eqnarray}
R =1-\frac{\chi{(\ve{Q}-\delta \ve{ Q_0} } )} {\chi(\ve{Q})} \, ,
\end{eqnarray}
with $\ve{Q}= (\pi, 0)$,  and $\delta \ve{Q_0}$ the smallest non-zero momentum  deviation from $\Q$ along the spin-3/2 Heisenberg chain. This is a renormalization group  invariant quantity that scales to unity in the ordered phase and to zero in the disordered phase. In the vicinity of a quantum critical point,  $R= f(\left[J_k -  J_k^{c} \right] L^{1/\nu}, L^{z}/\beta, L^{-\omega})$. Here, $\nu$ is the correlation length exponent, $z$ is the dynamical exponent, and $\omega$ captures corrections to scaling. Figure \ref{fig:R_vs_Jk_BetaL_sm} plots correlation ratio as a function of $J_k/t$ at $\beta t= L^2/4$. As clearly seen, the AFQMC data  show that $R$ approaches to one  in the AFM  phase and decreases towards zero in the under-screened Kondo phase with increasing system size. The two phases are separated by a continuous quantum critical point at $J^c_k/t\simeq2$ with the dynamical exponent $z \simeq 2$.

\begin{figure}[htbp]
 \includegraphics[clip, width=9.1cm]{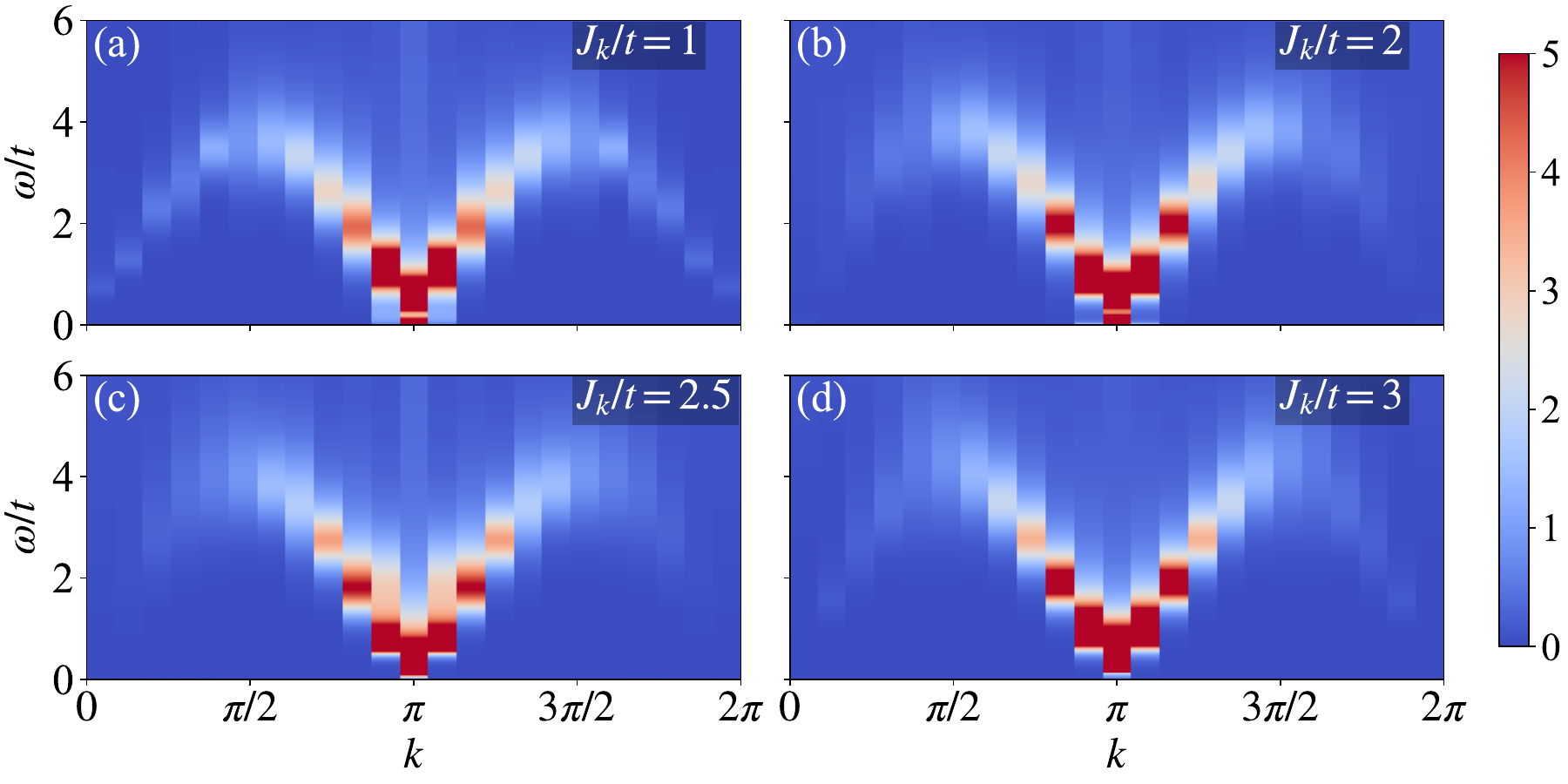}
 \caption{$S(k,\omega)$ as a function of  $k$ and $\omega/t$   along the  $S=3/2$ Heisenberg chain on a  2D metal  with PBC  at $\beta t = L^2/4$, $L =22$,  $J_h/t=1, D/t=0$ for various $J_k/t$ values: (a) $J_k/t=1$, (b) $J_k/t=2$, (c) $J_k/t=2.5$, (d) $J_k/t=3$.} %
 \label{fig:Skomega_vs_k_BetaL2by4_m_L22_D0}	
\end{figure}

Figure \ref{fig:Skomega_vs_k_BetaL2by4_m_L22_D0}  shows  the dynamical structure factor $S(k,\omega)$ along the spin-3/2 Heisenberg chain on a 2D metal for various $J_k/t$ values.  For $J_k/t\le2$, the data indicates the gapless spectrum but lacks the linear dispersion $\omega\propto k$ characteristic of $J_k=0$ limit.  Instead, it exhibits a quadratic-like dispersion $\omega\propto k^2$,  an instance of Landau-damped Goldstone mode as expected from  dissipative coupling to an Ohmic bath~\cite{Weber2021, Danu2022}. For  $J_k/t>2$,  we observe a depletion of spectral weight at low energies.  

While at  weak coupling the physics of  the spin chain on semi-metallic and  metallic surfaces differ considerably they are
similar at strong coupling.  In both cases, the relevant Hamiltonian  is that of Eq.~(\ref{largeJkef})  describing a spin-1 Haldane chain, 
ferromagnetically coupled to the metallic environment. The emergence of this state of matter occurs via partial Kondo 
screening,   and a Luttnger volume, as defined in Sec.~\ref{sec:kondo_phases} that includes  an additional degree of freedom. We now 
discuss if this state,  for the metallic surface is a Type I or  Type II heavy fermion  metal.
 \begin{figure}[htbp]
 \includegraphics[clip, width=8.9cm]{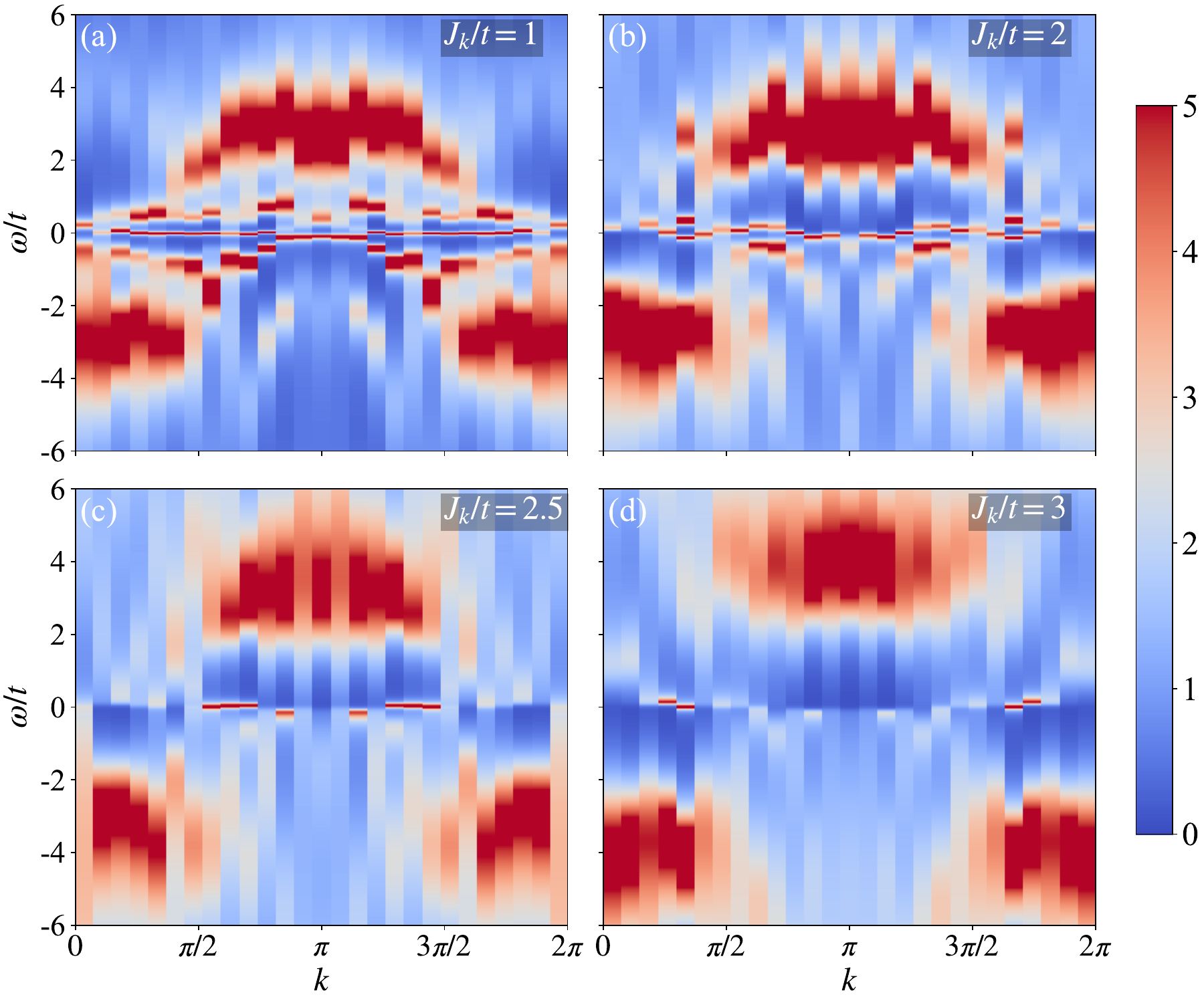}
 \caption{$A^\psi (\ve{k},\omega)$ as a function of $k$ and $\omega/t$   along   the $S=3/2$ Heisenberg chain on a  2D metal with PBC  at $\beta t = L =26, J_h/t=1, D/t=0$ for various $J_k/t$ values: (a) $J_k/t=1$, (b) $J_k/t=2$, (c) $J_k/t=2.5$, (d) $J_k/t=3$.} 
 \label{fig:Apsi_vs_k_BetaL_sm_m}	
\end{figure}

The composite fermion spectral function, $A^\psi(k,\omega)$, shown in Figure~\ref{fig:Apsi_vs_k_BetaL_sm_m},  reveals the continued existence  of a low-lying heavy fermion band in both phases. 
 \begin{figure}[htbp]
  \includegraphics[clip, width=8.9cm]{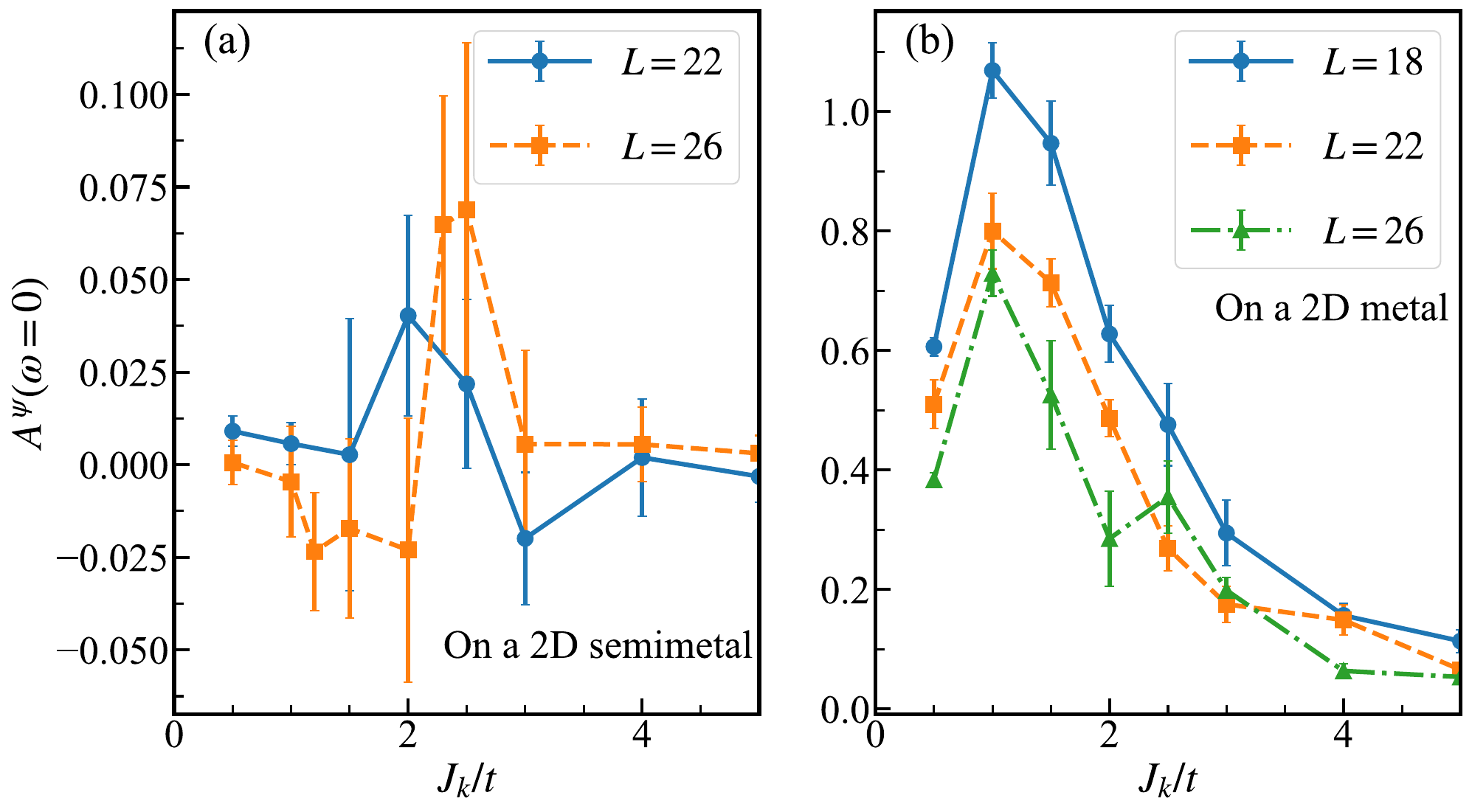}
 \caption{Composite fermion  zero frequency spectral weight,  $A^\psi(\omega =0)$, along the periodic spin chain as a function of $J_k/t$ at  $L=\beta t $ $J_h/t=1, D/t=0$. (a) $S=3/2$  Heisenberg chain on a 2D semi-metal. (b) The same on a 2D metal.}
 \label{fig:Aomega0_vs_k_BetaL_m_L22}	
\end{figure}

 Figure~\ref{fig:Aomega0_vs_k_BetaL_m_L22}(b) shows  $A^\psi (\omega=0)$  as defined in Eq.~(\ref{eq:dIdV}) versus $J_k/t$   for 
 the metallic surface.  The data  confirms the existence of low lying spectral weight in the composite fermion spectral function.  
 In the  partially screened Kondo  metallic phases  this  implies a Type I  Kondo metallic phase at  finite values of $J_k$. 
 The data also  confirms  that the heavy fermion quasiparticle is  equally present in the dissipation induced long ranged ordered phase. 
\subsection{Massless spin-$\frac{1} {2}$ edge modes}
\label{sec:edge}
\begin{figure}[htbp]
 \includegraphics[clip, width=9.cm]{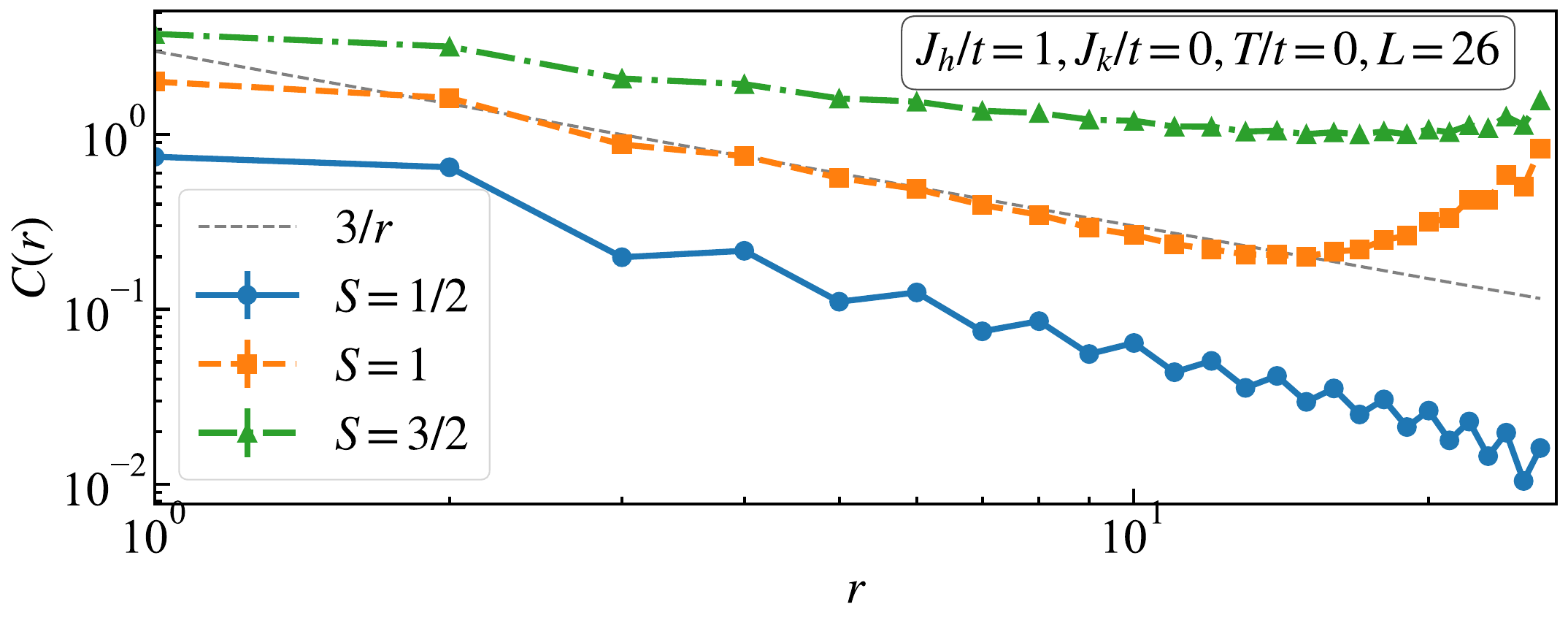}
 \caption{$S=1/2, 1, 3/2$ Heisenberg chain with OBC  at $L=26, T/t=0, J_k/t=0, J_h/t=1, D/t=0$. Real space spin-spin correlations $C(r)$ as a function of distance $r$.}
 \label{fig:Cr_vs_r_BetaL_sm_OPBC_L26_Jkp0}	
\end{figure}

An isolated spin-1 is characterized by  the non-local string order parameter~\cite{NijsPRB1989} that picks up hidden long-ranged antiferromagnetic order.  
Another characteristic of this  symmetry protected topological state of matter are localized spin 1/2 degrees of freedom on open manifolds \cite{AKLT1987}.  Here we will use the second 
criterion, and  carry out  simulations on open topologies as depicted in Figure~\ref{fig:Sketch_opbc_latt}. Let us first consider an isolated spin-1 chain at  $T=0$  and measure equal time  spin-spin correlations   $C(\ve{r})$ 
from one edge of  the chain  to a  position $\ve{r}$ along the  chain.  For an infinite chain    $C(\ve{r})$ will decay exponentially thereby confirming the localized nature of the spin 1/2 edge  
degree of freedom.  For any finite chain length, we  expect  the zero temperature spin-spin correlations   between the two edges to be substantial, the reason being the following. In  this case,  the  Hilbert space of  the degrees of freedom below the bulk spin gap,  will be that of the two edge  spin-1/2, $\ve{\hat{S}}_{L/R}$.   The only SU(2) symmetric Hamiltonian  on this  Hilbert  space reads  $J_{eff} \ve{\hat{S}}_L  \cdot  \ve{\hat{S}}_R $.    At zero 
temperature and for any value  of  $J_{eff}$  the correlation between the two spins will be substantial. Hence even  if the two edge states are very far apart,  we  expect a upturn of the 
spin-spin correlation function $C(\ve{r})$   when $\ve{r}$ links the opposite edges of the chain. This is indeed what is observed in Figure~\ref{fig:Cr_vs_r_BetaL_sm_OPBC_L26_Jkp0}  where we plot $C(\ve{r})$ for an isolated spin-1 chain of length $L=26$.
 \begin{figure}[htbp]
 \includegraphics[clip, width=8.9cm]{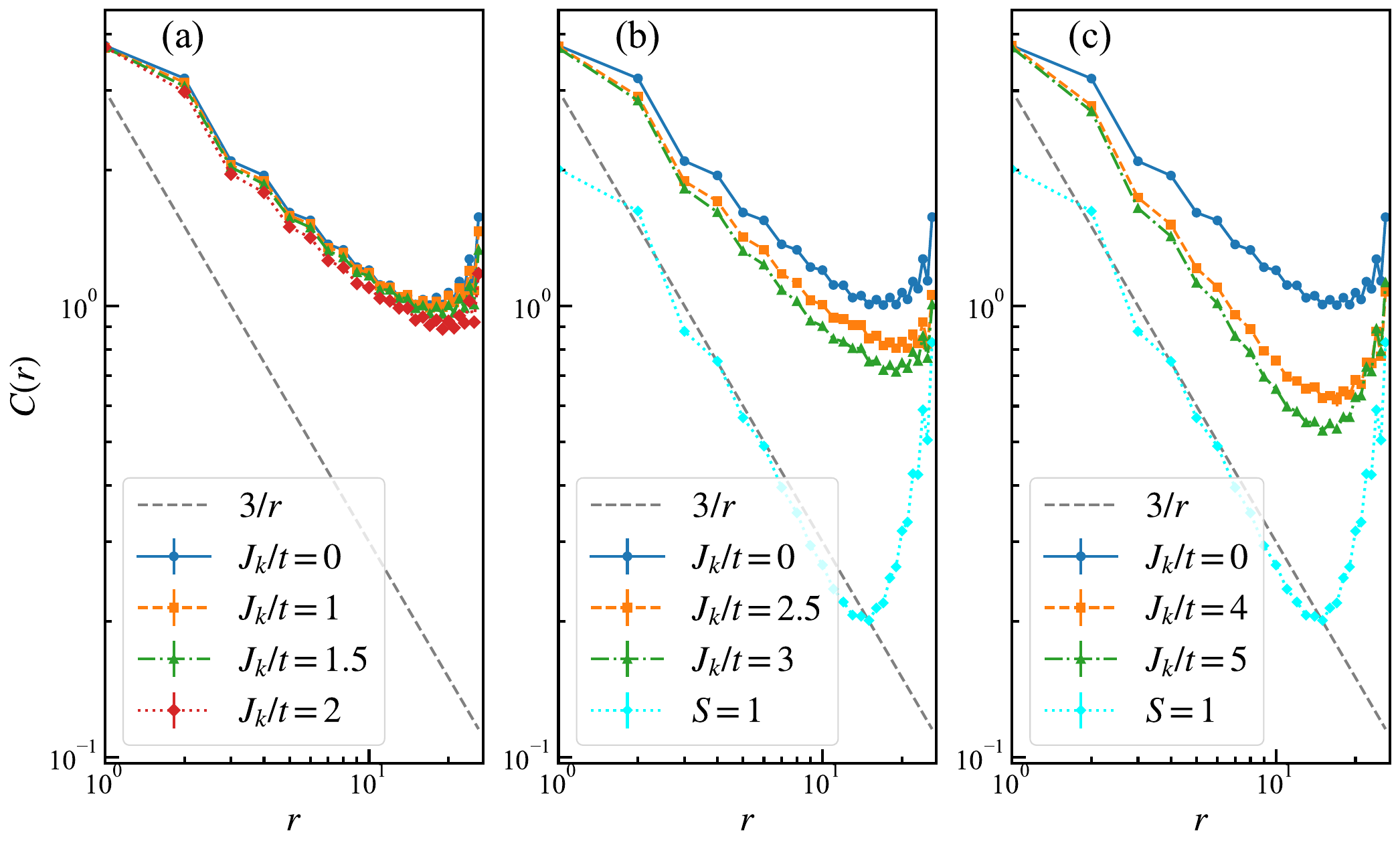}
 \caption{$S=3/2$ Heisenberg chain on a 2D semi-metal with OBC  at $L=26, T/t=0, J_h/t=1, D/t=0$.  Real space spin-spin correlations $C(r)$ as a function of distance $r$   at various $J_k/t$  along the spin  chain.  (a) KBD  phase at $J_k\le2$. (b) Kondo screening + emergent spin-1 chain at $J_k/t>2$.  (c) Residual spin-1 chain at $J_k/t>3$. The $S=1$ label corresponds to the decoupled spin-1 chain data at $L=26$. The grey dashed  line corresponds to $3/r$ 
 decay.}
 \label{fig:Cr_vs_r_BetaL_sm_OPBC_L26}	
\end{figure}

Let us now consider an isolated spin-3/2 chain. As mentioned previously,  in the long wavelength limit or infrared (IR) limit this Hamiltonian maps onto a  level-1 WZW model.    An understanding of this mapping  lies in
seeing the spin 3/2  chain in terms of a spin-1 and spin-1/2 chain.  A very similar understanding of the three leg ladder is put forward in Ref.~\cite{Rice97}.  Following this line of thought  the spin 3/2 chain should equally  possess edge states.  Importantly since the
spin-1/2 edge mode  of the spin-1 chain can  couple to the spin-1/2 degrees of freedom of the spin-1/2  chain, we do not expect it to  be localized. Numerical evidence for this picture
can be found  in Fig.~6 of Ref.~\cite{QinPRB1995},  where the  edge mode of the spin-3/2 chain is seen to be delocalized in comparison to that of the spin-1 chain. In Figure~\ref{fig:Cr_vs_r_BetaL_sm_OPBC_L26_Jkp0}  we
plot $C(\ve{r})$  for the isolated spin 3/2  chain. As apparent we see a notable increase in the spin-spin correlations at  distances corresponding to the length of the lattice.  As for the spin-1 chain, we interpret this
upturn as evidence for the presence of delocalized edge modes. Finally, Figure~\ref{fig:Cr_vs_r_BetaL_sm_OPBC_L26_Jkp0}  plots the spin-spin correlations for  the 
spin-1/2 chain. Here, we observe no upturn of the spin-spin correlation function at the largest distances.

With the above understanding of the limiting cases, we can now discuss the numerical data of the spin-3/2 chain coupled to  2D semi-metallic and metallic surfaces, see 
Figures~\ref{fig:Cr_vs_r_BetaL_sm_OPBC_L26}(a)-(c). 
In the weak-coupling regime ($J_k/t\leq 2$), Figure~\ref{fig:Cr_vs_r_BetaL_sm_OPBC_L26}(a) reveals that $C(\r)$ closely follows the correlation profile of the decoupled ($J_k/t = 0$) spin-3/2 chain. 
In this  parameter range  we are in the KBD  phase where the spin-3/2 chain 
 decouples from the semi-metal in the low energy limit.  The data suggests  that both the edge mode and the 
 bulk  of the spin-3/2  chain decoupled from the semi-metal in this phase.

In the strong $J_k/t$ limit, partial Kondo screening leads to a spin-1 chain ferromagnetically coupled to the 
semi-metallic surface. Since the ferromagnetic coupling cannot screen the spin-1/2 edge state, we expect to observe similar behaviour as for the isolated spin-1 chain. The data of Figures~\ref{fig:Cr_vs_r_BetaL_sm_OPBC_L26}(b)-(c) supports this 
point of view. However, there are important differences. While for the isolated spin-1 chain the edge states are localized, on the semi-metallic or metallic surfaces, spin-spin correlations measured from the edge will inherit the power-law decay of the host metal.
 \begin{figure}[htbp]
 \includegraphics[clip, width=8.6cm]{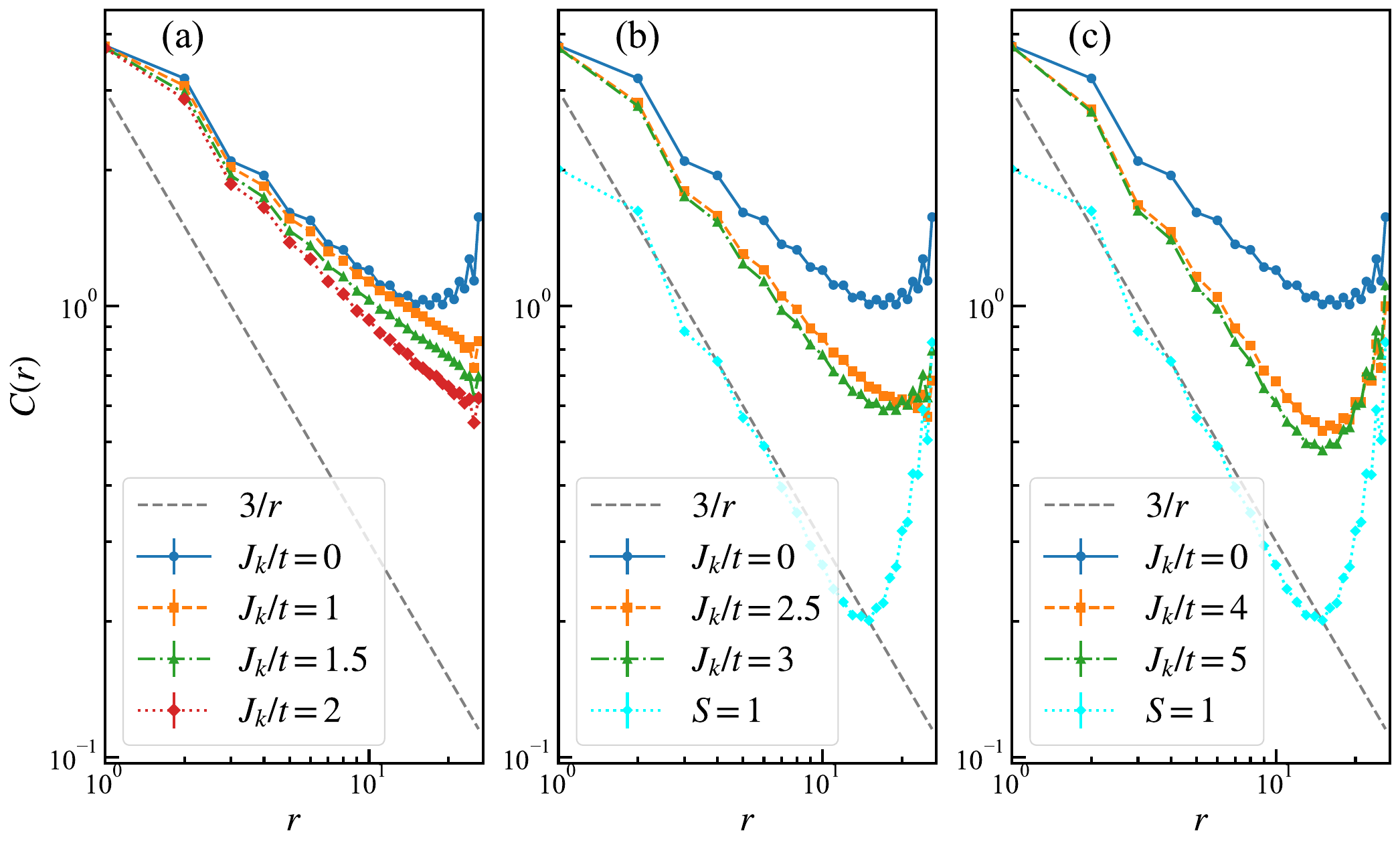}
 \caption{$S=3/2$ Heisenberg chain on a 2D metal  with OBC  at $L=26, T/t=0, J_h/t=1, D/t=0$.  Real space spin-spin correlations $C(r)$ as a function of distance $r$  at various $J_k/t$ along the spin chain.  (a) Dissipation induced AFM metallic phase at $J_k\le2$. (b) At $2\le J_k/t\le 3$.  (c) At $J_k/t\ge4$.  The $S=1$ label corresponds to decoupled Haldane $S=1$ chain  data at $L=26$.  The grey dashed  line corresponds to $3/r$ decay.}
 \label{fig:Cr_vs_r_BetaL_m_OPBC_L26}	
\end{figure} 

A qualitatively different behaviour is observed  for the  spin-3/2 chain on a metallic surface at weak coupling, see Figures~\ref{fig:Cr_vs_r_BetaL_m_OPBC_L26}(a)-(c). For $J_k/t \leq 2$, in the dissipation-induced AFM phase, Figure~\ref{fig:Cr_vs_r_BetaL_m_OPBC_L26}(a) shows no upturn at the largest distances. We interpret this in terms of the absence of edge modes in this phase.  As mentioned previously the metallic 
surface is  a marginally relevant  perturbation such that very large lattice sizes and low temperatures are
required to observe the long ranged magnetic ordering at \textit{small}  values of $J_k$. This provides an understanding of the apparent power-law decay. 
 In the intermediate regime ($J_k/t >2$), Figure~\ref{fig:Cr_vs_r_BetaL_m_OPBC_L26}(b) reveals the emergence of significant edge correlations characteristic of a residual spin-1 chain. At even stronger coupling ($J_k/t>3$), Figure~\ref{fig:Cr_vs_r_BetaL_m_OPBC_L26}(c) shows that the correlation profile closely follows the characteristic  bulk correlations and the emergent edge modes of a  finite spin-1 Heisenberg chain. In the strong coupling limit, the nature 
 of the metallic surface does not play a dominant role.

To further to confirm the evolution of  massless  spin-1/2 edge modes in strong Kondo coupling limit, we compute the local  spin-susceptibility at the edges of spin-3/2 Heisenberg chain with  OBC. Consider  the  Hamiltonian $\hat{H}(h) =  \hat{H}  + h \left( \sum_{l \in  \{1, L\}} \hat{S}^{z}_{l} \right) $, where $\hat{H}$ is given by Eq.~(\ref{ham_1}), and $h$ is a local magnetic field applied only at the edges of the chain. The local edge spin-susceptibility  at $D=0$  where SU(2) 
spin symmetry is present reads: 
  \begin{eqnarray}
\chi_e  =  & & - 3 \left. \frac{ \partial^2 F(h) }{\partial h^2  } \right|_{h=0}   \nonumber   \\ 
        =  & & \int^\beta_0  d \tau \sum_{l, l^\prime \in \{1, L\}} \Big\langle  {\S}_l(\tau)\cdot {\S_{l^\prime}}(0) \Big\rangle.
\end{eqnarray}
\begin{figure}[htbp]
  \includegraphics[clip, width=8.9cm]{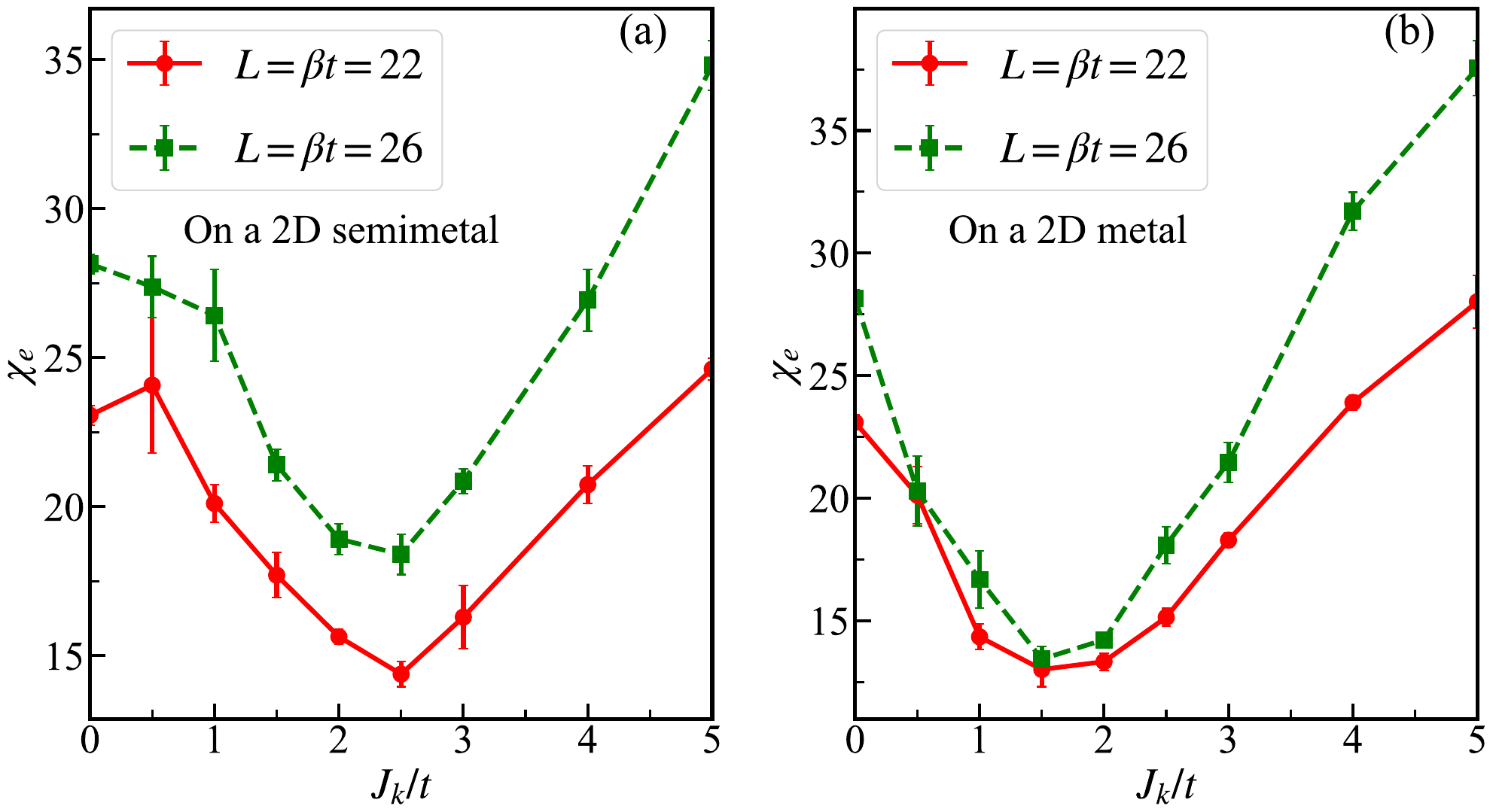}
 \caption{Local edge spin-susceptibility $\chi_e$ as a function of $J_k/t$ along the  spin chain  with OBC at $J_h/t=1, D/t=0, \beta t =L$. (a)  $S=3/2$  Heisenberg chain on a 2D semi-metal. (b)  The same on a 2D metal.}%
 \label{fig:endchiJk_vs_Jk_BetaL_sm_m}
\end{figure}
Figures \ref{fig:endchiJk_vs_Jk_BetaL_sm_m}(a)-(b) show $\chi_e$ as a function of $J_k/t$ on a 2D semi-metal and a 2D metal, respectively.  
As a function of the coupling constant $J_k/t$ and for both metallic surfaces, $\chi_e$ shows a dip behaviour, the dip being located in the 
vicinity of the quantum critical point. For small values of $J_k$, conduction electron spins align antiferromagnetically with the impurity spins, such that the net local magnetic field is reduced as $J_k$ increases. On the other hand, for large values of $J_k$, where partial Kondo screening reduces the spin-3/2 to a spin-1 degree of freedom, the coupling between the spin-1 and conduction electrons is ferromagnetic. This leads to an 
enhancement of the local magnetic field, and hence to an increase in the local spin-susceptibility. Therefore, the observed dip is a 
consequence of the change of sign of the magnetic coupling between effective local moments and conduction electrons.
 \begin{figure}[htbp]

 \includegraphics[clip, width=9.cm]{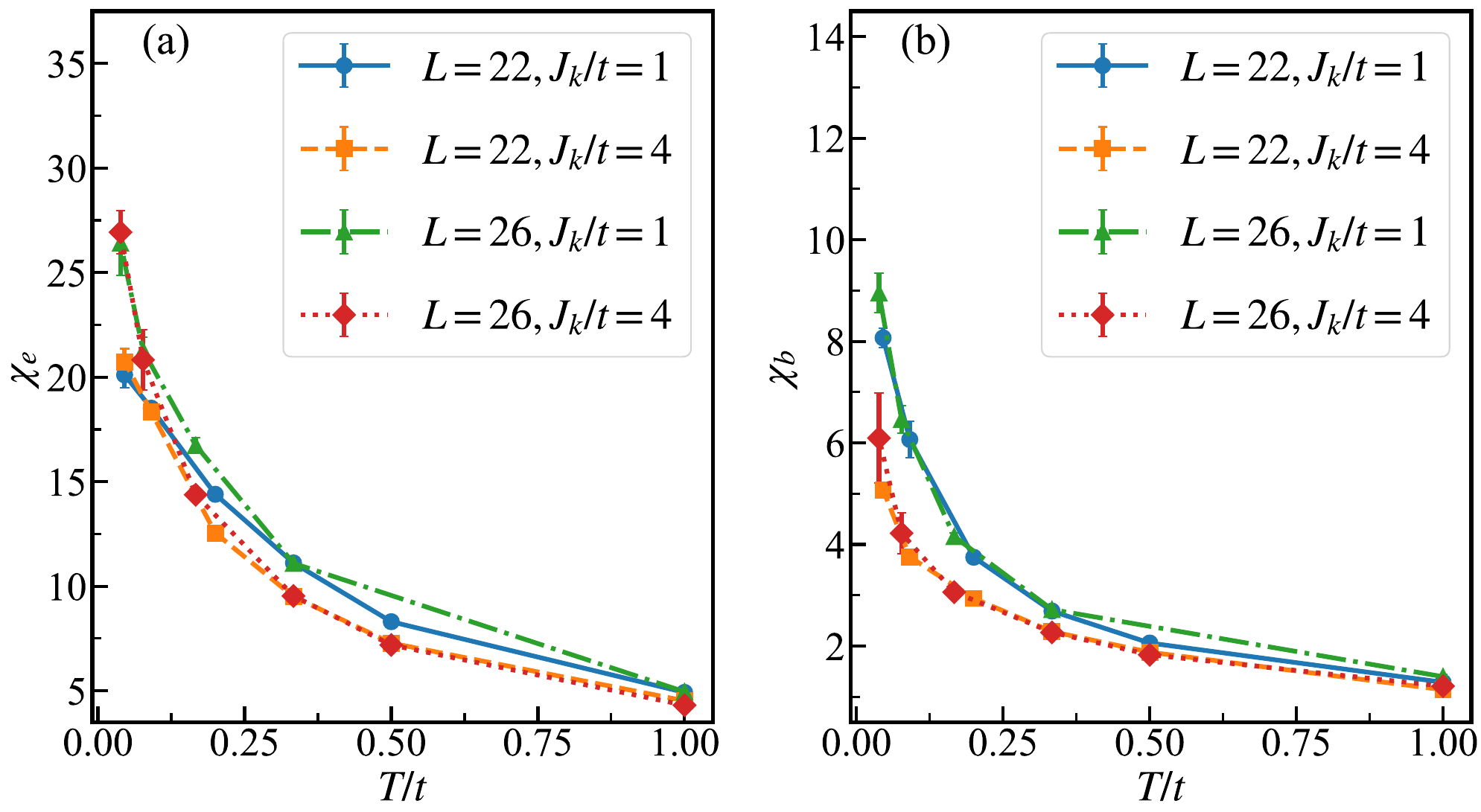}
 \caption{(a)  Local edge spin-susceptibility $\chi_e$ as a function of temperature ($T/t$) along the spin-3/2  Heisenberg  chain 2D semi-metal with OBC at $J_h/t=1, D/t=0$. (b) Local bulk spin-susceptibility  $\chi_b$ as a function of $T/t$ for same parameter specification as (a).  Here, $J_k/t=1$ corresponds to  the KBD phase and $J_k/t=4$ corresponds to the Kondo phase.}%
 \label{fig:chi_vs_T_sm}

~

  \includegraphics[clip, width=8.9cm]{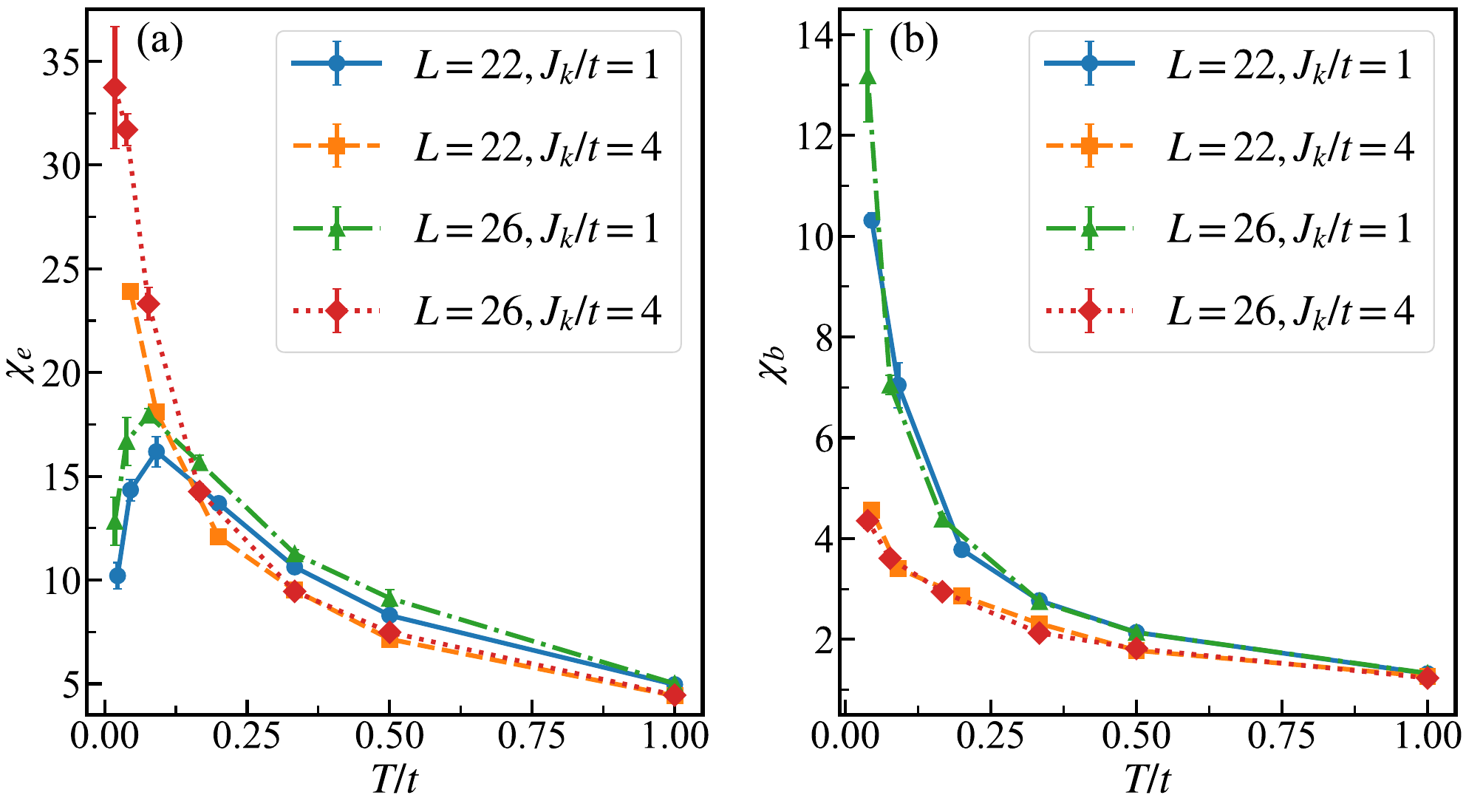}
 \caption{(a) Local edge spin-susceptibility $\chi_e$ as a function of $T/t$ along the spin-3/2  Heisenberg  chain 2D metal with OBC at $J_h/t=1, D/t=0$. (b) Local bulk spin-susceptibility  $\chi_b$ as a function of  $T/t$ for same parameter specification as (a). Here, $J_k/t=1$ corresponds to the dissipation induced AFM phase and $J_k/t=4$ corresponds to the Kondo phase.}
 \label{fig:chi_vs_T_m}
\end{figure}
We now concentrate on the temperature dependence. In the strong coupling limit, and for both metallic surfaces, we expect edge states of the spin-1
chain to show up in terms of a Curie law in $\chi_e$. A marked enhancement of $\chi_e$ as a function of decreasing temperature in this parameter range is
indeed observed in the AFQMC data, see Figures~\ref{fig:endchiJk_vs_Jk_BetaL_sm_m}-\ref{fig:chi_vs_T_m}.  For the  open boundary conditions  simulations 
the Manhattan distance of our  chain is odd, see Figure~\ref{fig:Sketch_opbc_latt}(b),  such that at a temperature lower then the ones considered in  the 
simulations we expect a  downturn of  $\chi_e$    due  to antiferromagnetic coupling between edges.  

At weak coupling, the AFQMC results are very much surface dependent. For the semi-metallic surface, $\chi_e$ grows substantially.  This corresponds to the 
KBD  phase, where we expect an isolated spin-3/2 chain in the IR limit,  and hence \textit{delocalized} edge modes.  In contrast on the metallic surface, we 
observe  dissipation induced long ranged AFM order.   In Figure~\ref{fig:chi_vs_T_m}(a) the data shows a downturn as a function of decreasing temperature. We interpret this 
in terms of the onset of antiferromagnetic correlations between the two edges of the chain.

We  conclude our study of by  considering the bulk local susceptibility. 
  \begin{eqnarray}
\chi_b  =  \int^\beta_0  d \tau  \Big\langle  {\S}_{L/2}(0)\cdot {\S_{L/2}}(\tau) \Big\rangle.
\end{eqnarray}
This quantity is depicted in Figures~\ref{fig:chi_vs_T_sm}-\ref{fig:chi_vs_T_m} on 2D semi-metals and metals.
At strong  coupling  and for both surfaces  this quantity is strongly suppressed with respect to the  edge susceptibility.  We note that on finite lattices, 
there will always an overlap between the edge  and local bulk modes. This provides an understanding of the upturn in the data as a function of lowering temperature. 

At weak couplings  we  observe that $\chi_b$ is much enhanced on the metallic surface in comparison to the semi-metallic one. On  the semi-metallic surface and in
the  KBD phase,  local spin-spin correlations decay as  $1/\tau$ such that a log divergence,  $\log(1/T)$ is expected. On the other hand, on the metallic  surface,
the  local spin-spin correlations are  expected to a $1/\sqrt{\tau}$  law akin to dissipation induced long ranged order  thereby leading to a $1/\sqrt{T}$ divergence of $\chi_b$.

\subsection{Influence of single ion anisotropy $D\ne0$}
The presence of single-ion anisotropy breaks SU(2) spin symmetry down to U(1). In the large-$|D|$ limit, a perturbative analysis of the spin-3/2 chain maps the Hamiltonian in Eq.(\ref{ham_1}) to an effective spin-1/2 XXZ chain coupled to a 2D electrons system (see Section \ref{perturb_cal}). Specifically,  for $D>0$, the effective low energy  description is given by Eq.(\ref{model_efham_PD}),  whereas for $D<0$,  it corresponds to Eq.(\ref{model_ham_nxz_ND}).  

Another important limiting case is $J_k/t\rightarrow \infty$, where the system reduces to an effective residual spin-1 chain with single-ion anisotropy coupled to the 2D electronic bath, as described by Eq.~(\ref{largeJkef}). In this limit, the resulting residual spin-1 chain exhibits massless spin-1/2 edge states that are expected to remain stable against weak positive or negative values of the anisotropy parameter $D$. 

The  case of positive $D>0$  with  $J^\perp_h \simeq4 J^z_h$ is relevant for chain of Co adatoms on a Cu$_2$N/Cu(100) surface~\cite{Otte2008, Spinelli2015, Toskovic2016}. In this regime, for $J^\perp_h > J^z_h$,  at the decoupled fixed point, the Kondo coupling  is irreverent (relevant) on a semi metallic (metallic) surface for transverse  spin-spin correlation decays, and remains irrelevant for the longitudinal spin-spin correlations of the  decoupled spin-1/2 XXZ chain. For  negative  $D<0$, recent experimental studies have reported the emergence of novel spinaron quasiparticle  in  setups involving  single spin-3/2 Co adatom on Cu(111) and Au(111) surfaces~\cite{Bouaziz2020, Friedrich2024}.  

To incorporate realistic experimental parameters relevant to Co adatoms on Cu$_2$N/Cu(100) surface, the Heisenberg interaction $J_h=0.24$ meV and  the anisotropy energy  $D=2.25$ meV and the Kondo temperature of a single Co adatom $T^\text{Co}_{K}=2.6 {\text K}\approx0.22$ meV~\cite{Spinelli2015, Toskovic2016}.  On the other hand, Co adatoms  Ag(111) and Cu(111) surfaces carry a large spin magnetic moment of 2 $\mu_B$ and an orbital moment of around $0.47 \mu_B$.  These moments tend to point out of the surface plane, a direction favored by a significant magnetic anisotropy energy of approximately  4.3meV~\cite{Bouaziz2020, Friedrich2024}. Based on large-$|D|$ associated with Co adatoms, we present numerical results for the system as a function of $D/t\in[-4, 4]$ for various $J_k/t$ values. 
 \begin{figure}[htbp]
  \includegraphics[clip, width=8.9cm]{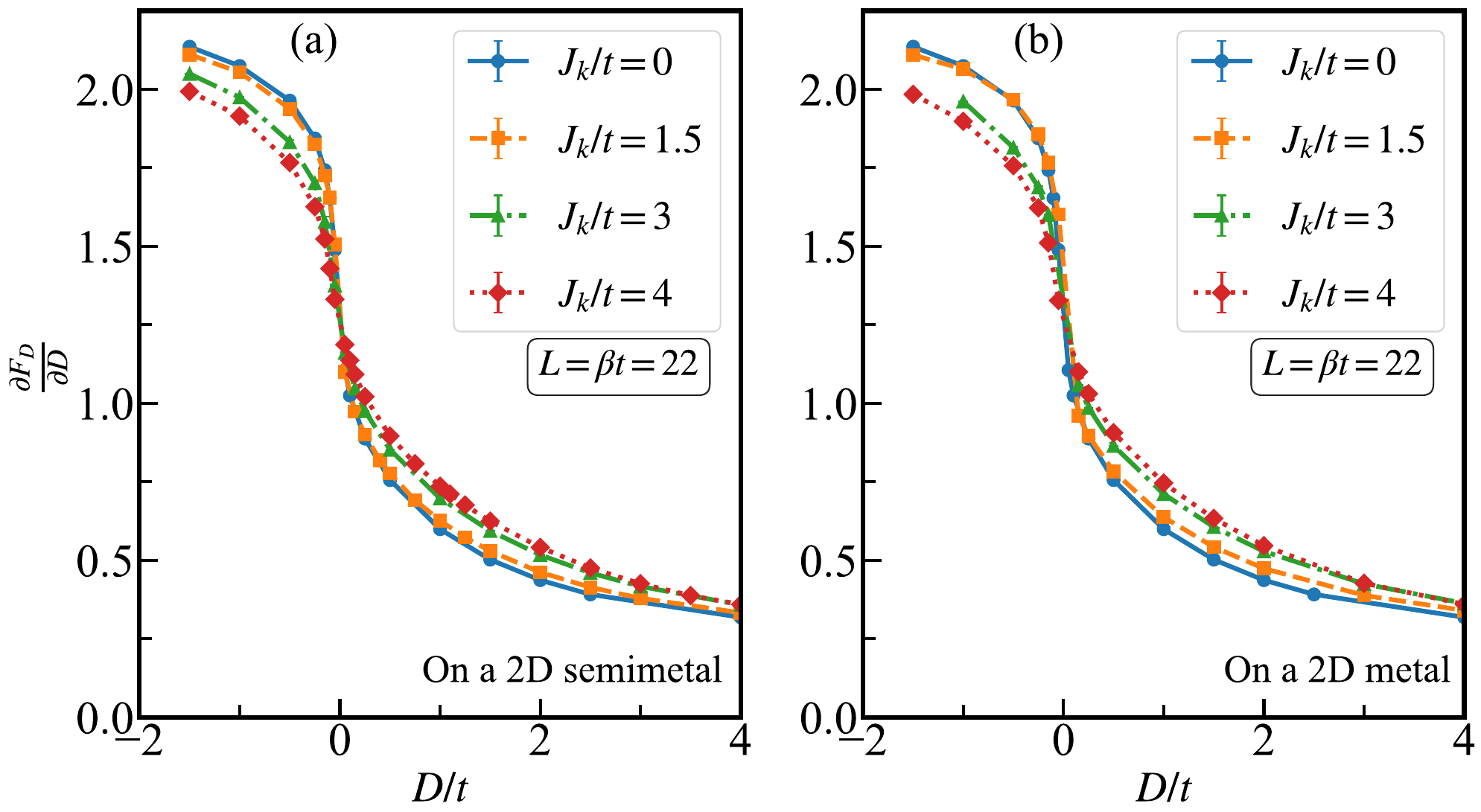}
 \caption{ First derivative  $\partial F_D/\partial D$  versus $D/t$  along the spin-3/2 chain  with PBC  at  $J_h/t=1, L=\beta t =22$. (a) On a  2D semi-metal. (b) On a  2D metal.}  %
 \label{fig:dFbydD_vs_D_m}	
\end{figure}

To investigate whether the system undergoes a QPT  as a function of 
$D/t$ at fixed $J_k/t$, we compute the first derivative of the free energy with respect to $D$ defined as:
\begin{eqnarray}
(1/L){\partial F_D}/ {\partial D}= (1/L)  \sum_{r}  \Big\langle \big({\hat S}^z_{ r}\big) ^2 \Big\rangle.
\end{eqnarray}
 Figures~\ref{fig:dFbydD_vs_D_m}(a)-(b) show  $\partial F_D/\partial D$ as a function of $D/t$ for spin-3/2 chain with PBC coupled to  2D semi-metallic and metallic surfaces.  This quantity is bounded by $S^2$  with $S=3/2$  in the limit $D/t \rightarrow -\infty$ and $S=0$ in the limit $D/t \rightarrow +\infty$. 
For the isolated $S=3/2$ chain, we expect the transtion from the Ising ($D<0$)  to XY ($D>0$) to be in the same universality class as for the  spin-1/2 XXZ model,  the Berezinskii-Kosterlitz-Thouless (BKT) universality class \cite{giamarchi2003}. While the Ising order parameter shows a jump at the transition, $\partial F_D/\partial D$  does not diverge.  In Figure~\ref{fig:dFbydD_vs_D_m}(a) we observe that for the semi-metal, $\partial F_D/\partial D$ show very similar behaviours at $J_k=0$ (isolated spin-3/2 chain)  and in the Kondo breakdown phase at $J_k/t=1.5$.  
In contrast for the metallic-surface at $J_k/t=1.5$ we  observe that the slope at $D=0$ is enhance in comparison the isolated chain. This is consistent  with 
the notion put forward in Sec.~\ref{sec:Jk_small} of a spin-flop transition associated with the dissipation induced long-range order. Note that the dissipation induced 
ordered phase is characterised by a $z=2$ dynamical exponent such that our temperature scaling $L=\beta t$  corresponds to \textit{high} temperatures. We hence expect 
the observed \textit{jump} to increase as the   temperature is lowered.  At strong coupling, $J_k/t=4$,  the data for both surfaces  shows a smooth behaviour across $D=0$ consistent with the presence of a spin-1 Haldane phase. Within our resolution, we do not observe significant features away from  $D=0$ corresponding to  transitions to the large-$D$ or Ising phases (see Sec.~\ref{RG_analysis}).

 To further investigate phase transitions as a function of $D$ physics, we compute space- and time-displaced transverse and longitudinal spin-spin correlations along the spin-3/2 chain with PBC  on both  2D semi-metallic and metallic hosts  for  few $J_k/t$ values representing the weak and strong Kondo coupling  phases.
  The transverse component of  the spin structure factor computed  as: $S^{\perp}(\ve{k})  = \frac {1}{L} \sum_r e^{-i \ve{k}\cdot \ve{r}} \big\langle   \big({\hat S}^+ (\ve{r}) {\hat S}^-(0)  +\text{H.c.} \big)  \big\rangle$,
and the corresponding spin susceptibility computes as:
\begin{eqnarray}
\chi^{\perp}(\ve{k})= \int^\beta_0  d\tau  \sum_{\ve{r}} e^{-i \ve{k}\cdot \ve{r}} \Big\langle  \big({\hat S}^+_{\ve{r}} (\tau) {\hat S}^-_0(0)  +\text{H.c.} \big) \Big\rangle.
\label{Sktauperp}
\end{eqnarray}
The longitudinal components are defined similarly via, $ S^{zz}(\ve{k}) =\frac{1}{L} \sum_{\ve{r}} e^{-i \ve{k}\cdot \ve{r} } \big\langle  {\hat S}^z(\ve{r}) {\hat S}^z(0) \big\rangle$, and:
\begin{eqnarray}
\chi^{zz}(\ve{k})= \int^\beta_0  d\tau  \sum_{\ve{r}} e^{-i \ve{k}\cdot \ve{r}} \left<  {\hat S}^z_{\ve{r}} (\tau)  {\hat S}^z_0(0)   \right>.
\label{Sktauzz}
\end{eqnarray}
Note that we have omitted the background term, $\langle {\hat S}^z(\ve{r})(\tau)\rangle \langle {\hat S}^z(0)\rangle$, in the definition of $\chi^{zz}(\ve{k})$ since it vanishes on finite lattices.
\begin{figure}[htbp]
 \includegraphics[clip, width=8.9cm]{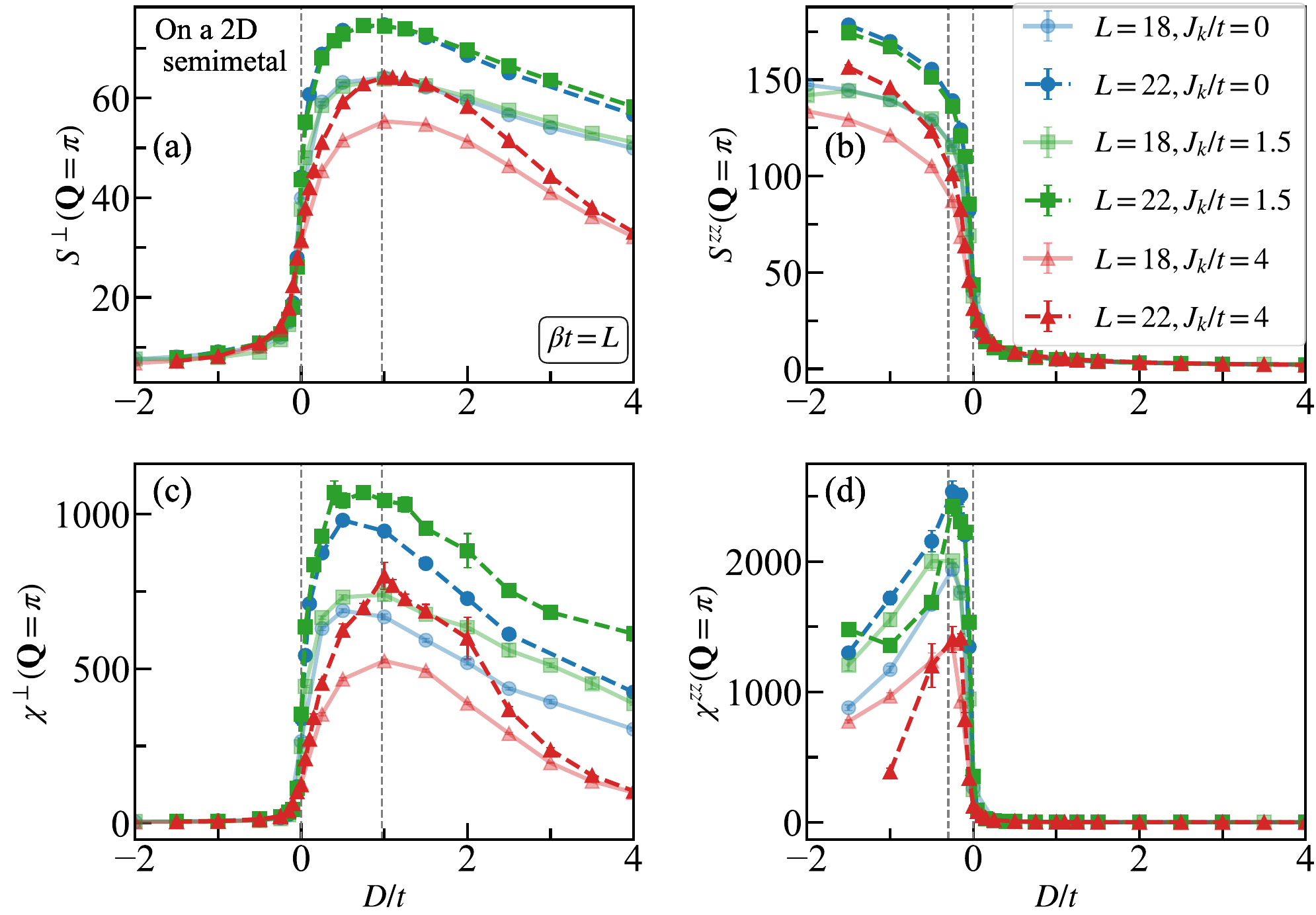}  
  \caption{Spin-3/2 chain on a 2D semi-metal with PBC for various $J_k/t$ at $J_h/t=1$ and $L=\beta t=18,22$. (a) Transverse  $S^{\perp}(\Q=\pi)$ and (b) longitudinal  $S^{zz}(\Q=\pi)$ spin structure factors  as a function of $D/t$.
  (c) Transverse $\chi^{\perp}(\Q=\pi)$) and (d) longitudinal $\chi^{zz}(\Q=\pi)$ spin susceptibilities along the spin chain  as a function of $D/t$. The grey  vertical dashed lines mark  critical points $D^+_c=0.97$ (Gaussian) and   $D^-_c=-0.3$ (Ising),  corresponding  to a  decoupled $S=1$ chain, and $D_c=0$  correspond to the  XY to Ising critical point of the $S=3/2$ chain.}
    \label{fig:Skpi_vs_D_L18_2dsm_Jkscan_Dscan}
\end{figure}
\begin{figure}[htbp]
 \includegraphics[clip, width=8.9cm]{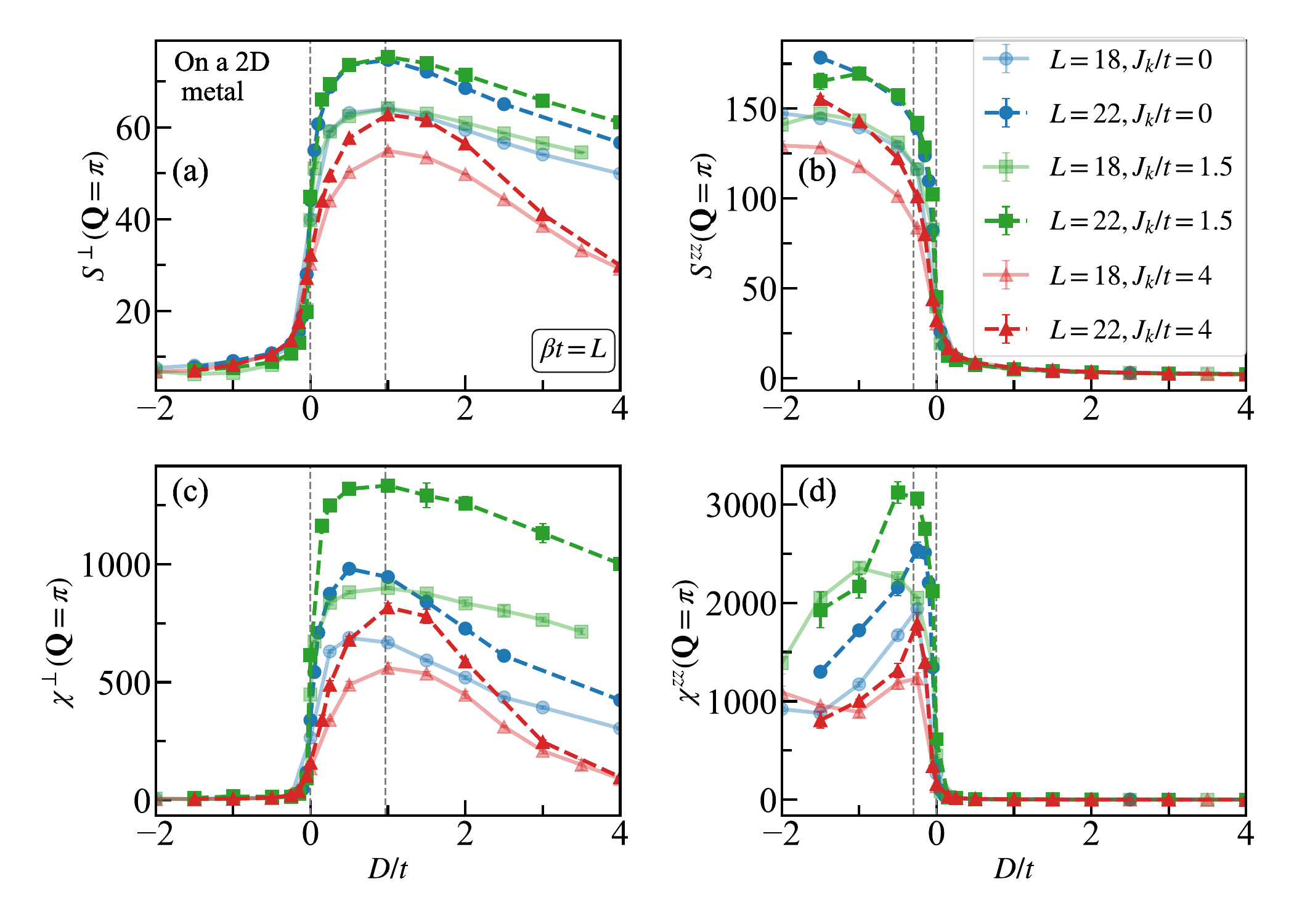}
    \caption{Spin-3/2 chain on a 2D metal with PBC for various $J_k/t$ at $J_h/t=1$ and $L=\beta t=18,22$. (a) Transverse  $S^{\perp}(\Q=\pi)$ and (b) longitudinal  $S^{zz}(\Q=\pi)$  and spin structure factors   and (c) Transverse $\chi^{\perp}(\Q=\pi)$) and (d) longitudinal $\chi^{zz}(\Q=\pi)$ spin susceptibilities along the spin chain  as a function of $D/t$. The grey vertical dashed lines are specified in Figure~\ref{fig:Skpi_vs_D_L18_2dsm_Jkscan_Dscan}.}
    \label{fig:Skpi_vs_D_L18_L22_2dm_Jkscan_Dscan}
\end{figure}

Figures~\ref{fig:Skpi_vs_D_L18_2dsm_Jkscan_Dscan}(a)-(b)  and (c)-(d) respectively present the transverse and longitudinal spin structure factors $S^{\perp,zz}(\Q)$ and dynamical spin susceptibilities $\chi^{\perp,zz}(\Q)$ at the antiferromagnetic wave vector $\Q=\pi$ along spin-3/2 chain on a 2D semi-metal as a function of $D/t$ for  various  $J_k/t$ values. 
As is apparent from the data, at intermediate coupling ($J_k/t=1.5$), which corresponds to the KBD phase, the results at small $D/t$ closely follow those of the decoupled spin-3/2 Heisenberg chain at $J_k/t=0$, confirming, as discussed in Sec.~\ref{sec:Jk_small} that in  KBD phase remains stable against weak single-ion anisotropy. That is, even at finite $D$ spin and conduction electrons decouple.  As argued previously we see the $S=3/2$  chain to be  described by the combination of  a $S=1$  and $S=1/2$ chain. Following this point of view we  expect  a transition
between the small and large $D$ limits akin to the spin-1 chain.   We ascribe the  peak and large increase in the  transverse (longitudinal) spin susceptibility  as a function of positive (negative)
values of $D$ for  $J_k = 0, 1.5$  to this transition.  In the strong coupling limit (here $J_k = 4$)  partial Kondo screening leads to a spin-1  chain ferromagnetically coupled to the conduction electrons. The peaks observed  again in 
both the longitudinal and transverse susceptibilities  as a function of $D$  are then associated with the transition between the Haldane and large-$D$ phase (for $D>0$) and the Ising phase (for $D<0$) of the spin-1 chain. We note that
the size effects are strong  at the transition points.  In the large positive $D$ limit, and at small values of $J_k$, the system maps 
onto a spin-1/2 XXZ in the XY phase. The growth of the transverse spin susceptibility as a function of increasing lattice size is consistent with this point of view.  At $J_k=4$  and still at large values of $D$ the transverse spin susceptibility  saturates with increasing lattice size. We understand this as a signature of the  large-$D$ and Kondo phases. 
Finally, in the large negative $D$ limit, the system maps onto an Ising chain, where fluctuations are suppressed. In this case, the longitudinal 
spin susceptibility shows a saturation as a function of increasing lattice size and is suppressed as a function of decreasing values of $D$. We understand the large error bars in this phase to be a
consequence of the long autocorrelation times associated with restoring Ising symmetry.

For comparison, Figures~\ref{fig:Skpi_vs_D_L18_L22_2dm_Jkscan_Dscan} (a)-(d) present AFQMC results for $S^{\perp,zz}(\Q=\pi)$ and $\chi^{\perp,zz}(\Q=\pi)$ on a 2D metal both weak and strong Kondo coupling values. 
In this case, at weak coupling we observe  dissipation induced long ranged order and as a consequence (see Sec.~\ref{sec:Jk_small}) a fist order spin flop transition. The AFQMC data at $J_k = 1.5$ support this.  In comparison the the $J_k = 0 $ we observe 
a very strong  increase  of the transverse ($D>0$) and longitudinal ($D<0$) spin susceptibilities at weak values of  the single ion anisotropy $D$.   In the strong coupling limit, we observer very similar results as for the semi-metallic surface.  
In Sec.~\ref{sec:Jk_big}  we have argued that the metallic surface is a relevant perturbation to the single ion anisotropy induced phase transition in the $S=1$ Haldane chain.   At the temperature and size scales  considered here, the nature of the 
metallic surface does not seem to play a role.  Further investigations will be  required  to clarify this point.  

\section{Summary and outlook}\label{summary_outlook}
We have presented a theoretical framework, supported by exact numerical simulations,  for a spin-3/2 chain of Co adatoms placed on 2D metallic and semi-metallic surfaces.  Our results are summarized in the phase 
diagrams of Figures~\ref{fig:RG_cartoon_sm} and~\ref{fig:RG_cartoon_m},  and demonstrate the richness  of the physics realized in these \textit{ simple } systems.

Our work provides a concrete roadmap and motivation for designing quantum simulator architectures based on magnetic adatoms on metallic and semi-metallic surfaces. Our calculations open a promising direction of study for emergent phases, order-disorder transition in metals, Kondo breakdown, topological states, and dissipation-induced phenomena. We hope that this will motivate experimental studies of Kondo nanostructures on metallic surfaces arranged in a circular geometry so as to avoid boundary effects. Our simulations involve a maximum of $L=38$ ($L=26$) adatoms on semimetallic (metallic) surfaces. Already on these \textit{small} system sizes we can observe clear signs of the phases depicted in Figures~\ref{fig:RG_cartoon_sm} and~\ref{fig:RG_cartoon_m}.

As a concrete example, our investigation of a spin-3/2 chain on a 2D metal is directly relevant to STM experiments of Co adatoms chains on a Cu$_2$N/Cu(100) surface. STM technologies offer remarkable control over tuning the Kondo coupling. For instance, Co adatoms on Cu(100) substrate are coupled via several  Cu$_2$N monolayers, which effectively suppresses the Kondo interaction--providing an ideal platform to access the physics discuss in this paper.  Thus the  setup presents a promising route to realise massless spin-1/2 edge modes by continuously tuning the Kondo interactions.

Similarly, our study of the spin-3/2 chain on a 2D semi-metal motivates adatom-based experiments on graphene, where the interplay between Dirac fermions and local magnetic moments could enable exploration of KBD physics coexisting with topological spin-1/2 edge states. These investigations may pave the way for discovering new classes of surface-bound quantum states and exotic correlated behaviour in dimensionality mismatch Kondo systems.

A comprehensive understanding of the full phase diagram in the $D/t$ versus $J_k/t$ plane, from both theoretical and numerical perspectives, remains an important direction for future research. In particular, it would be interesting to explore how the quantum critical behaviour on metallic and semi-metallic surfaces evolves with increasing spin magnitude $S$ and under generalized SU($N$) symmetries. The other important question which we  believed  is interesting is to explore the underlying physics of spin-1 dimensional mismatch Kondo-Heisenberg systems.

\section{Acknowledgments}  
We thank T. Grover, F. Mila, M. Vojta for  work on related subjects and M. Raczkowski, S. Biswas, and G. Pan for fruitful discussions.
The authors gratefully acknowledge the Gauss Centre for Supercomputing e.V. (www.gauss-centre.eu) for funding this project by providing computing time on the GCS Supercomputer SUPERMUC-NG at Leibniz Supercomputing Centre
(www.lrz.de) as well as through the John von Neumann Institute for Computing (NIC) on the GCS Supercomputer JUWELS  at the  J\"ulich Supercomputing Centre (JSC).  We  also gratefully acknowledge the scientific support and HPC resources provided by  the Erlangen National High Performance Computing Center (NHR@FAU) of the Friedrich-Alexander-Universität Erlangen-Nürnberg (FAU) under NHR project 80069 provided by federal and Bavarian state authorities. NHR@FAU hardware is partially funded by the German Research Foundation (DFG) through grant 440719683.  B.D. acknowledges financial support from the German Research Foundation (DFG) under the grant DA 2805/2 (Project number 528834426).  B.D.  also acknowledges  travel support  from the ct.qmat for  the Conference on ``Fractionalization and Emergent Gauge Fields in Quantum Matter (smr 3834)" at ICTP. F.F.A. acknowledges financial support from the German Research Foundation (DFG) under the grant AS 120/16-1 (Project number 493886309) that is part of the collaborative research project SFB QMS funded by the Austrian Science Fund (FWF) F 86. 
\bibliography{Ref_main}
\clearpage
 \end{document}